\documentclass[%
 reprint,
 amsmath,amssymb,
 aps,
]{revtex4-2}

\usepackage{graphicx}
\usepackage[utf8]{inputenc}
\usepackage{setspace}
\usepackage{enumitem}
\usepackage{latexsym,amsfonts}
\usepackage{etoolbox}
\usepackage{amsmath}
\usepackage{amssymb}
\usepackage[version=4]{mhchem}
\usepackage{caption}
\usepackage{subcaption}
\usepackage{xr}

\usepackage{graphicx}
\usepackage{dcolumn}
\usepackage{bm}
\usepackage[usenames,dvipsnames]{color}
\usepackage{stmaryrd} 

\usepackage{cancel}
\usepackage[normalem]{ulem}
\usepackage{filecontents}

\begin{document}
\setlength\parindent{0pt}

\preprint{AIP/123-QED}

\title[]{Thermal Isomerization Rates in Retinal Analogues using Ab-Initio Molecular Dynamics}%
\author{Simon Ghysbrecht}
\affiliation{ 
Freie Universität Berlin, Department of Biology, Chemistry and Pharmacy, Arnimallee 22, 14195 Berlin
}

\author{Bettina G.~Keller}%
\email{bettina.keller@fu-berlin.de}
\affiliation{ 
Freie Universität Berlin, Department of Biology, Chemistry and Pharmacy, Arnimallee 22, 14195 Berlin
}%

\date{\today}

%
%
\begin{abstract}
For a detailed understanding of chemical processes in nature and industry, we need accurate models of chemical reactions in complex environments.
While Eyring transition state theory is commonly used for modeling chemical reactions, it is most accurate for small molecules in the gas phase.
A wide range of alternative rate theories exist that can better capture reactions involving complex molecules and environmental effects. 
However, they require that the chemical reaction is sampled by molecular dynamics simulations. 
This is a formidable challenge since the accessible simulation timescales are many orders of magnitude smaller than typical timescales of chemical reactions.
To overcome these limitations, rare event methods involving enhanced molecular dynamics sampling are employed.
In this work, thermal isomerization of retinal is studied using tight-binding density functional theory.
Results from transition state theory are compared to those obtained from enhanced sampling.
Rates obtained from dynamical reweighting using infrequent metadynamics simulations were in close agreement with those from transition state theory. 
Meanwhile, rates obtained from application of Kramers' rate equation to a sampled free energy profile along a torsional dihedral reaction coordinate were found to be up to three orders of magnitude higher.
This discrepancy raises concerns about applying rate methods to one-dimensional reaction coordinates in chemical reactions.
\end{abstract}

\maketitle

\section{Introduction}
\label{chapter:Introduction}
Precises models of chemical reactions, encompassing reaction mechanisms and precise rate constants, are critical for a nuanced understanding of reactions occurring in nature, laboratory experiments, and industrial processes.
Yet, computational models of chemical reactions remain challenging, because the transition state of a reaction has to be calculated using a highly accurate model of the Born-Oppenheimer potential energy surface (PES).
This usually involves evaluating the electronic structure and explicitly calculating the electronic energy at each nuclear configuration. 
Thus, the current standard to calculate reaction rate constants remains Eyring transition state theory (TST) \cite{eyring1935activated}, which requires only calculations at a few select points along the PES.
However, Eyring TST makes strong assumptions for the shape of the PES and the dynamics on this PES and is therefore limited to small to medium sized systems in the gas phase. 
Eyring TST defines the transition state as a saddle point on the PES, assumes an equilibrium between reactant state $A$ and transition state $TS$ and models the PES at $A$ and $TS$ by a harmonic approximation \cite{peters2017reaction}. 
The dynamics of the nuclei are treated quantum mechanically.
The most prominent error source is the accuracy of the energy barrier, which enters the equation for the rate exponentially.
But the assumptions are easily violated if
($i$) the saddle point of the PES does not coincide with the bottleneck of the reaction (i.e.~the free energy maximum along the optimal reaction coordinate), 
($ii$) the PES at $A$ or at $TS$ is anharmonic,
($iii$) the reaction coordinate has a strong curvature in the configurational space, or
($iv$) the reactant state exhibits multiple minima.
These violations occur in systems with many degrees of freedom, in particular if these degrees of freedom are very mobile.
Then the reactant state may comprise multiple molecular conformations and various vibrational modes may couple to the reactive vibrational mode.
The situation is further complicated if the reaction occurs in solution or if the reactants otherwise strongly interacts with their environment, e.g.~in a catalysed reaction. 
To model reactions for these systems, one shifts to a classical description of the nuclear dynamics and samples the reaction using molecular dynamics (MD) simulations \cite{frenkel2023understanding}.
The simplest estimator for a reaction rate constant from a MD simulation is to count the number of transition from reactant state $A$ to product state $B$.
However, since the accessible simulation times are orders of magnitude smaller than the mean first passage times even of very fast reactions, one uses enhanced sampling protocols to increase the statistics.
A wide variety of methods to recover accurate rate constants and mechanisms from these sped-up simulations have been proposed \cite{hanggi1990reaction,peters2017reaction}. 
They broadly fall into two categories: 
($i$) dynamical reweighting methods \cite{kieninger2020dynamical} sample the reaction on a biased PES and reweight the transition count, 
($ii$) reaction coordinate based methods define a one-dimensional reaction coordinate $s$ \cite{peters2016reaction} and calculate the rate constant from an effective dynamics on this reaction coordinate. 
Besides the definition of a reaction coordinate, the second approach involves the calculation of the free energy surface (FES) \cite{henin2022enhanced} and diffusion profile \cite{hummer2005position} via MD simulations. 
Kramers' rate theory \cite{kramers1940brownian} is the most prominent example for this second approach. 
It assumes separation of timescales and models the FES at $A$ and $TS$ by a harmonic approximation. 
These two assumptions may be relaxed by using Pontryagin's rate theory \cite{pontryagin1933statistical}.
In both cases, the dynamics are modelled by a stochastic classical equation of motion. 
Infrequent metadynamics \cite{tiwary2013metadynamics} is an example for the first approach. 
The method assumes separation of timescales, but does not use a harmonic approximation. 
The dynamics are modelled and simulated in the full configurational space using either a deterministic or a stochastic classical equation of motion.
Owing to recent progress in electronic structure calculations \cite{gaus2011dftb3,pracht2020automated, schade2023breaking} and quantum mechanics/molecular mechanics approaches \cite{senn2009qm}, the development of reactive force fields \cite{senftle2016reaxff}, and the emergence of neural network potentials \cite{gkeka2020machine}, chemical reactions will increasingly be modeled through simulations rather than through Eyring TST. 
Thus, models of chemical reactions in large molecular systems with complex environments come within reach.
However, moving from Eyring TST to sampling-based rate estimates involves a considerable reconstruction of the theoretical foundation through which the reaction is modelled. 
Most importantly, the quantum mechanical description of the nuclear degrees of freedom is replaced by a classical approximation. 
Furthermore, the search for a transition state $TS$ is replaced by a statistical estimate of the transition count (first approach) or by the search for an optimal reaction coordinate.
It is not obvious how these changes influence the accuracy with which the reaction rate constant can be determined.
The first aim of this study is to explore and to quantify the influence of these approximations on the estimate of a reaction rate constant. 
As test reactions we choose the thermal cis-trans isomerization in two analogues of retinal: pSb5 and pSb1 (Fig.~\ref{fig:retinal_structures}). 
The thermal cis-trans isomerization over a C=C double bond in vacuum fulfills the assumptions of Eyring TST well and thus the rate constant from Eyring TST is a suitable reference. 
Additionally, it is an unimolecular reaction, so that the encounter complex of the reactants does not need to be modelled.
On the other hand, the reaction energy barrier is high and the two molecules are large enough for non-trivial coupling between vibrational modes, so that the two test systems pose a reasonable challenge for sampling-based approaches. 
Extensive literature has addressed the precise modeling of the potential energy surface (PES) for the cis-trans isomerization in retinal \cite{bondar2011ground}, as well as for retinal analogues \cite{tavan1985effect,tajkhorshid1999influence,gozem2012dynamic}. 
We here model the PES by self-consistent-charge density-functional tight-binding
method with a third-order expansion of the total energy around a reference
density (DFTB3) \cite{elstner2006scc,gaus2011dftb3} and include density functional theory (DFT) \cite{lee1988development,becke1988density} as a reference. 
This allows us to explore the sensitivity of the reaction rate constant to variations in both the rate model and the underlying PES.
Thus, as a second aim of the study, we ask whether the precision of the activation energy is indeed the most pivotal parameter when calculating a reaction rate constant.

\section{Theory}
\label{chapter:Theory}
We here summarize the rate theories used in this study.
For a more in-depth exploration of rate theories, please refer to Refs.~\citenum{peters2017reaction}, \citenum{hanggi1990reaction}, and section \ref{supp-chapter:Theoretical Background} of the supplementary material.

\begin{table*}
    \centering
    \begin{tabular}{|l| r ||c|c|| c|c|| c|c|c|| c|} 
    \hline\hline
    & Eq. & activated  &separat. of     &harmonic &high $T$ &QM vs.       &TS vs.  &sampling    &further\\
    &     & complex    &timescales      &approx.   &        &CM    &RC  &            &assumptions\\    
    \hline\hline
    Eyring TST      &\ref{eq:EyringTST_1}   &\checkmark &- &  \checkmark &  -          &QM          &TS          &-&\\
    high $T$ TST    &\ref{eq:high_T_TST}    &\checkmark &- &  \checkmark &  \checkmark &QM/CM        &TS          &-&\\
    \hline
    InMetaD         &\ref{eq:InMetaD}       &\checkmark &\checkmark&  -          &  \checkmark &CM &  (RC)   &\checkmark &Poisson statistics\\
                    &&                      &           &&             &             &                                      &&no bias on TS\\
    \hline
    Pontryagin      &\ref{eq:Pontryagin}    &  -        &\checkmark&  \checkmark &  \checkmark &CM &  RC &\checkmark&high friction\\
    Kramers         &\ref{eq:Kramers}       &  -        &\checkmark&  -          &  \checkmark &CM &  RC &\checkmark&medium-to-high friction\\
    \hline\hline
    \end{tabular}
    \caption{Overview of the model assumptions for different reaction rate models.
    QM = quantum mechanics, 
    CM = classical mechanics, 
    TS = transition state, 
    RC = reaction coordinate. 
    }
    \label{tab:model_assumptions}
\end{table*}

%
%
\subsection{Eyring TST}
The cis-trans isomerization around a C=C double bond 
is a unimolecular reaction
\begin{eqnarray}
\label{eq:A_to_B}
	A \ce{->[k_{AB}]} B \, ,
\end{eqnarray}
which, according to the theory of the activated complex, is modeled as
\begin{equation}
    	A \,\ce{<=>}\, AB^{\ddagger} \rightarrow B
\label{eq:activatedComplex}     
\end{equation}
where $A$ is the reactant state, $B$ is the product state and $AB^{\ddagger}$ is the activated complex.
The critical assumption in eq.~\ref{eq:activatedComplex} is that reactant and transition state configurations are in equilibrium.
Eyring TST \cite{eyring1935activated} models this equilibrium by statistical thermodynamics and arrives at the following equation for the reaction rate 
\cite{hanggi1990reaction,peters2017reaction}:
\begin{subequations}
\begin{eqnarray}
    k_{AB}^\mathrm{Eyr} 
    &=& \frac{RT}{h} \frac{\widetilde{q}_{\mathrm{AB}^\ddagger}}{q_{A}} \left(-\frac{E_b}{RT}\right) \label{eq:EyringTST_2}\\
    &=& \frac{RT}{h} \exp \left(-\frac{\Delta F^\ddagger}{RT}\right) \label{eq:EyringTST_1} 
\end{eqnarray}    
\label{eq:EyringTST}
\end{subequations}
where
$R$ is the ideal gas constant, 
$T$ is the temperature, and
$h$ is the Planck constant in molar units.
See section \ref{supp-chapter:Theoretical Background} of the supplementary material.

The free energy difference $\Delta F^\ddagger$ between the $AB^{\ddagger}$ and $A$ can be calculated from the molecular partition function at the transition state $\widetilde{q}_{AB^\ddagger}$ and the molecular partition function at the reactant state $q_{A}$:
\begin{equation}
\label{eq:deltaG_Eyr}
    \Delta F^\ddagger = E_b - RT\ln\left(\frac{\widetilde{q}_{\mathrm{AB}^\ddagger}}{q_{A}}\right)
\end{equation}
where $E_b$ is the potential energy barrier, i.e.~the potential energy difference between the reactant minimum and the maximum of the energy barrier. 
The partition functions are calculated relative to the electronic ground state energy of the respective configurations.
The tilde symbol in $\widetilde{q}_{AB^\ddagger}$ indicates that, for $AB^{\ddagger}$, the vibrational contribution corresponding to the reaction coordinate is excluded in the free energy calculation.
We calculate and report potential and free energies in units of J/mol, correspondingly the thermal energy is also reported as a molar quantity: $RT$. 
If units of energy are used for potential and free energies, $R$ should be replaced by the Boltzmann constant $k_B = R/N_A$ in eqs.~\ref{eq:EyringTST} and ~\ref{eq:deltaG_Eyr} and all of the following equations. 
$N_A$ is the Avogadro constant.
The molecular partition functions are determined by separating their translational, rotational, vibrational and electronic degrees of freedom.
Each part is treated using appropriate quantum mechanical models, i.e.~particle-in-a-box for translational, rigid rotor for rotational and harmonic oscillator for vibrational degrees of freedom.
In the case of a unimolecular reaction, the contributions of the translational degrees of freedom to the free energy difference in eq.~\ref{eq:deltaG_Eyr} will cancel.
We assume $q_\mathrm{el}=1$ for all situations, i.e.~the electronic ground state energy level is non-degenerate, and any contributions from higher electronic states can be ignored.
With these approximations, the free energy difference (eq.~\ref{eq:deltaG_Eyr}) of the cis-trans isomerization can be decomposed as
\begin{eqnarray}
\label{eq:deltaG_Eyr_2}
    \Delta F^\ddagger &=& E_b + \Delta F_{\mathrm{rot}} + \Delta F_{\mathrm{vib}}
\end{eqnarray}
where 
\begin{subequations}
\begin{eqnarray}
\Delta F_{\mathrm{rot}} &=&- RT \ln \left[ \frac{q_{AB^{\ddagger}, \mathrm{rot}}}{q_{A, \mathrm{rot}}}\right] \label{eq:DeltaF_rot} \\
\Delta F_{\mathrm{vib}} &=&- RT \ln \left[ \frac{\widetilde{q}_{AB^{\ddagger}, \mathrm{vib}}}{q_{A, \mathrm{vib}}}\right]\label{eq:DeltaF_vib}
\end{eqnarray}
\end{subequations}
define the rotational and vibrational free energy difference.
$q_{A, \mathrm{rot}}$ and $q_{AB^{\ddagger}, \mathrm{rot}}$ are the rotational partition functions of $A$ and $AB^\ddagger$.
$q_{A, \mathrm{vib}}$ and $\widetilde{q}_{AB^{\ddagger}, \mathrm{vib}}$ are the vibrational partition functions of $A$ and $AB^\ddagger$, 
where the tilde symbol indicates that the reactive vibrational mode has been excluded from vibrational partition function of $AB^\ddagger$.
See section \ref{supp-chapter:Theoretical Background} of the supplementary material.
%
%

%
%
\subsection{High-temperature TST}

If the thermal energy $RT$ is large compared to the energy difference of the vibrational states, the following high-temperature approximation to Eyring TST may be used
\begin{equation}
    k_{AB}^\mathrm{Eyr}  \approx
    k_{AB}^\mathrm{ht} = \frac{\prod_{k=1}^{3N-6} \nu_{A,k}}{\prod_{k=1,k\neq r}^{3N-6} \nu_{AB^\ddagger,k}}\exp\left(-\frac{E_b}{RT}\right)
\label{eq:high_T_TST}
\end{equation}
where the frequencies $\nu_{A,k}$ and $\nu_{AB^\ddagger,k}$ correspond to the harmonic frequencies at the reactant and transition state respectively. 
Note that the frequency of the reactive vibrational mode $\nu_{AB^\ddagger,r}$ is excluded from the product.
Eq.~\ref{eq:high_T_TST} can be brought into the form of eq.~\ref{eq:EyringTST_1} by setting
\begin{equation}
 \begin{aligned}
\label{eq:deltaG_high_T}
    \Delta F^\ddagger \approx 
    \Delta F^{\ddagger, \mathrm{ht}} 
    &= E_b - RT\ln\left(\frac{\widetilde{q}_{\mathrm{AB}^\ddagger, \mathrm{vib}}^{\mathrm{ht}}}{q_{A, \mathrm{vib}}^{\mathrm{ht}}}\right)  \\
    &= E_b + \Delta F_{\mathrm{vib}}^{\mathrm{ht}}
 \end{aligned}
\end{equation}
where $\widetilde{q}_{\mathrm{AB}^\ddagger, \mathrm{vib}}^{\mathrm{ht}}$ and $q_{A, \mathrm{vib}}^{\mathrm{ht}}$ are the high-temperature approximations to the vibrational partition functions of $A$ and $AB^\ddagger$.
In deriving eqs.~\ref{eq:high_T_TST} and \ref{eq:deltaG_high_T}, one assumes that the moments of inertia of the reactant and TS configuration are approximately the same and thus the rotational contribution to the $\Delta F^\ddagger$ is negligible. 
Additionally, one neglects the vibrational zero-point energy and takes the continuum limit of the vibrational partition function
See section \ref{supp-chapter:Theoretical Background} of the supplementary material.
The high-temperature TST can also be derived by treating the partition functions in eqs.~\ref{eq:EyringTST_2} and \ref{eq:deltaG_Eyr} classically and using a harmonic approximation for the PES.
See section \ref{supp-chapter:Theoretical Background}  of the supplementary material.
Eqs.~\ref{eq:high_T_TST} and \ref{eq:deltaG_high_T} thus constitute the link between a quantum mechanical and a classical treatment of the activated complex. 
%

%
%
\subsection{Infrequent metadynamics}
A statistical estimate for reaction rate constant is obtained via the mean first-passage time $\tau_{AB}$
\begin{eqnarray}
    k_{AB} &=& \frac{1}{\tau_{AB}} \, ,
\label{eq:MFPT}    
\end{eqnarray}
where $\tau_{AB}$ is the average time it takes for the system to reach the product state $B$ from the reactant state $A$. 
The relation between $k_{AB}$ and $\tau_{AB}$ stated in eq.~\ref{eq:MFPT} relies on a separation of timescales between the timescale of equilibration within $A$ and the much slower timescale of equilibration between $A$ and $B$.
From MD simulations on the PES $V(x)$, where $x$ is the molecular configuration, $\tau_{AB}$ can be calculated as the arithmetic mean of the first-passage times from $A$ to $B$ \cite{peters2017reaction}. 
However, a better statistical accuracy is obtained by fitting a the cumulative distribution function of a Poisson process \cite{salvalaglio2014assessing}
\begin{equation}
\label{eq:TCDF}
    P(\tau_{AB,i}) = 1 - \exp\left(-\frac{\tau_{AB,i}}{ \tau_{AB}}\right)
\end{equation} 
to the cumulative distribution histogram of these fist passage times. 
In eq.~\ref{eq:TCDF}, 
$\tau_{AB,i}$ is the $i$ first-passage time observed in the simulation and $\tau_{AB}$ is the MFPT and acts as a fitting parameter, which is inserted into eq.~\ref{eq:MFPT} to obtain the reaction rate.
Infrequent metadynamics \cite{tiwary2013metadynamics} is a method to calculate transition times for systems in which the mean first-passage times is larger than the accessible simulation time.
The molecular system is prepared in the reactant state $A$ and a time dependent bias function $U(x,t)$ is introduced that increases in strength as the simulation proceeds and pushes the system over the barrier into state $B$.
One terminates the simulation and records the biased transition time $t_{A\rightarrow B}^{(i)}$, where $i$ is the index of the infrequent metadynamics simulation.
Each accelerated first-passage time is then reweighted to the corresponding physical first-passage time by a discretized time-integral over the length of the trajectory \cite{grubmuller1995predicting,voter1997hyperdynamics,tiwary2013metadynamics, khan2020fluxional}
\begin{eqnarray}
    \tau_{AB,i} = \Delta t \sum_{k=1}^{T_i}  \exp\left(\frac{U(x_{i,k}, k\Delta t)}{R T} \right)\, .
\label{eq:InMetaD}   
\end{eqnarray}
where $\Delta t$ is the time step of the trajectory, $T_i$ is the total number of time steps in the $i$th trajectory, $x_{i,k}$ is the $k$th configuration in this trajectory, and $t=k\Delta t$ is the corresponding time.
This reweighting assumes that no bias has been deposited on the transition state, which is approximately ensured by the slow deposition of the infrequent metadynamics protocol \cite{barducci2008well}.

Eq.~\ref{eq:InMetaD} is derived from the eq.~\ref{eq:EyringTST_1}, i.e.~the method assumes that the reaction proceeds via an activated complex. 
In contrast to Eyring TST, partition functions $q_A$ and $q_{AB^\ddagger}$ are treated classically. 
The derivation considers a statistical estimate of $q_{AB^\ddagger}/q_A$ from MD simulation data, which has the advantage that no harmonic approximation is needed. 
See section \ref{supp-chapter:Theoretical Background} of the supplementary material.

%
%
\subsection{Reaction coordinate based rate theories}
In reaction coordinate based rate theories, one assumes that the system evolves according to a diffusive dynamics along a reaction coordinate $s(x)$.
This approach requires the free energy surface (FES) along $s(x)$, which is defined as \cite{peters2017reaction}:
\begin{equation}
    F(s) = -RT \ln \pi (s)
    \label{eq:FES}
\end{equation}
where $\pi(s)$ is the equilibrium distribution in $s$:
\begin{equation}
    \pi(s) =  Z_x^{-1} \int_{\Gamma_x} \mathrm{d}x \exp\left(-\frac{V(x)}{RT}\right)  \delta\left(s(x) - s\right) \, .
    \label{eq:Boltzmanns}
\end{equation}
$Z_x$ the configurational partition function, $\Gamma_x$ is the configurational space, and $\delta\left(s(x) - s\right)$ is the Dirac delta function.
The interaction of the internal degrees of freedom with the effective dynamics along $s$ is modelled as a thermal bath, i.e.~by a friction and random force which are balance by the Einstein relation.
The friction force can be scaled by a friction coefficient or collision rate $\xi$ (with units time$^{-1}$).
The following two rate theories assume separation of timescales which can be quantified in terms of the FES as $F_{AB}^{\ddagger} \gg RT$, 
where $F^\ddagger_{AB} = F(s_{AB^\dagger}) - F(s_A)$ is the difference of the FES between the FES minimum in the reactant state and the maximum of the free energy barrier.
We remark that the location of the free energy maximum $s_{AB^\dagger}$ does not necessarily coincides with a saddle point in the PES.
In Kramers' rate theory \cite{kramers1940brownian, hanggi1990reaction,peters2017reaction}, the reactant and product state, as well as the maximum of the FES are modelled using a harmonic approximation of the FES around these extrema.
In the medium-to-high friction regime, one obtains the following analytical expression for the reaction rate constant
\begin{align}
k_{AB}^\mathrm{Kra}
&=
\frac{\xi}{\omega_\ddagger} \left(\sqrt{\frac{1}{4} + \frac{\omega_\ddagger^2}{\xi^2}} - \frac{1}{2} \right)
\frac{\omega_A}{2 \pi} \exp\left(-\frac{F_{AB}^{\ddagger}}{RT} \right)
\label{eq:Kramers}
\end{align}
where $\omega_A$ is the angular frequency of the harmonic approximation in the reactant state $A$, 
and $\omega^\ddagger$ is the angular frequency of the harmonic approximation at the maximum of the free energy barrier.
By relaxing the harmonic approximation one obtains Pontryagin's expression for the rate constant \cite{pontryagin1933statistical}:
\begin{align}
k_{AB}^{\mathrm{Pon}}  = \left\lbrace 
\int_{s_A}^{s_B}\mathrm{d} s'\left[ \frac{1}{D(s')} e^{\beta F(s')}
\int_{-\infty}^{s'} \mathrm{d}s'' \, e^{-\beta F(s'')}
\right]
\right\rbrace^{-1}
\label{eq:Pontryagin}
\end{align} 
where $\beta = 1/RT$, and  $D(s) = \frac{RT}{\mu_q\xi(s)}$.
$D(s)$ is the position dependent diffusion profile, which arises from the position dependent friction coefficient $\xi(s)$.
$\mu_q$ is a effective molar mass.
$D(s)$ can be estimated form MD simulations following Ref.~\citenum{hummer2005position}.
Note that, while eq.~\ref{eq:Kramers} is valid in the intermediate and in the high friction regime, eq.~\ref{eq:Pontryagin} is only valid in the high friction regime, where the effective dynamics can be modelled by overdamped Langevin dynamics. 
Eq.~\ref{eq:Pontryagin} is often quite generically referred to as the formula for mean first-passage time (MFPT) for diffusion over a barrier (which is inverted to get the rate) or the escape rate. For the sake of clarity, we shall refer to it as the Pontryagin rate equation after Ref.~\citenum{pontryagin1933statistical}.

The assumptions of the reaction rate models introduced in this section are summarized in Table \ref{tab:model_assumptions}.
We remark that all sampling-based approaches use the high-temperature approximation, and that infrequent metadynamics needs a reaction coordinate to apply the bias, but not for the actual estimate for the rate constant.

\section{Model Systems and Potential Energy Surface}
\label{chapter:Potential Energy Landscape}

\begin{figure}[t]
\includegraphics[scale=1]{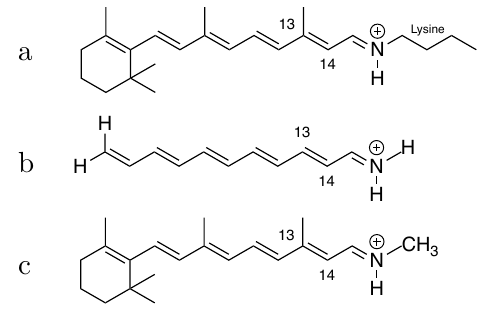}
\caption{
Structures of retinal compounds: \textbf{a:} retinal cofactor attached to lysine chain \textbf{b:} pSb5 and \textbf{c:} pSb1.}
\label{fig:retinal_structures}
\end{figure}

pSb5 and pSb1  (Fig.~\ref{fig:retinal_structures}.b and c) are model compounds for retinal.
In proteins, retinal is covalently linked to a lysine side chain via a protonated Schiff base (Fig.~\ref{fig:retinal_structures}.a).
In pSb5 (naming following Ref.~\citenum{bondar2011ground}), the $\beta$-ionone ring and methyl substituents as well as the lysine chain have been removed. 
In pSb1 (naming following Ref.~\citenum{bondar2011ground}), the $\beta$-ionone ring and methyl substituents remain but the lysine chain has been replaced by a methyl group. 
Both compounds have been used as models for retinal in previous studies \cite{tavan1985effect, baudry1997quantum,baudry1999simulation,tajkhorshid1999dielectric,tajkhorshid1999influence,tajkhorshid1999role,zhou2002performance,bondar2011ground} 
Our goal is to evaluate various rate theories for two model compounds on a specific potential energy surface. Here, we outline our selection of the electronic structure method for PES calculation. 
Even though most computational models emphasize photo-isomerization in electronically excited states, our focus centers on the thermal isomerization within the electronic ground state.

A critical point in modelling the thermal isomerization is the highly correlated $\pi$-electron system along the retinal polyene chain, which allows for two possible pathways for the cis-trans isomerization.
In the first pathway, the double bond is broken homolytically when the torsion angle reaches ca.~90 degrees, creating a transition state with diradical character. 
In the second pathway, cis-trans isomerization over the double bond occurs through charge transfer, with the electrons moving towards the protonated imine in the transition state. 
From quantum chemical methods that account for dynamic electron correlation,  there is little consensus as to whether cis-trans isomerization in molecules with three conjugated double bonds proceeds via a charge-transfer or a diradical mechanism  \cite{de2002reaction, gozem2012dynamic,gozem2017theory,zen2015quantum}.
However, DFT studies of retinal and related systems \cite{tajkhorshid1999dielectric,tajkhorshid1999influence,tajkhorshid1999role,zhou2002performance,bondar2011ground} conclude that the isomerization over double bonds in the polyene chains proceeds through a charge-transfer pathway if the Schiff-base is protonated. 
Since both pSb5 and pSb1 feature a protonated Schiff-base, the charge-transfer pathway seems to be a reasonable assumption for the isomerization of the C$_{13}$=C$_{14}$ double bond in our model compounds.
In a charge-transfer pathway, electrons stay paired (closed-shell) during isomerization, and we thus do not necessarily need an electronic structure method that models unpaired electrons.
Ab-initio MD simulations of the thermal isomerization in retinal at the level of DFT are limited to simulation times in the order 1 ns to 10 ns, which is not enough to converge a free energy surface. 
An alternative is the self-consistent-charge tight-binding density-functional method (DFTB) \cite{elstner2006scc,gaus2011dftb3}, whose computational cost is 2-3 orders of magnitude lower than DFT, thus giving access to much longer simulation timescales.
DFTB is an approximation to DFT based on expansion of the total energy around a reference density \cite{elstner2006scc}, where DFTB3 \cite{gaus2011dftb3} includes the third order of the expansion.
Even though spin polarization has been introduced for DFTB \cite{aradi2007dftb+,hourahine2007self,melix2016spin}, most applications are based on restricted DFT and cannot model unpaired electrons. 

For retinal compounds, DFTB-predicted structures are in good agreement with NMR experiments \cite{sugihara2002}. 
Relative to DFT, DFTB yields a reasonable description of the torsional properties of retinal not only in the gas phase \cite{zhou2002performance}, but also in the protein environment \cite{bondar2004mechanism,bondar2011ground,elghobashi2018catalysis}. 
Torsional barriers for the C$_{13}$=C$_{14}$ bond in retinal compounds are slightly underestimated (about $2\,\mathrm{kcal/mol}$) when using DFTB as compared to DFT/B3LYP. 
More approximate potential energy functions, including semi-empirical methods such as AM1 or PM3, overestimate the delocalization more dramatically than DFTB \cite{zhou2002performance}. 
In empirical force fields, the delocalization can be modelled by imposing the bond lengths along the polyene chain.
But since these potential energy functions use fixed partial atomic charges, they are not well suited to describe the charge-shift in the polyene chain during the isomerization, and consequently the isomerization is highly sensitive to the choice of these charges. 
Since DFTB strikes a suitable balance between the accuracy of the potential energy function and the computational cost of conducting ab-initio MD simulations, we use will use it in our simulations. 
For rate theories that do not require sampling, we include calculations at the level of unrestricted DFT/B3LYP for comparison to a higher level of theory. 

\section{Results}
\label{chapter:Results}

\subsection{Free energy surface and diffusion profile}
\begin{figure*}
\centering
\includegraphics[scale=1]{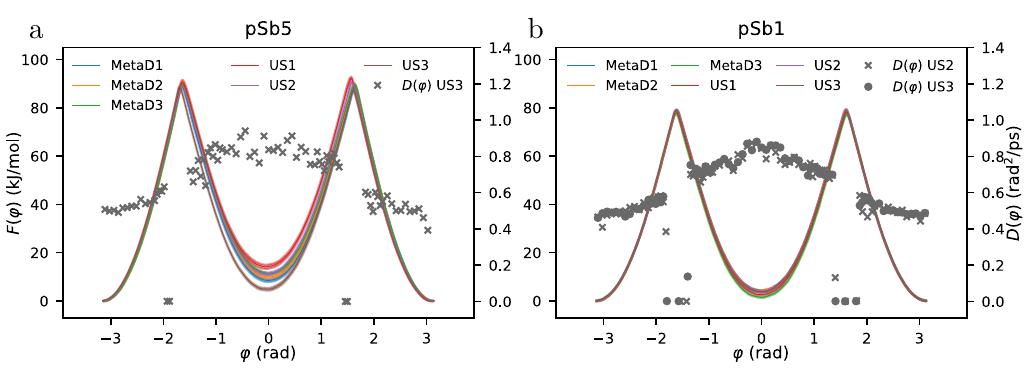}
\caption{
Free energy surfaces $F(\varphi)$ and diffusion profiles $D(\varphi)$ along C$_{13}$=C$_{14}$ dihedral angle $\varphi$ for \textbf{a:} pSb5  and \textbf{b:} pSb1 from metadynamics (MetaD) and umbrella sampling (US) using DFTB3.
Parameters for MetaD  and US simulations are reported in Tab.~\ref{tab:metad_parameters} and \ref{tab:us_parameters}.
Free energy curves are filled between plus and minus one standard error. 
}
\label{fig:FES_diff}
\centering
\end{figure*}

The potential energy functions of pSb5 and pSb1 are high-dimensional functions of 72 and 156 internal degrees of freedom, respectively. 
With MD simulations, one can characterize these high-dimensional energy functions in more manageable, lower-dimensional collective variable spaces using free energy surfaces (FES) and diffusion profiles. 
Fig.~\ref{fig:FES_diff} shows the FES (eq.~\ref{eq:FES}) along the C$_{13}$=C$_{14}$ torsion angle $\varphi$, as estimated from umbrella sampling \cite{torrie1977nonphysical} (US) and well-tempered metadynamics \cite{laio2002escaping,barducci2008well} (MetaD) simulations using ab-initio MD with the DFTB3 method.
The line thickness shows the statistical error in the estimated FES.
A full rotation around $\varphi$ yields two barriers which, as expected, have the same absolute height.
The vertical rotational barriers are $F^\ddagger_{t\rightarrow c}\approx 89\,\mathrm{kJ/mol}$ and $F^\ddagger_{c\rightarrow t}\approx 81\,\mathrm{kJ/mol}$ for pSb5 and $F^\ddagger_{t\rightarrow c}\approx 79\,\mathrm{kJ/mol}$ and $F^\ddagger_{c\rightarrow t}\approx 75\,\mathrm{kJ/mol}$ for pSb1 (both from MetaD1 in Fig.~\ref{fig:FES_diff}).
The rotational barrier in pSb5 is slightly higher than in pSb1, because the tertiary C$_{13}$ in pSb1 stabilizes the positive charge at the charge-transfer transition state better than the secondary C$_{13}$ in pSb5 \cite{tavan1985effect,tajkhorshid1999influence}.
As usual for a carbon double bond, the trans state at $\varphi=\pi\,\mathrm{rad}$  is slightly more stable than the cis state at $\varphi=0\,\mathrm{rad}$, however the stabilization is larger in pSb5 ($9.43 \pm 1.20$ kJ/mol) than in pSb1 ($2.70 \pm 1.23$, kJ/mol). 
A possible explanation might be that in the trans state of pSb1, the methyl group at C$_{13}$ sterically interacts with the hydrogens at C$_{15}$, which destabilizes this conformation.
The effective dynamics along a reaction coordinate are suitably modelled by stochastic dynamics with position dependent diffusion constant.
The dependence of the diffusion constant on the collective variable is due to the dynamics in the orthogonal degrees of freedom and due to the curvature of the collective variable. 
The diffusion profile in Fig.~\ref{fig:FES_diff} show the form that is expected for a dihedral angle rotations. 
Note that the estimate of the diffusion constant fails in the barrier region, because of the sharpness of the barriers. 

\begin{figure}
\includegraphics[scale=1]{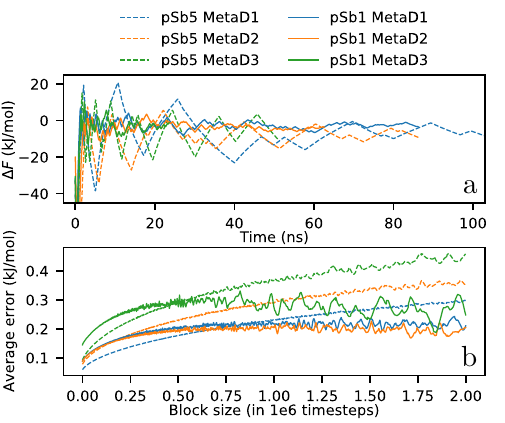} 
\caption{
\textbf{a}: Convergence of free energy difference $\Delta F=F_\mathrm{cis}-F_\mathrm{trans}$ from metadynamics bias as a function of simulation time for pSb5 (dashed) and pSb1 (full).
\textbf{b}: Convergence of the average errors from block averaging analysis as a function of block size for the same simulations as above. 
}
\label{fig:MetaD_convergence}
\end{figure}

In pSb1, umbrella sampling and metadynamics yield essentially the same FES for various parameter settings (Fig.~\ref{fig:FES_diff}.b). 
By contrast, estimate of the FES for pSb5, and especially the relative stability of the cis state, depends on the method that is used to construct the FES and on the parameters settings (Fig.~\ref{fig:FES_diff}.a). 
Additionally, the metadynamics simulations converge much slower for pSb5 than for pSb1. 
Convergence of the metadynamics simulation can be checked by monitoring the estimated free-energy difference between cis and trans state $\Delta F$ as function of simulation time (Fig.~\ref{fig:MetaD_convergence}.a), 
or by monitoring average errors in block analysis \cite{bussi2019analyzing} as a function of block size (Fig.~\ref{fig:MetaD_convergence}.b).
The kinks in the lines in Fig.~\ref{fig:FES_diff}.a correspond to transitions between cis and trans state during the metadynamics build-up.
The larger and less frequent kinks in the simulations for pSb5 compared to those for pSb1 imply that the bias builds up within one state longer before moving to the other.
Slow convergence can be caused by correlated motion in degrees of freedom orthogonal to the biased coordinate, which in this case is the C$_{13}$=C$_{14}$ dihedral $\varphi$. 

\subsection{Correlated degrees of freedom}
\label{sec:correlatedDOF}
\begin{figure*}
\includegraphics[scale=1]{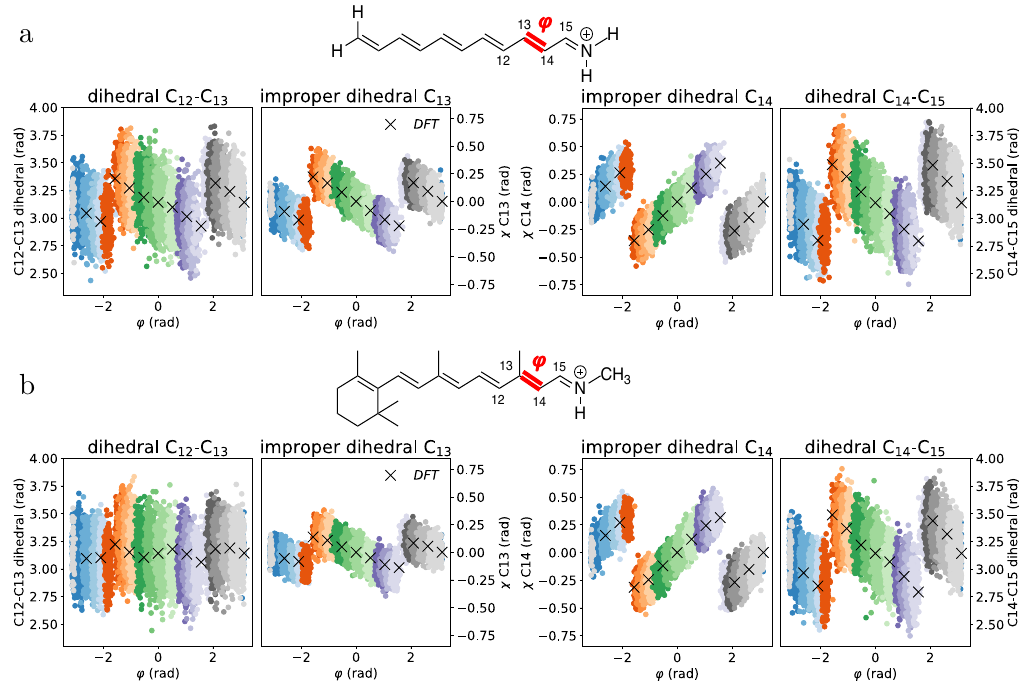}
\caption{
Correlations in pSb5 and pSb1.
\textbf{a:} Umbrella sampling simulations using DFTB3 (set US1) for pSb5 showing correlation between C$_{13}$=C$_{14}$ dihedral $\varphi$ and dihedral C$_{12}$-C$_{13}$, the improper dihedral on C$_{13}$, the dihedral C$_{14}$-C$_{15}$ and the improper dihedral on C$_{14}$.
Different colors represent different umbrella windows.
Black crosses represent constrained optimizations along $\varphi$ using unrestricted DFT/B3LYP.
\textbf{b:} Same analysis for pSb1.
}
\label{fig:main_correlation}
\end{figure*}

Using MD simulations, correlation between various collective variables can be assessed. 
In the case of pSb1, previous research in Ref.~\citenum{bondar2011ground} documented correlations between the C$_{13}$=C$_{14}$ dihedral angle and adjacent dihedral angles along the minimum energy path, i.e.~at $0\,\mathrm{K}$. 
In Fig.~\ref{fig:main_correlation}, the correlations at $300\,\mathrm{K}$  are presented for both pSb5 and pSb1.
These correlation plots were generated from  US simulations, with each color in the plot representing a different umbrella potential. 
Crosses represent the minimum energy path calculated at unrestricted DFT/B3LYP. 
The MD simulations at DFTB3 follow closely the DFT/B3LYP minimum energy path, which substantiates that DFTB faithfully represents the DFT-PES of these two molecules. 
However, there is considerable thermal fluctuations around the minimum energy path. 
For the single-bond dihedrals C$_{12}$-C$_{13}$ and C$_{14}$-C$_{15}$, the spread is $\pm 0.4\,\mathrm{rad}$ ($\approx \pm 23\,\mathrm{degrees}$), whereas for the improper torsions the spread is $\pm 0.2\,\mathrm{rad}$ ($\approx \pm 11\,\mathrm{degrees}$).
Overall, we find that the correlation extends to the neighboring single bond, but not to the improper dihedrals at C$_{12}$ and C$_{15}$.
We observed a certain level of correlation to the C$_{15}$-N double bond along the minimum energy paths. However, this correlation is obscured by thermal fluctuations at $300\,\mathrm{K}$ (see Fig.~\ref{supp-fig:rest_correlation} in the supplement).

\begin{figure}[h!]
\includegraphics[width=7cm]{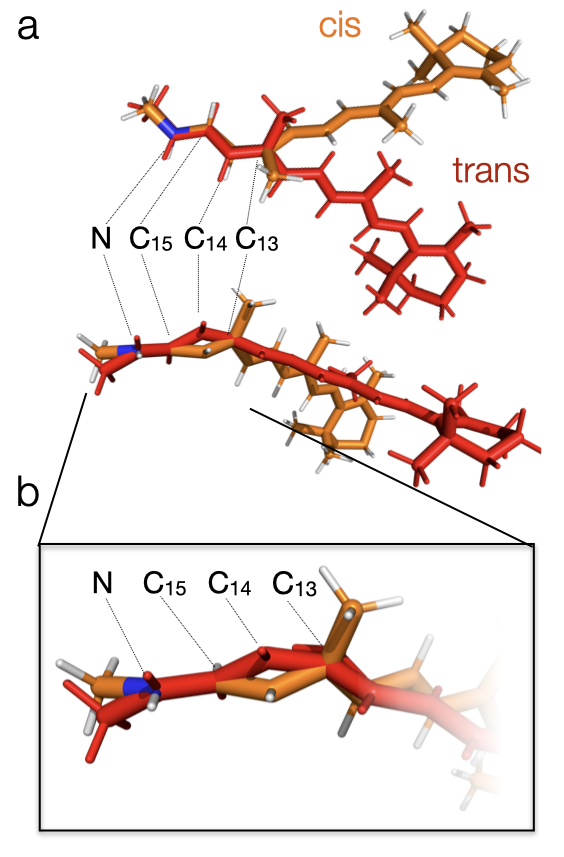} 
\caption{
\textbf{a:} Minimum energy structures from constrained optimizations using unrestricted DFT/B3LYP calculations on pSb1 at $\varphi=-60\,\mathrm{degrees}$ (cis, orange structure) and $\varphi=-120\,\mathrm{degrees}$ (trans, red structure). 
Structures are aligned along the N, C$_{15}$ and C$_{13}$ atoms.
\textbf{b:} zoom  on the reaction center.
}
\label{fig:aligned_structures}
\end{figure}
The most remarkable feature of the correlations plots are the sudden jumps at the transition states ($\varphi = + \pi/2\,\mathrm{rad}$ and $\varphi = -\pi/2\,\mathrm{rad}$). 
The improper dihedral angles at C$_{13}$ and C$_{14}$ represent the planarity at these sp$^2$-carbon atoms, with $\chi=0\,\mathrm{rad}$ representing a planar conformation. 
Consider the improper torsion at C$_{14}$ when approaching the transition state at $\varphi=-\pi/2\,\mathrm{rad}$ from the trans-state, the local conformation at C$_{14}$ bends out of plane up to $20\,\mathrm{degrees}$ ($0.35\,\mathrm{rad}$, minimum energy value).
At the transition state, it suddenly inverts to an out-of-plane distortion of $-20\,\mathrm{degrees}$. 
On top of the out-of-plane wagging at C$_{14}$, the substituent at N (H for pSb5 and CH$_3$ for pSb1) slightly rotates. 
The concerted motion is illustrated in Fig.~\ref{fig:aligned_structures}, where structures of pSb1 nearing the transition state ($\varphi\approx -90\,\mathrm{degrees}$) from cis (orange structure, $\varphi=-60\,\mathrm{degrees}$) and from trans (red structure, $\varphi=-120\,\mathrm{degrees}$) are aligned along the C$_{13}$, C$_{15}$ and N atoms.
Note that the out-of-plane wagging at C$_{14}$ contributes to the correlation between C$_{13}$=C$_{14}$ dihedral and C$_{14}$-C$_{15}$ dihedral.
C$_{13}$ shows a similar out-of-plane wagging as C$_{14}$. 
However, while at C$_{14}$ we do not find any difference between pSb5 and pSb1, the correlation of the C$_{13}$=C$_{14}$ dihedral to the improper dihedral at C$_{13}$ and the C$_{13}$-C$_{12}$ torsion is less pronounced in pSb1 than in pSb5.
Presumably, the methyl substituent hinders the out-plane motion at C$_{13}$ in pSb1 compared to C$_{14}$ in the same molecule and thus interrupts the correlation. 
%

\subsection{Rates for the DFTB3 potential energy surface}
With a model of the free energy surface of pSb5 and pSb1 and good understanding of the reaction mechanism, we are ready to discuss the reaction rate constants for the thermal isomerization at the level of DFTB3 (Tables \ref{tab:pSb5_rates} and \ref{tab:pSb1_rates}).
Transition states for both compounds were optimized using the Nudged Elastic Band (NEB) method. 
NEB optimization converged well for pSb5, but was very sensitive to the choice of the NEB parameters (spring constants, maximal force, amount of nodes) for pSb1.
\paragraph{Eyring TST}
In pSb5, the potential energy barrier $E_b$ for trans$\rightarrow$cis reaction is $112.2 \,\mathrm{kJ/mol}$, which is in good agreement with the previously reported value of  $27.5\,\mathrm{kcal/mol}=115.1\,\mathrm{kJ/mol}$ \cite{zhou2002performance}.
The barrier for cis $\rightarrow$ trans reaction is about $7.7\,\mathrm{kJ/mol}$ lower, which implies that the cis reactant state is slightly higher in energy than the trans state. 
This aligns closely  with the free energy difference of $8$ to $10\,\mathrm{kJ/mol}$ between cis and trans states in pSb5 (Fig.~\ref{fig:FES_diff}).
For pSb1, the potential energy barriers have about equal height (93.7 kJ/mol and 91.1 kJ/mol) and are about 10 kJ/mol lower than for pSb5. 
Again, this aligns closely with the FES along $\varphi$ for this molecule.
In each of the four reactions, the free energy difference $\Delta F^{\ddagger}$ at $T=300\,\mathrm{K}$ is about 8 to 12 kJ/mol lower than $E_b$ due to the vibrational and rotational contribution to the free energy difference.
For pSb5, the Eyring TST rates are $8.30 \cdot 10^{-6}\,\mathrm{s}^{-1}$ for the trans $\rightarrow$ cis reaction and $1.99 \cdot 10^{-4}\,\mathrm{s}^{-1}$ for the reverse reaction.
In pSb1, the lower energy barrier $E_b$  leads to considerably faster rates, namely $1.42\cdot 10^{-2}\, \mathrm{s}^{-1}$ for the trans$\rightarrow$cis transition and $2.06\cdot 10^{-1}\, \mathrm{s}^{-1}$ for the reverse reaction.
\paragraph{High-temperature TST}
The high-temperature approximation approximates the free energy contribution to the rates by neglecting the contribution due to the rotational degrees of freedom and by making a classical approximation for the harmonic vibrational partition function.
Tables \ref{tab:pSb5_rates} and \ref{tab:pSb1_rates} show that for pSb5 and pSb1 the rotational contribution is less than 1 kJ/mol, and thus neglecting this contribution is well justified.
In our systems, the vibrational component contributes negatively to the free energy difference, thereby reducing the overall free energy difference $\Delta F^\ddagger$ in comparison to the potential energy barrier $E_b$.
This effect is slightly underestimated in the classical approximation. 
Consequently, $\Delta F^{\ddagger, \mathrm{ht}}$ in high-temperature TST appears higher than $\Delta F^{\ddagger}$ in Eyring TST, and the high-temperature TST rates are slower than Eyring TST rates. 
The effect amounts to about 3 kJ/mol which lowers the rate by about a factor of two.
Thus, the high-temperature approximation is suitable for our two systems.

\paragraph{Infrequent metadynamics}
The high-temperature approximation constitutes the link between a quantum partition function and the classical partition functions. 
Models based on classical partition functions can be sampled by MD simulations to obtain a statistical estimate of the rate. 
One method to do this is infrequent metadynamics, in which Gaussian bias function are deposited in the potential energy well of the reactant state, and the enhanced reaction rate constant is subsequently reweighted to the unbiased reaction rate constant.
Rate constants for two different deposition paces of the Gaussian bias functions are shown in Tables \ref{tab:pSb5_rates} and \ref{tab:pSb1_rates}.
The obtained rate constants were insensitive with regards to doubling the pace of deposition (Table \ref{supp-tab:InMetaD_parameters}), indicating that the deposition rate is infrequent enough for rate constants to be reliable.
The infrequent metadynamics simulations passed the Kolmogorov-Smirnoff test \cite{salvalaglio2014assessing} which serves as indication whether the assumptions of TST are violated.
The rate constants from infrequent metadynamics are in very good agreement with the rate constants from high-temperature TST. 
Only in pSb1, the rate constant for the cis $\rightarrow$ trans reaction is slightly underestimated by infrequent metadynamics.
This confirms that a sampling-based approach is suitable for calculation of reaction rates of chemical reaction and the required MD simulation times are accessible when using DFTB3 to calculate the PES.

\paragraph{Reaction coordinate based rate theories}
We calculated the rate constants from Kramers' (eq.~\ref{eq:Kramers}) and Pontryagin's (eq.~\ref{eq:Pontryagin}) rate theories using the free energy surfaces and diffusion profiles in Fig.~\ref{fig:FES_diff} (Tables \ref{tab:pSb5_rates} and \ref{tab:pSb1_rates}).
Values for the parameters in eq.~\ref{eq:Kramers} are reported in Tables \ref{supp-tab:RC_based_rates_pSb5} and \ref{supp-tab:RC_based_rates_pSb1} in the supplement.
Surprisingly, these rates are orders of magnitude higher than rates from TST or from infrequent metadynamics.
In pSb5, the results from Kramers' rate theory overestimate the high-temperature TST rate constant by three orders of magnitude.
This is caused by the low free energy barriers $F_{AB}^\ddagger$ in the FES compared to the energy difference $\Delta F^\ddagger$, as calculated within TST.
The slow convergence of the metadynamics simulations for the FES in pSb5, as well as the sensitivity of the FES to parameters collectively raise concerns about the optimality of the chosen reaction coordinate.
With a sub-optimal reaction coordinate, the free-energy barriers are underestimated and thus Kramers' rate theory overestimates the rate constants.
Although $\varphi$ appears an intuitive choice for the reaction coordinate, the correlated motions in orthogonal degrees of freedom described above suggests that these motions need to be taken into account to construct a sufficiently accurate reaction coordinate. 
On the other hand, while pSb1 exhibits the same correlated motions as pSb5, the discrepancy between $F_{AB}^\ddagger$ and $\Delta F^\ddagger$ is much smaller.
Consequently Kramers' rate theory overestimates the rate constants from high-temperature TST for pSb1 only by a factor of 40 for the trans $\rightarrow$ cis reaction and by a factor of 17 for the reverse reaction.
Additionally, the calculation of the FES converges quickly for pSb1.
It is not obvious, why the C$_{13}$=C$_{14}$ dihedral angle $\varphi$ would be a poor reaction coordinate for pSb5 but a reasonably accurate reaction coordinate in pSb1.
Pontryagin's rate theory yields even higher rate constants than Kramers' rate theory and this points to a second effect that might be at play. 
Pontryagin's rate theory assumes overdamped Langevin dynamics along the reaction coordinate and would overestimate the rates if the effective dynamics actually falls into the intermediate or friction regime. 
In the weak friction regime, also Kramers' rate theory for intermediate friction (eq.~\ref{eq:Kramers}) would overestimate the rate constant. 
The friction regime is in part determined by the ``sharpness" of the free energy barriers as measured by $\omega_{\ddagger}$, the angular frequency of the harmonic approximation of the FES maximum.
Both systems in fact exhibit very sharp barriers and thus high values of $\omega_{\ddagger}$, which might shift the effective dynamics into the weak-to-intermediate friction regime.

\begin{table*}
    \centering
    \begin{tabular}{|l| c 
    |>{\raggedleft}p{2cm}|>{\raggedleft}p{2cm}|
    |>{\raggedleft}p{2cm}|>{\raggedleft}p{2cm}|
    |>{\raggedleft}p{2cm}|r|
    } 
    \hline
    \hline
    &       &\multicolumn{2}{c||}{\textbf{DFTB3}} 
            & \multicolumn{2}{c||}{\textbf{DFT}} 
            & \multicolumn{2}{c|}{\textbf{DFT-D3}}  \\
    &Eq.    &trans $\rightarrow$ cis &cis $\rightarrow$ trans 
            &trans $\rightarrow$ cis &cis $\rightarrow$ trans 
            &trans $\rightarrow$ cis &cis $\rightarrow$ trans \\
    \hline
    \multicolumn{8}{c}{\rule{0pt}{12pt} \textbf{Free energy difference between $A$ and $AB^\ddagger$ in Eyring TST [kJ/mol]}}\\
    \hline
    $E_b$  && 
    112.2 & 104.5 & 123.6 & 108.5 & 123.6 & 109.3\\
    $\Delta F_{\mathrm{rot}}$ &\ref{eq:DeltaF_rot}& 
    $-0.5$ & $0.2$ & $ -0.4 $ & $ 0.1 $ & $-0.4$ & $0.1$\\
    $\Delta F_{\mathrm{vib}}$ &\ref{eq:DeltaF_rot}& 
    $-7.3$ & $-8.3$ & $ -6.2 $ & $ -6.5 $ & $-6.5$ & $-6.6$\\
    \hline
    $\Delta F^{\ddagger}$  &\ref{eq:deltaG_Eyr_2}& 
    $104.4$ & $96.5$  & $117.0$ & $102.2$ & $116.8$ & $102.8$\\    
    \hline 
    \multicolumn{8}{c}{\rule{0pt}{12pt} \textbf{Free energy difference betweeen $A$ and $AB^\ddagger$ in high-temperature TST [kJ/mol]}}\\    
    \hline
    $E_b$ && 
    112.2 & 104.5 & 123.6 & 108.5 & 123.6 & 109.3\\ 
    $\Delta F_{\mathrm{vib}}^{\mathrm{ht}}$ &\ref{eq:deltaG_high_T}& 
    $-4.6$ & $-5.2$ & $ -3.6 $ & $ -3.6 $ & $-3.7$ & $-3.7$\\
    \hline 
    $\Delta F^{\ddagger, \mathrm{ht}}$ &\ref{eq:deltaG_high_T}&  
    $107.6$ & $99.3$  & $120.0$ & $104.8$ & $119.9$ & $105.6$ \\   
    \hline 
    \multicolumn{8}{c}{\rule{0pt}{12pt} \textbf{Free energy barrier $F_{AB}^\ddagger$ [kJ/mol]}}\\    
    \hline
    $F_{AB}^\ddagger$ &&  
    $89.0$& $80.5$ & n/a & n/a & n/a & n/a \\ 
    \hline
    \multicolumn{8}{c}{\rule{0pt}{12pt} \textbf{Rates [s$^{-1}$]}}\\
    \hline
    Eyring TST      & \ref{eq:EyringTST}        & $8.30\times10^{-6}$ & $1.99\times10^{-4}$ & $5.40\times10^{-8}$ &  $2.04\times10^{-5}$  & $5.81\times10^{-8}$ & $1.56\times10^{-5}$\\
    high $T$ TST    & \ref{eq:high_T_TST}     & $2.28\times10^{-6}$ & $6.39\times10^{-5}$ & $1.62\times10^{-8}$ &  $7.00\times10^{-6}$ & $1.65\times10^{-8}$ & $5.19\times10^{-6}$\\
    \hline        
    InMetaD 1        & \ref{eq:InMetaD}          & $1.94\times10^{-6}$ & $1.04\times10^{-4}$ & n/a & n/a & n/a & n/a \\
    InMetaD 2        & \ref{eq:InMetaD}          & $2.63\times10^{-6}$ & $9.22\times10^{-5}$ & n/a & n/a & n/a & n/a \\
    \hline
    Kramers         & \ref{eq:Kramers}          & $4.06\times10^{-3}$ & $1.74\times10^{-1}$ & n/a & n/a & n/a & n/a \\
    Pontryagin      & \ref{eq:Pontryagin}       & $1.14\times10^{-2}$ & $2.74\times10^{-1}$ & n/a & n/a & n/a & n/a \\
    \hline\hline
    \end{tabular}
    \caption{Rates for thermal cis-trans isomerization around the C$_{13}$=C$_{14}$ double bond in \textbf{pSb5}
    }
    \label{tab:pSb5_rates}
\end{table*}
\begin{table*}
    \centering
    \begin{tabular}{|l| c 
    |>{\raggedleft}p{2cm}|>{\raggedleft}p{2cm}|
    |>{\raggedleft}p{2cm}|>{\raggedleft}p{2cm}|
    |>{\raggedleft}p{2cm}|r|
    } 
    \hline\hline
    &       &\multicolumn{2}{c||}{\textbf{DFTB3}} 
            & \multicolumn{2}{c||}{\textbf{DFT}} 
            & \multicolumn{2}{c|}{\textbf{DFT-D3}}  \\
    &Eq.    &trans $\rightarrow$ cis &cis $\rightarrow$ trans 
            &trans $\rightarrow$ cis &cis $\rightarrow$ trans 
            &trans $\rightarrow$ cis &cis $\rightarrow$ trans \\
    \hline
    \multicolumn{8}{c}{\rule{0pt}{12pt} \textbf{Free energy difference between $A$ and $AB^\ddagger$ in Eyring TST [kJ/mol]}}\\
    \hline    
    $E_b$ & & 
    93.7 & 91.1 & 98.0 & 90.1 & 98.2 & 90.4\\
    $\Delta F_{\mathrm{rot}}$ &\ref{eq:DeltaF_rot}& 
    $-0.1$ & $0.0$& $ -0.1 $ & $ -0.0 $ & $-0.1$ & $-0.0$\\
    $\Delta F_{\mathrm{vib}}$ &\ref{eq:DeltaF_rot}&
    $-7.7$  & $-12.0$ & $ -2.4 $ & $ -3.8 $ & $-4.1$ & $-3.7$\\
    \hline
    $\Delta F^{\ddagger}$ &\ref{eq:deltaG_Eyr_2}& 
    $85.8$& $79.2$ & $95.4$ & $86.2$ & $94.0$ & $86.7$\\
    \hline 
    \multicolumn{8}{c}{\rule{0pt}{12pt} \textbf{Free energy difference $\Delta F^\ddagger$ between $A$ and $AB^\ddagger$ in high-temperature TST [kJ/mol]}}\\    
    \hline
    $E_b$ & & 
    93.7 & 91.1 & 98.0 & 90.1 & 98.2 & 90.4\\
    $\Delta F_{\mathrm{vib}}^{\mathrm{ht}}$ &\ref{eq:deltaG_high_T}& 
    $-5.1$ & $-9.4$ & $ 0.5 $ & $ -1.1 $ & $-1.1$ & $-0.9$\\
    \hline 
    $\Delta F^{\ddagger, \mathrm{ht}}$ &\ref{eq:deltaG_high_T}&  
    $88.5$& $81.7$ & $98.4$ & $89.0$ & $97.0$ & $89.5$ \\   
    \hline 
    \multicolumn{8}{c}{\rule{0pt}{12pt} \textbf{Free energy barrier $F_{AB}^\ddagger$ [kJ/mol]}}\\    
    \hline
    $F_{AB}^\ddagger$ &&  
    $78.9$& $75.0$ & n/a & n/a & n/a & n/a \\   
    \hline
    \multicolumn{8}{c}{\rule{0pt}{12pt} \textbf{Rates [s$^{-1}$]}}\\
    \hline
    Eyring TST      & \ref{eq:EyringTST}    & $1.42\times10^{-2}$ & $2.06\times10^{-1}$  & $3.09\times10^{-4}$ &  $1.24\times10^{-2}$ & $5.35\times10^{-4}$ & $9.92\times10^{-3}$ \\
    high $T$ TST    & \ref{eq:high_T_TST}   & $4.81\times10^{-3}$ & $7.44\times10^{-2}$ & $9.21\times10^{-5}$ &  $4.12\times10^{-3}$ & $1.58\times10^{-4}$ & $3.31\times10^{-3}$ \\
    \hline    
    InMetaD1        & \ref{eq:InMetaD}      & $2.97\times10^{-3}$ & $9.30\times10^{-3}$ & n/a & n/a & n/a & n/a\\
    InMetaD2        & \ref{eq:InMetaD}      & $3.09\times10^{-3}$ & $1.22\times10^{-2}$ & n/a & n/a & n/a & n/a\\
    \hline    
    Kramers         & \ref{eq:Kramers}      & $1.92\times10^{-1}$ & $1.30\times10^{0}$ & n/a & n/a & n/a & n/a\\
    Pontryagin      & \ref{eq:Pontryagin}   & $5.43\times10^{-1}$ & $2.07\times10^{0}$  & n/a & n/a & n/a & n/a\\
    \hline\hline
    \end{tabular}
    \caption{Rates for thermal cis-trans isomerization around the C$_{13}$=C$_{14}$ double bond in \textbf{pSb1}.}
    \label{tab:pSb1_rates}
\end{table*}
%

\subsection{Comparison across different PES}
Tables \ref{tab:pSb5_rates} and \ref{tab:pSb1_rates} compare
the energy barriers and the rotational and vibrational contribution to the free energy differences  at the level of DFTB3 and to those at unrestricted DFT/B3LYP.
The potential energy barriers $E_b$ from DFT/B3LYP calculations closely aligning with literature-reported values [2, 20, 21].
We used different algorithms for the transition state search for DFTB3 and unrestricted DFT/B3LYP (nudge elastic band vs. synchronous transit-guided quasi-Newton), but this likely causes only small discrepancies, and the differences in Tables \ref{tab:pSb5_rates} and \ref{tab:pSb1_rates} can be attributed to the underlying PES.

Compared to unrestricted DFT/B3LYP, DFTB3 tends to underestimate the barrier heights, as has been reported previously \cite{zhou2002performance,bondar2004mechanism,elghobashi2018catalysis}.
Energies of constrained optimizations along $\varphi$ for DFT/B3LYP and DFTB3 are shown in Fig.~\ref{supp-fig:D3_comparison} in section \ref{supp-chapter:supplementary_figures} of the supplementary material.
The discrepancy between DFT/B3LYP and DFTB3 is larger for the trans $\rightarrow$ cis transitions than in the cis $\rightarrow$ trans transitions, with a discrepancy as high as $11.4\,\mathrm{kJ/mol}$ in pSb5.
The exception to this trend is the cis $\rightarrow$ trans transition in pSb1, for which the DFT/B3LYP is $1\,\mathrm{kJ/mol}$ lower than the DFTB3 barrier.
The rotational contributions to total free energy difference are  nearly identical at DFT/B3LYP and at DFTB3, which can be attributed to the rigid molecular scaffold.
The vibrational motion lowers the total free energy barrier in all four reactions. 
However, the effect is smaller at the level of DFT/B3LYP than with DFTB3. 
The discrepancy between DFT/B3LYP an DFTB3 for the vibrational contribution can be as large as the discrepancy for the potential energy barrier $E_b$
(see e.g. cis $\rightarrow$ trans reaction in pSb1).
This highlights the need to consider not only the potential energy barrier but also the vibrational free energy of the reactant and transition state when comparing different PES.
Overall, we find that the total free energy difference $\Delta F^\ddagger$ is 6 to 13 kJ/mol larger in DFT/B3LYP than in DFTB3. 
Consequently, the Eyring TST rate constants at the level of DFT/B3LYP are one to two orders of magnitude lower than at the level of DFTB3.
As with DFTB3, the high-temperature rate constant is slightly lower than the Eyring TST rate constant, because the reduction of the total free energy difference due to the vibrational contribution is underestimated when using the high-temperature limit.
Rates from infrequent metadynamics or reaction-coordinate based methods are not available, because they require ab-initio MD simulations.
The necessary simulation time to converge these rate estimates is challenging to attain at this level of theory.

The last two columns in Tables \ref{tab:pSb5_rates} and \ref{tab:pSb1_rates} report the influence of the the D3 dispersion correction for DFT \cite{grimme2010consistent} on the energy barriers and the free energy contributions to the rates.
The effect is less than $1\,\mathrm{kJ/mol}$ (only exception: $\Delta F_{\mathrm{vib}}$ for trans $\rightarrow$ cis in pSb1). 
The difference in the energies for constrained optimizations along $\varphi$ (Fig.~\ref{supp-fig:D3_comparison} in the supplement) is equally small.
As a result, there is a minimal difference in the rates when calculated with and without D3 correction.
We suspect that the small influence of the D3 correction on the PES stems from the rigid structure of the two molecules.
Due to the polyene scaffold, most pairwise distances do not change during the reaction.
The few distances that do change significantly are situated at opposite ends of the molecule and have minimal interaction through dispersion interactions.

\section{Methods}
Calculations for DFTB3 were carried out with the DFTB+ software package \cite{hourahine2020dftb+} using the 3ob-3-1 Slater-Koster parameter set \cite{gaus2013parametrization}.
Energy minimizations, constrained optimizations and Nudged Elastic Band (NEB) calculations were done by interfacing DFTB+ with the Atomic Simulation Environment (ASE) \cite{larsen2017atomic} and using the Broyden-Fletcher-Goldfarb-Shanno (BFGS)  algorithm \cite{fletcher2000practical} for numerical optimization.
Vibrational analysis of the optimized structures was done using DFTB+ to obtain the vibrational frequencies, while rotational moments of inertia were calculated by entering the optimized configuration into the Gaussian 16 software \cite{g16}.
From these data we calculated rates for Eyring TST and high-temperature TST.
Ab-initio MD simulations were performed using DFTB+ using the velocity-Verlet integrator with a time step of 1 fs.
Before simulations, energy minimization was done, followed by temperature equilibration in two steps.
In a first equilibration run, the Berendsen thermostat \cite{berendsen1984molecular} is employed, while in a second equilibration run a Nos\'e-Hoover chain setup \cite{nose1984unified,hoover1985canonical,martyna1992nose} of chain length 3 is used.
For production runs, the same thermostat setup was used as for the second equilibration runs.
Well-tempered metadynamics \cite{barducci2008well} and umbrella sampling \cite{torrie1977nonphysical} were carried out by plugging the PLUMED \cite{tribello2014plumed} software package with DFTB+.
Parameter sets for metadynamics and umbrella sampling sets can be found in Tables \ref{tab:metad_parameters} and \ref{tab:us_parameters} respectively.
Sets of runs for infrequent metadynamics were set up by equilibrating in the reactant state, after which a metadynamics bias is applied until a transition is registered.
The transition times were reweighted using the acceleration factor which was directly calculated by PLUMED.
The set of reweighted transition times was fitted to the TCDF of a Poisson distribution to obtain a mean first-passage time and corresponding rate.
Diffusion profiles were calculated using the method from Ref~\onlinecite{hummer2005position}.
Effective masses of the reactant states were calculated by measuring the average squared velocity along the dihedral angle and using the equipartition theorem.
Frequencies of the harmonic approximations of the reactant wells and transition state barriers were calculated from spring constants obtained by harmonically fitting the corresponding wells or barriers.
Free energy barriers $F^\ddagger_{AB}$ are measured from the FES directly.
One-dimensional rate methods (Kramers and Pontryagin) can then be applied straightforwardly.
\begin{table}[h!]
    \centering
\begin{tabular}{|l  c  c  r  c  c  c |}
\hline
\hline
  & height      & width     & pace  & bias   &  time &runs\\
  & [kJ/mol]    & [rad]     & [ps]  & factor &  [ns] & \\
\hline
\textbf{pSb5}   & \multicolumn{6}{c|}{}   \\
MetaD1 & 1.3 & 0.15 & 0.5 & 16 & 152 &n/a\\
MetaD2 & 1.3 & 0.10 & 0.5 & 16 & 87 &n/a\\
MetaD3 & 0.75 & 0.075 & 0.25 & 25 & 51 &n/a\\
InMetaD1 & 1.3 & 0.05 & 5.0 & 16 & n/a   &25\\
InMetaD2 & 1.3 & 0.05 & 10.0 & 16 & n/a &30\\
\hline
\textbf{pSb1}   & \multicolumn{6}{c|}{}   \\
MetaD1 & 1.3 & 0.10 & 0.5 & 16 & 86 &n/a\\
MetaD2 & 1.3 & 0.05 & 0.5 & 16 & 63 &n/a\\
MetaD3 & 1.3 & 0.10 & 0.5 & 16 & 24 &n/a\\
InMetaD1 & 1.3 & 0.05 & 5.0 & 16 & n/a   &25\\
InMetaD2 & 1.3 & 0.05 & 10.0 & 16 & n/a &30\\
\hline
\hline
\end{tabular}
\caption{Parameters for metadynamics and infrequent metadynamics simulations for pSb5 and pSb1 using DFTB3.
}
\label{tab:metad_parameters}
\end{table}
\begin{table}[h]
    \centering
    \begin{tabular}{|c|c|c|c|}
    \hline\hline
        windows & biased region  & interval & force  constant\\
                & [rad]    & [rad]    & [$\mathrm{kJ}/(\mathrm{mol} \cdot \mathrm{rad}^{2)}$]\\
    \hline
    63 &$[-3.1, +3.1]$     &0.1    &500\\
    10 &$[-1.95, -1.05]$   &0.1    &500\\
    10 &$[+1.05, +1.95]$   &0.1    &500\\
    \hline\hline    
    \end{tabular}
    \caption{Parameters for umbrella sampling using DFTB3. 
    Each umbrella sampling set was run with 83 windows positioned as shown here.
    In total, three sets were run for pSb5 and three for pSb1 (Fig.~\ref{fig:FES_diff}).}
   \label{tab:us_parameters}
\end{table}
Calculations at the DFT level were performed using the Gaussian 16 software  \cite{g16} using unrestricted DFT with the B3LYP functional \cite{lee1988development,becke1988density} and the 6-31G* basis set.
Full geometry optimizations as well as constrained optimizations were done using the Berny optimization algorithm \cite{schlegel1982optimization} as implemented in Gaussian.
Transition state search was performed using the Synchronous Transit-guided Quasi-Newton (STQN) method \cite{peng1993combining,peng1996using} as implemented in Gaussian, where the reactant and product state input configurations were chosen to be the geometry optimized structures in the trans and cis states. 
Gaussian performs a full thermochemical analysis including calculation of the translational, rotational and vibrational partition functions and corresponding energies and
entropies \cite{ochterski2000thermochemistry}. 

This allows for straightforward calculation of rates for Eyring TST.
Vibrational frequencies were obtained from Gaussian separately \cite{ochterski1999vibrational} and used to calculate rates for high-temperature TST.

A complete overview of the computational details is given in section B of the Supporting Information.

\section{Conclusions}

\begin{figure}
    \centering
    \includegraphics[width=8cm]{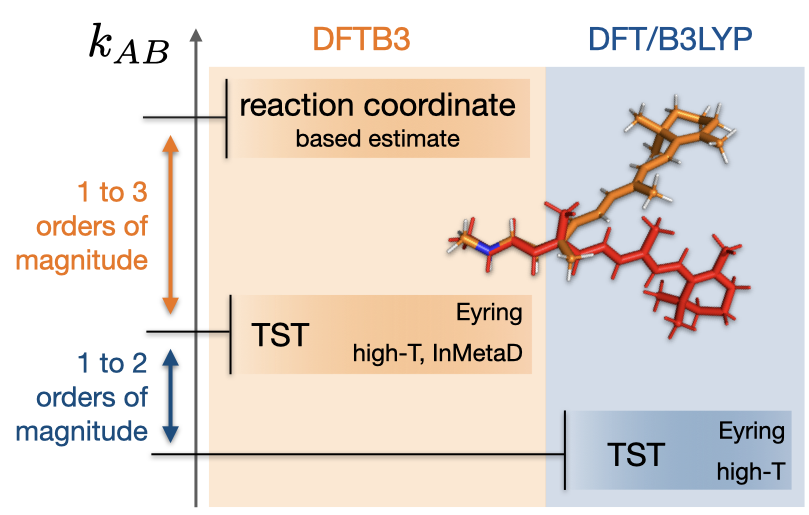}
    \caption{Effect of rate model and PES on the estimated reaction rate constant.}
    \label{fig:summary}
\end{figure}
We studied the thermal cis-trans isomerization in two retinal analogues at $300\,\mathrm{K}$ in the gas phase. 
This reaction falls well within the approximation of TST, and thus accurate values for the reaction rate constant can be obtained from this theoretical framework. 
However, reactions in molecules with numerous flexible degrees of freedom or reactions in complex environments may not be accurately modeled by TST.
We therefore explored whether accurate estimates of the reaction rate constant can be obtained from MD simulations. 
The impact of changing the theoretical framework for modelling the reaction rate must be assessed relative to the error in the potential energy surface (PES), which is often considered to be the primary error source in rate modeling. 
To gauge the effect of changing the PES, we compared TST rate constants at the level of  at the DFTB3 and the DFT/B3LYP level. 
Fig.~\ref{fig:summary} summarizes our results.
Reaction rate constants at DFTB3 are larger than those at DFT/B3LYP, with a difference of one to two orders magnitude. 
However only in pSb5 the increase in reaction rate can be attributed to a lower potential energy barrier. 
In pSb1, the change of the vibrational free energy has an equally strong (trans $\rightarrow$ cis) or even larger (cis $\rightarrow$ trans) contribution to the increase of the reaction rate constant.
Thus, reducing the comparison of different PES to the height of the potential energy barrier and neglecting entropic effects may be misleading.  
For cis-trans isomerization in polyene chains,  DFT/B3LYP underestimates the torsional barrier compared to  CASSCF \cite{paizs1999electronic} as well as compared to experimental data \cite{okada2004retinal,buda2000density,bondar2010mechanism}.
Thus, the true reaction rate constant might be even lower than our DFT/B3LYP estimates.
An important assumption when estimating reaction rate constants from MD simulations is that classical mechanics are used to model the dynamics of the nuclei and to approximate the partition functions. 
The classical limit is well justified for rugged potential energy landscapes with broad minima and relatively low energy barriers, but might not be appropriate for modelling a chemical reaction.
In both of our molecules, the classical limit (high-temperature TST) yields a lower rate constants than Eyring TST, but the effect is less than an order of magnitude. 
Thus, the error due to the classical limit is small compared to the uncertainty due to the model of the PES. 
Since the classical limit is justified for our systems, one should in principle be able to estimate reaction rate constants from MD simulations. 
Indeed, the infrequent metadynamics results are in excellent agreement with results from high-temperature TST. 
The length of the simulations (on a biased PES) were in the nanosecond regime, the mean-first passage times $\tau_{AB} = 1/k_{AB}$ are in the regime of hundreds to hundred thousands of seconds.
This is a enormous speedup, with the largest acceleration factors being of the order of $10^{14}$ .
By contrast, our results from Kramers and Pontryagin's rate theory overestimate the rate constant by multiple orders of magnitude.
These two methods rely on a reaction coordinate - in our case the C$_{13}$=C$_{14}$ torsion angle $\varphi$ - but the FES can be very sensitive to the choice of this reaction coordinate.
In fact, we found that the improper dihedrals of the substituents on the C$_{13}$ and C$_{14}$ atoms correlate with the reaction coordinate $\varphi$.
We hypothesized that this was indicative of a isomerization mechanism consisting of a concerted motion where the C$_{14}$ atom temporarily nods out of the polyene plain, before the isomerization is completed.
Thus, even though the torsion angle $\varphi$ is a very intuitive reaction coordinate, it might not be optimal enough to yield accurate results for Kramers and Pontryagin's rate theory.
This places us in a challenging position.
Both MD based-approaches, infrequent metadynamics and reaction coordinate based rate models, come with a high computational cost.
Our simulations required approximately 10 to 100 ns for each rate estimation.
However, since infrequent metadynamics is derived from TST, it is particularly suitable for chemical reactions that align well with Eyring TST.
By contrast, reaction coordinate based rate theories have the potential to model systems deviating from the harmonic approximation and the separation of timescales. 
However, their robustness is compromised due to sensitivity to the reaction coordinate and the friction regime.
Fortunately, several alternatives exist. 
Within reaction coordinate-based rate models, one can optimize the reaction coordinate \cite{vanden2010transition, roux2021string}, potentially utilizing neural networks \cite{gkeka2020machine, jung2023machine}. 
Another avenue extends these models to encompass effective dynamics within multidimensional collective variable spaces \cite{Lie2013, donati2021markov}. 
Alternatively, one can opt for transition path sampling \cite{bolhuis2002transition, zuckerman2017weighted, palacio2022free} and leverage dynamical reweighting techniques that are not based on TST \cite{donati2018girsanov, kieninger2020dynamical, shmilovich2023girsanov}. 
In summary, while sampling chemical reactions in complex systems poses a formidable challenge, there is optimism that this wide variety of ideas will allow us to solve this task.

%
%
\section{Acknowledgements}

This research has been funded by Deutsche Forschungsgemeinschaft (DFG) 
through grant SFB 1114 "Scaling Cascades in Complex Systems" - project number 235221301.
S.G.~acknowledges funding by the Einstein Center of Catalysis/BIG-NSE.

%
%
\section{References}
\bibliography{literature}

\begin{thebibliography}{76}%
\makeatletter
\providecommand \@ifxundefined [1]{%
 \@ifx{#1\undefined}
}%
\providecommand \@ifnum [1]{%
 \ifnum #1\expandafter \@firstoftwo
 \else \expandafter \@secondoftwo
 \fi
}%
\providecommand \@ifx [1]{%
 \ifx #1\expandafter \@firstoftwo
 \else \expandafter \@secondoftwo
 \fi
}%
\providecommand \natexlab [1]{#1}%
\providecommand \enquote  [1]{``#1''}%
\providecommand \bibnamefont  [1]{#1}%
\providecommand \bibfnamefont [1]{#1}%
\providecommand \citenamefont [1]{#1}%
\providecommand \href@noop [0]{\@secondoftwo}%
\providecommand \href [0]{\begingroup \@sanitize@url \@href}%
\providecommand \@href[1]{\@@startlink{#1}\@@href}%
\providecommand \@@href[1]{\endgroup#1\@@endlink}%
\providecommand \@sanitize@url [0]{\catcode `\\12\catcode `\$12\catcode `\&12\catcode `\#12\catcode `\^12\catcode `\_12\catcode `\%12\relax}%
\providecommand \@@startlink[1]{}%
\providecommand \@@endlink[0]{}%
\providecommand \url  [0]{\begingroup\@sanitize@url \@url }%
\providecommand \@url [1]{\endgroup\@href {#1}{\urlprefix }}%
\providecommand \urlprefix  [0]{URL }%
\providecommand \Eprint [0]{\href }%
\providecommand \doibase [0]{https://doi.org/}%
\providecommand \selectlanguage [0]{\@gobble}%
\providecommand \bibinfo  [0]{\@secondoftwo}%
\providecommand \bibfield  [0]{\@secondoftwo}%
\providecommand \translation [1]{[#1]}%
\providecommand \BibitemOpen [0]{}%
\providecommand \bibitemStop [0]{}%
\providecommand \bibitemNoStop [0]{.\EOS\space}%
\providecommand \EOS [0]{\spacefactor3000\relax}%
\providecommand \BibitemShut  [1]{\csname bibitem#1\endcsname}%
\let\auto@bib@innerbib\@empty
\bibitem [{\citenamefont {Eyring}(1935)}]{eyring1935activated}%
  \BibitemOpen
  \bibfield  {author} {\bibinfo {author} {\bibfnamefont {H.}~\bibnamefont {Eyring}},\ }\bibfield  {title} {\bibinfo {title} {The activated complex in chemical reactions},\ }\href@noop {} {\bibfield  {journal} {\bibinfo  {journal} {The Journal of Chemical Physics}\ }\textbf {\bibinfo {volume} {3}},\ \bibinfo {pages} {107} (\bibinfo {year} {1935})}\BibitemShut {NoStop}%
\bibitem [{\citenamefont {Peters}(2017)}]{peters2017reaction}%
  \BibitemOpen
  \bibfield  {author} {\bibinfo {author} {\bibfnamefont {B.}~\bibnamefont {Peters}},\ }\href@noop {} {\emph {\bibinfo {title} {Reaction rate theory and rare events}}}\ (\bibinfo  {publisher} {Elsevier},\ \bibinfo {year} {2017})\BibitemShut {NoStop}%
\bibitem [{\citenamefont {Frenkel}\ and\ \citenamefont {Smit}(2023)}]{frenkel2023understanding}%
  \BibitemOpen
  \bibfield  {author} {\bibinfo {author} {\bibfnamefont {D.}~\bibnamefont {Frenkel}}\ and\ \bibinfo {author} {\bibfnamefont {B.}~\bibnamefont {Smit}},\ }\href@noop {} {\emph {\bibinfo {title} {Understanding molecular simulation: from algorithms to applications}}}\ (\bibinfo  {publisher} {Elsevier},\ \bibinfo {year} {2023})\BibitemShut {NoStop}%
\bibitem [{\citenamefont {H{\"a}nggi}\ \emph {et~al.}(1990)\citenamefont {H{\"a}nggi}, \citenamefont {Talkner},\ and\ \citenamefont {Borkovec}}]{hanggi1990reaction}%
  \BibitemOpen
  \bibfield  {author} {\bibinfo {author} {\bibfnamefont {P.}~\bibnamefont {H{\"a}nggi}}, \bibinfo {author} {\bibfnamefont {P.}~\bibnamefont {Talkner}},\ and\ \bibinfo {author} {\bibfnamefont {M.}~\bibnamefont {Borkovec}},\ }\bibfield  {title} {\bibinfo {title} {Reaction-rate theory: fifty years after {K}ramers},\ }\href@noop {} {\bibfield  {journal} {\bibinfo  {journal} {Reviews of Modern Physics}\ }\textbf {\bibinfo {volume} {62}},\ \bibinfo {pages} {251} (\bibinfo {year} {1990})}\BibitemShut {NoStop}%
\bibitem [{\citenamefont {Kieninger}\ \emph {et~al.}(2020)\citenamefont {Kieninger}, \citenamefont {Donati},\ and\ \citenamefont {Keller}}]{kieninger2020dynamical}%
  \BibitemOpen
  \bibfield  {author} {\bibinfo {author} {\bibfnamefont {S.}~\bibnamefont {Kieninger}}, \bibinfo {author} {\bibfnamefont {L.}~\bibnamefont {Donati}},\ and\ \bibinfo {author} {\bibfnamefont {B.~G.}\ \bibnamefont {Keller}},\ }\bibfield  {title} {\bibinfo {title} {Dynamical reweighting methods for {M}arkov models},\ }\href@noop {} {\bibfield  {journal} {\bibinfo  {journal} {Current Opinion in Structural Biology}\ }\textbf {\bibinfo {volume} {61}},\ \bibinfo {pages} {124} (\bibinfo {year} {2020})}\BibitemShut {NoStop}%
\bibitem [{\citenamefont {Peters}(2016)}]{peters2016reaction}%
  \BibitemOpen
  \bibfield  {author} {\bibinfo {author} {\bibfnamefont {B.}~\bibnamefont {Peters}},\ }\bibfield  {title} {\bibinfo {title} {Reaction coordinates and mechanistic hypothesis tests},\ }\href@noop {} {\bibfield  {journal} {\bibinfo  {journal} {Annual Review of Physical Chemistry}\ }\textbf {\bibinfo {volume} {67}},\ \bibinfo {pages} {669} (\bibinfo {year} {2016})}\BibitemShut {NoStop}%
\bibitem [{\citenamefont {H{\'e}nin}\ \emph {et~al.}(2022)\citenamefont {H{\'e}nin}, \citenamefont {Leli{\`e}vre}, \citenamefont {Shirts}, \citenamefont {Valsson},\ and\ \citenamefont {Delemotte}}]{henin2022enhanced}%
  \BibitemOpen
  \bibfield  {author} {\bibinfo {author} {\bibfnamefont {J.}~\bibnamefont {H{\'e}nin}}, \bibinfo {author} {\bibfnamefont {T.}~\bibnamefont {Leli{\`e}vre}}, \bibinfo {author} {\bibfnamefont {M.}~\bibnamefont {Shirts}}, \bibinfo {author} {\bibfnamefont {O.}~\bibnamefont {Valsson}},\ and\ \bibinfo {author} {\bibfnamefont {L.}~\bibnamefont {Delemotte}},\ }\bibfield  {title} {\bibinfo {title} {Enhanced sampling methods for molecular dynamics simulations},\ }\href@noop {} {\bibfield  {journal} {\bibinfo  {journal} {Living Journal of Computational Molecular Science}\ }\textbf {\bibinfo {volume} {4}},\ \bibinfo {pages} {1583} (\bibinfo {year} {2022})}\BibitemShut {NoStop}%
\bibitem [{\citenamefont {Hummer}(2005)}]{hummer2005position}%
  \BibitemOpen
  \bibfield  {author} {\bibinfo {author} {\bibfnamefont {G.}~\bibnamefont {Hummer}},\ }\bibfield  {title} {\bibinfo {title} {Position-dependent diffusion coefficients and free energies from {B}ayesian analysis of equilibrium and replica molecular dynamics simulations},\ }\href@noop {} {\bibfield  {journal} {\bibinfo  {journal} {New Journal of Physics}\ }\textbf {\bibinfo {volume} {7}},\ \bibinfo {pages} {34} (\bibinfo {year} {2005})}\BibitemShut {NoStop}%
\bibitem [{\citenamefont {Kramers}(1940)}]{kramers1940brownian}%
  \BibitemOpen
  \bibfield  {author} {\bibinfo {author} {\bibfnamefont {H.~A.}\ \bibnamefont {Kramers}},\ }\bibfield  {title} {\bibinfo {title} {Brownian motion in a field of force and the diffusion model of chemical reactions},\ }\href@noop {} {\bibfield  {journal} {\bibinfo  {journal} {Physica}\ }\textbf {\bibinfo {volume} {7}},\ \bibinfo {pages} {284} (\bibinfo {year} {1940})}\BibitemShut {NoStop}%
\bibitem [{\citenamefont {Pontryagin}\ \emph {et~al.}(1933)\citenamefont {Pontryagin}, \citenamefont {Andronov},\ and\ \citenamefont {Vitt}}]{pontryagin1933statistical}%
  \BibitemOpen
  \bibfield  {author} {\bibinfo {author} {\bibfnamefont {L.}~\bibnamefont {Pontryagin}}, \bibinfo {author} {\bibfnamefont {A.}~\bibnamefont {Andronov}},\ and\ \bibinfo {author} {\bibfnamefont {A.}~\bibnamefont {Vitt}},\ }\bibfield  {title} {\bibinfo {title} {On the statistical investigation of dynamic systems},\ }\href@noop {} {\bibfield  {journal} {\bibinfo  {journal} {Journal of Experimental and Theoretical Physics}\ }\textbf {\bibinfo {volume} {3}},\ \bibinfo {pages} {165} (\bibinfo {year} {1933})}\BibitemShut {NoStop}%
\bibitem [{\citenamefont {Tiwary}\ and\ \citenamefont {Parrinello}(2013)}]{tiwary2013metadynamics}%
  \BibitemOpen
  \bibfield  {author} {\bibinfo {author} {\bibfnamefont {P.}~\bibnamefont {Tiwary}}\ and\ \bibinfo {author} {\bibfnamefont {M.}~\bibnamefont {Parrinello}},\ }\bibfield  {title} {\bibinfo {title} {From metadynamics to dynamics},\ }\href@noop {} {\bibfield  {journal} {\bibinfo  {journal} {Physical Review Letters}\ }\textbf {\bibinfo {volume} {111}},\ \bibinfo {pages} {230602} (\bibinfo {year} {2013})}\BibitemShut {NoStop}%
\bibitem [{\citenamefont {Gaus}\ \emph {et~al.}(2011)\citenamefont {Gaus}, \citenamefont {Cui},\ and\ \citenamefont {Elstner}}]{gaus2011dftb3}%
  \BibitemOpen
  \bibfield  {author} {\bibinfo {author} {\bibfnamefont {M.}~\bibnamefont {Gaus}}, \bibinfo {author} {\bibfnamefont {Q.}~\bibnamefont {Cui}},\ and\ \bibinfo {author} {\bibfnamefont {M.}~\bibnamefont {Elstner}},\ }\bibfield  {title} {\bibinfo {title} {{DFTB3}: Extension of the self-consistent-charge density-functional tight-binding method ({SCC-DFTB})},\ }\href@noop {} {\bibfield  {journal} {\bibinfo  {journal} {Journal of Chemical Theory and Computation}\ }\textbf {\bibinfo {volume} {7}},\ \bibinfo {pages} {931} (\bibinfo {year} {2011})}\BibitemShut {NoStop}%
\bibitem [{\citenamefont {Pracht}\ \emph {et~al.}(2020)\citenamefont {Pracht}, \citenamefont {Bohle},\ and\ \citenamefont {Grimme}}]{pracht2020automated}%
  \BibitemOpen
  \bibfield  {author} {\bibinfo {author} {\bibfnamefont {P.}~\bibnamefont {Pracht}}, \bibinfo {author} {\bibfnamefont {F.}~\bibnamefont {Bohle}},\ and\ \bibinfo {author} {\bibfnamefont {S.}~\bibnamefont {Grimme}},\ }\bibfield  {title} {\bibinfo {title} {Automated exploration of the low-energy chemical space with fast quantum chemical methods},\ }\href@noop {} {\bibfield  {journal} {\bibinfo  {journal} {Physical Chemistry Chemical Physics}\ }\textbf {\bibinfo {volume} {22}},\ \bibinfo {pages} {7169} (\bibinfo {year} {2020})}\BibitemShut {NoStop}%
\bibitem [{\citenamefont {Schade}\ \emph {et~al.}(2023)\citenamefont {Schade}, \citenamefont {Kenter}, \citenamefont {Elgabarty}, \citenamefont {Lass}, \citenamefont {K{\"u}hne},\ and\ \citenamefont {Plessl}}]{schade2023breaking}%
  \BibitemOpen
  \bibfield  {author} {\bibinfo {author} {\bibfnamefont {R.}~\bibnamefont {Schade}}, \bibinfo {author} {\bibfnamefont {T.}~\bibnamefont {Kenter}}, \bibinfo {author} {\bibfnamefont {H.}~\bibnamefont {Elgabarty}}, \bibinfo {author} {\bibfnamefont {M.}~\bibnamefont {Lass}}, \bibinfo {author} {\bibfnamefont {T.~D.}\ \bibnamefont {K{\"u}hne}},\ and\ \bibinfo {author} {\bibfnamefont {C.}~\bibnamefont {Plessl}},\ }\bibfield  {title} {\bibinfo {title} {Breaking the exascale barrier for the electronic structure problem in ab-initio molecular dynamics},\ }\href@noop {} {\bibfield  {journal} {\bibinfo  {journal} {The International Journal of High Performance Computing Applications}\ }\textbf {\bibinfo {volume} {37}},\ \bibinfo {pages} {530} (\bibinfo {year} {2023})}\BibitemShut {NoStop}%
\bibitem [{\citenamefont {Senn}\ and\ \citenamefont {Thiel}(2009)}]{senn2009qm}%
  \BibitemOpen
  \bibfield  {author} {\bibinfo {author} {\bibfnamefont {H.~M.}\ \bibnamefont {Senn}}\ and\ \bibinfo {author} {\bibfnamefont {W.}~\bibnamefont {Thiel}},\ }\bibfield  {title} {\bibinfo {title} {{QM/MM} methods for biomolecular systems},\ }\href@noop {} {\bibfield  {journal} {\bibinfo  {journal} {Angewandte Chemie International Edition}\ }\textbf {\bibinfo {volume} {48}},\ \bibinfo {pages} {1198} (\bibinfo {year} {2009})}\BibitemShut {NoStop}%
\bibitem [{\citenamefont {Senftle}\ \emph {et~al.}(2016)\citenamefont {Senftle}, \citenamefont {Hong}, \citenamefont {Islam}, \citenamefont {Kylasa}, \citenamefont {Zheng}, \citenamefont {Shin}, \citenamefont {Junkermeier}, \citenamefont {Engel-Herbert}, \citenamefont {Janik}, \citenamefont {Aktulga} \emph {et~al.}}]{senftle2016reaxff}%
  \BibitemOpen
  \bibfield  {author} {\bibinfo {author} {\bibfnamefont {T.~P.}\ \bibnamefont {Senftle}}, \bibinfo {author} {\bibfnamefont {S.}~\bibnamefont {Hong}}, \bibinfo {author} {\bibfnamefont {M.~M.}\ \bibnamefont {Islam}}, \bibinfo {author} {\bibfnamefont {S.~B.}\ \bibnamefont {Kylasa}}, \bibinfo {author} {\bibfnamefont {Y.}~\bibnamefont {Zheng}}, \bibinfo {author} {\bibfnamefont {Y.~K.}\ \bibnamefont {Shin}}, \bibinfo {author} {\bibfnamefont {C.}~\bibnamefont {Junkermeier}}, \bibinfo {author} {\bibfnamefont {R.}~\bibnamefont {Engel-Herbert}}, \bibinfo {author} {\bibfnamefont {M.~J.}\ \bibnamefont {Janik}}, \bibinfo {author} {\bibfnamefont {H.~M.}\ \bibnamefont {Aktulga}}, \emph {et~al.},\ }\bibfield  {title} {\bibinfo {title} {The {ReaxFF} reactive force-field: development, applications and future directions},\ }\href@noop {} {\bibfield  {journal} {\bibinfo  {journal} {npj Computational Materials}\ }\textbf {\bibinfo {volume} {2}},\ \bibinfo {pages} {1} (\bibinfo {year} {2016})}\BibitemShut {NoStop}%
\bibitem [{\citenamefont {Gkeka}\ \emph {et~al.}(2020)\citenamefont {Gkeka}, \citenamefont {Stoltz}, \citenamefont {Barati~Farimani}, \citenamefont {Belkacemi}, \citenamefont {Ceriotti}, \citenamefont {Chodera}, \citenamefont {Dinner}, \citenamefont {Ferguson}, \citenamefont {Maillet}, \citenamefont {Minoux} \emph {et~al.}}]{gkeka2020machine}%
  \BibitemOpen
  \bibfield  {author} {\bibinfo {author} {\bibfnamefont {P.}~\bibnamefont {Gkeka}}, \bibinfo {author} {\bibfnamefont {G.}~\bibnamefont {Stoltz}}, \bibinfo {author} {\bibfnamefont {A.}~\bibnamefont {Barati~Farimani}}, \bibinfo {author} {\bibfnamefont {Z.}~\bibnamefont {Belkacemi}}, \bibinfo {author} {\bibfnamefont {M.}~\bibnamefont {Ceriotti}}, \bibinfo {author} {\bibfnamefont {J.~D.}\ \bibnamefont {Chodera}}, \bibinfo {author} {\bibfnamefont {A.~R.}\ \bibnamefont {Dinner}}, \bibinfo {author} {\bibfnamefont {A.~L.}\ \bibnamefont {Ferguson}}, \bibinfo {author} {\bibfnamefont {J.-B.}\ \bibnamefont {Maillet}}, \bibinfo {author} {\bibfnamefont {H.}~\bibnamefont {Minoux}}, \emph {et~al.},\ }\bibfield  {title} {\bibinfo {title} {Machine learning force fields and coarse-grained variables in molecular dynamics: application to materials and biological systems},\ }\href@noop {} {\bibfield  {journal} {\bibinfo  {journal} {Journal of Chemical Theory and Computation}\ }\textbf {\bibinfo {volume} {16}},\ \bibinfo {pages}
  {4757} (\bibinfo {year} {2020})}\BibitemShut {NoStop}%
\bibitem [{\citenamefont {Bondar}\ \emph {et~al.}(2011)\citenamefont {Bondar}, \citenamefont {Knapp-Mohammady}, \citenamefont {Suhai}, \citenamefont {Fischer},\ and\ \citenamefont {Smith}}]{bondar2011ground}%
  \BibitemOpen
  \bibfield  {author} {\bibinfo {author} {\bibfnamefont {A.-N.}\ \bibnamefont {Bondar}}, \bibinfo {author} {\bibfnamefont {M.}~\bibnamefont {Knapp-Mohammady}}, \bibinfo {author} {\bibfnamefont {S.}~\bibnamefont {Suhai}}, \bibinfo {author} {\bibfnamefont {S.}~\bibnamefont {Fischer}},\ and\ \bibinfo {author} {\bibfnamefont {J.~C.}\ \bibnamefont {Smith}},\ }\bibfield  {title} {\bibinfo {title} {Ground-state properties of the retinal molecule: from quantum mechanical to classical mechanical computations of retinal proteins},\ }\href@noop {} {\bibfield  {journal} {\bibinfo  {journal} {Theoretical Chemistry Accounts}\ }\textbf {\bibinfo {volume} {130}},\ \bibinfo {pages} {1169} (\bibinfo {year} {2011})}\BibitemShut {NoStop}%
\bibitem [{\citenamefont {Tavan}\ \emph {et~al.}(1985)\citenamefont {Tavan}, \citenamefont {Schulten},\ and\ \citenamefont {Oesterhelt}}]{tavan1985effect}%
  \BibitemOpen
  \bibfield  {author} {\bibinfo {author} {\bibfnamefont {P.}~\bibnamefont {Tavan}}, \bibinfo {author} {\bibfnamefont {K.}~\bibnamefont {Schulten}},\ and\ \bibinfo {author} {\bibfnamefont {D.}~\bibnamefont {Oesterhelt}},\ }\bibfield  {title} {\bibinfo {title} {The effect of protonation and electrical interactions on the stereochemistry of retinal {S}chiff bases},\ }\href@noop {} {\bibfield  {journal} {\bibinfo  {journal} {Biophysical Journal}\ }\textbf {\bibinfo {volume} {47}},\ \bibinfo {pages} {415} (\bibinfo {year} {1985})}\BibitemShut {NoStop}%
\bibitem [{\citenamefont {Tajkhorshid}\ and\ \citenamefont {Suhai}(1999{\natexlab{a}})}]{tajkhorshid1999influence}%
  \BibitemOpen
  \bibfield  {author} {\bibinfo {author} {\bibfnamefont {E.}~\bibnamefont {Tajkhorshid}}\ and\ \bibinfo {author} {\bibfnamefont {S.}~\bibnamefont {Suhai}},\ }\bibfield  {title} {\bibinfo {title} {Influence of the methyl groups on the structure, charge distribution, and proton affinity of the retinal {S}chiff base},\ }\href@noop {} {\bibfield  {journal} {\bibinfo  {journal} {The Journal of Physical Chemistry B}\ }\textbf {\bibinfo {volume} {103}},\ \bibinfo {pages} {5581} (\bibinfo {year} {1999}{\natexlab{a}})}\BibitemShut {NoStop}%
\bibitem [{\citenamefont {Gozem}\ \emph {et~al.}(2012)\citenamefont {Gozem}, \citenamefont {Huntress}, \citenamefont {Schapiro}, \citenamefont {Lindh}, \citenamefont {Granovsky}, \citenamefont {Angeli},\ and\ \citenamefont {Olivucci}}]{gozem2012dynamic}%
  \BibitemOpen
  \bibfield  {author} {\bibinfo {author} {\bibfnamefont {S.}~\bibnamefont {Gozem}}, \bibinfo {author} {\bibfnamefont {M.}~\bibnamefont {Huntress}}, \bibinfo {author} {\bibfnamefont {I.}~\bibnamefont {Schapiro}}, \bibinfo {author} {\bibfnamefont {R.}~\bibnamefont {Lindh}}, \bibinfo {author} {\bibfnamefont {A.~A.}\ \bibnamefont {Granovsky}}, \bibinfo {author} {\bibfnamefont {C.}~\bibnamefont {Angeli}},\ and\ \bibinfo {author} {\bibfnamefont {M.}~\bibnamefont {Olivucci}},\ }\bibfield  {title} {\bibinfo {title} {Dynamic electron correlation effects on the ground state potential energy surface of a retinal chromophore model},\ }\href@noop {} {\bibfield  {journal} {\bibinfo  {journal} {Journal of Chemical Theory and Computation}\ }\textbf {\bibinfo {volume} {8}},\ \bibinfo {pages} {4069} (\bibinfo {year} {2012})}\BibitemShut {NoStop}%
\bibitem [{\citenamefont {Elstner}(2006)}]{elstner2006scc}%
  \BibitemOpen
  \bibfield  {author} {\bibinfo {author} {\bibfnamefont {M.}~\bibnamefont {Elstner}},\ }\bibfield  {title} {\bibinfo {title} {The {SCC-DFTB} method and its application to biological systems},\ }\href@noop {} {\bibfield  {journal} {\bibinfo  {journal} {Theoretical Chemistry Accounts}\ }\textbf {\bibinfo {volume} {116}},\ \bibinfo {pages} {316} (\bibinfo {year} {2006})}\BibitemShut {NoStop}%
\bibitem [{\citenamefont {Lee}\ \emph {et~al.}(1988)\citenamefont {Lee}, \citenamefont {Yang},\ and\ \citenamefont {Parr}}]{lee1988development}%
  \BibitemOpen
  \bibfield  {author} {\bibinfo {author} {\bibfnamefont {C.}~\bibnamefont {Lee}}, \bibinfo {author} {\bibfnamefont {W.}~\bibnamefont {Yang}},\ and\ \bibinfo {author} {\bibfnamefont {R.~G.}\ \bibnamefont {Parr}},\ }\bibfield  {title} {\bibinfo {title} {Development of the {C}olle-{S}alvetti correlation-energy formula into a functional of the electron density},\ }\href@noop {} {\bibfield  {journal} {\bibinfo  {journal} {Physical Review B}\ }\textbf {\bibinfo {volume} {37}},\ \bibinfo {pages} {785} (\bibinfo {year} {1988})}\BibitemShut {NoStop}%
\bibitem [{\citenamefont {Becke}(1988)}]{becke1988density}%
  \BibitemOpen
  \bibfield  {author} {\bibinfo {author} {\bibfnamefont {A.~D.}\ \bibnamefont {Becke}},\ }\bibfield  {title} {\bibinfo {title} {Density-functional exchange-energy approximation with correct asymptotic behavior},\ }\href@noop {} {\bibfield  {journal} {\bibinfo  {journal} {Physical Review A}\ }\textbf {\bibinfo {volume} {38}},\ \bibinfo {pages} {3098} (\bibinfo {year} {1988})}\BibitemShut {NoStop}%
\bibitem [{\citenamefont {Salvalaglio}\ \emph {et~al.}(2014)\citenamefont {Salvalaglio}, \citenamefont {Tiwary},\ and\ \citenamefont {Parrinello}}]{salvalaglio2014assessing}%
  \BibitemOpen
  \bibfield  {author} {\bibinfo {author} {\bibfnamefont {M.}~\bibnamefont {Salvalaglio}}, \bibinfo {author} {\bibfnamefont {P.}~\bibnamefont {Tiwary}},\ and\ \bibinfo {author} {\bibfnamefont {M.}~\bibnamefont {Parrinello}},\ }\bibfield  {title} {\bibinfo {title} {Assessing the reliability of the dynamics reconstructed from metadynamics},\ }\href@noop {} {\bibfield  {journal} {\bibinfo  {journal} {Journal of Chemical Theory and Computation}\ }\textbf {\bibinfo {volume} {10}},\ \bibinfo {pages} {1420} (\bibinfo {year} {2014})}\BibitemShut {NoStop}%
\bibitem [{\citenamefont {Grubm{\"u}ller}(1995)}]{grubmuller1995predicting}%
  \BibitemOpen
  \bibfield  {author} {\bibinfo {author} {\bibfnamefont {H.}~\bibnamefont {Grubm{\"u}ller}},\ }\bibfield  {title} {\bibinfo {title} {Predicting slow structural transitions in macromolecular systems: conformational flooding},\ }\href@noop {} {\bibfield  {journal} {\bibinfo  {journal} {Physical Review E}\ }\textbf {\bibinfo {volume} {52}},\ \bibinfo {pages} {2893} (\bibinfo {year} {1995})}\BibitemShut {NoStop}%
\bibitem [{\citenamefont {Voter}(1997)}]{voter1997hyperdynamics}%
  \BibitemOpen
  \bibfield  {author} {\bibinfo {author} {\bibfnamefont {A.~F.}\ \bibnamefont {Voter}},\ }\bibfield  {title} {\bibinfo {title} {Hyperdynamics: Accelerated molecular dynamics of infrequent events},\ }\href@noop {} {\bibfield  {journal} {\bibinfo  {journal} {Physical Review Letters}\ }\textbf {\bibinfo {volume} {78}},\ \bibinfo {pages} {3908} (\bibinfo {year} {1997})}\BibitemShut {NoStop}%
\bibitem [{\citenamefont {Khan}\ \emph {et~al.}(2020)\citenamefont {Khan}, \citenamefont {Dickson},\ and\ \citenamefont {Peters}}]{khan2020fluxional}%
  \BibitemOpen
  \bibfield  {author} {\bibinfo {author} {\bibfnamefont {S.~A.}\ \bibnamefont {Khan}}, \bibinfo {author} {\bibfnamefont {B.~M.}\ \bibnamefont {Dickson}},\ and\ \bibinfo {author} {\bibfnamefont {B.}~\bibnamefont {Peters}},\ }\bibfield  {title} {\bibinfo {title} {How fluxional reactants limit the accuracy/efficiency of infrequent metadynamics},\ }\href@noop {} {\bibfield  {journal} {\bibinfo  {journal} {The Journal of Chemical Physics}\ }\textbf {\bibinfo {volume} {153}},\ \bibinfo {pages} {054125} (\bibinfo {year} {2020})}\BibitemShut {NoStop}%
\bibitem [{\citenamefont {Barducci}\ \emph {et~al.}(2008)\citenamefont {Barducci}, \citenamefont {Bussi},\ and\ \citenamefont {Parrinello}}]{barducci2008well}%
  \BibitemOpen
  \bibfield  {author} {\bibinfo {author} {\bibfnamefont {A.}~\bibnamefont {Barducci}}, \bibinfo {author} {\bibfnamefont {G.}~\bibnamefont {Bussi}},\ and\ \bibinfo {author} {\bibfnamefont {M.}~\bibnamefont {Parrinello}},\ }\bibfield  {title} {\bibinfo {title} {Well-tempered metadynamics: a smoothly converging and tunable free-energy method},\ }\href@noop {} {\bibfield  {journal} {\bibinfo  {journal} {Physical Review Letters}\ }\textbf {\bibinfo {volume} {100}},\ \bibinfo {pages} {020603} (\bibinfo {year} {2008})}\BibitemShut {NoStop}%
\bibitem [{\citenamefont {Baudry}\ \emph {et~al.}(1997)\citenamefont {Baudry}, \citenamefont {Crouzy}, \citenamefont {Roux},\ and\ \citenamefont {Smith}}]{baudry1997quantum}%
  \BibitemOpen
  \bibfield  {author} {\bibinfo {author} {\bibfnamefont {J.}~\bibnamefont {Baudry}}, \bibinfo {author} {\bibfnamefont {S.}~\bibnamefont {Crouzy}}, \bibinfo {author} {\bibfnamefont {B.}~\bibnamefont {Roux}},\ and\ \bibinfo {author} {\bibfnamefont {J.~C.}\ \bibnamefont {Smith}},\ }\bibfield  {title} {\bibinfo {title} {Quantum chemical and free energy simulation analysis of retinal conformational energetics},\ }\href@noop {} {\bibfield  {journal} {\bibinfo  {journal} {Journal of Chemical Information and Computer Sciences}\ }\textbf {\bibinfo {volume} {37}},\ \bibinfo {pages} {1018} (\bibinfo {year} {1997})}\BibitemShut {NoStop}%
\bibitem [{\citenamefont {Baudry}\ \emph {et~al.}(1999)\citenamefont {Baudry}, \citenamefont {Crouzy}, \citenamefont {Roux},\ and\ \citenamefont {Smith}}]{baudry1999simulation}%
  \BibitemOpen
  \bibfield  {author} {\bibinfo {author} {\bibfnamefont {J.}~\bibnamefont {Baudry}}, \bibinfo {author} {\bibfnamefont {S.}~\bibnamefont {Crouzy}}, \bibinfo {author} {\bibfnamefont {B.}~\bibnamefont {Roux}},\ and\ \bibinfo {author} {\bibfnamefont {J.~C.}\ \bibnamefont {Smith}},\ }\bibfield  {title} {\bibinfo {title} {Simulation analysis of the retinal conformational equilibrium in dark-adapted bacteriorhodopsin},\ }\href@noop {} {\bibfield  {journal} {\bibinfo  {journal} {Biophysical Journal}\ }\textbf {\bibinfo {volume} {76}},\ \bibinfo {pages} {1909} (\bibinfo {year} {1999})}\BibitemShut {NoStop}%
\bibitem [{\citenamefont {Tajkhorshid}\ and\ \citenamefont {Suhai}(1999{\natexlab{b}})}]{tajkhorshid1999dielectric}%
  \BibitemOpen
  \bibfield  {author} {\bibinfo {author} {\bibfnamefont {E.}~\bibnamefont {Tajkhorshid}}\ and\ \bibinfo {author} {\bibfnamefont {S.}~\bibnamefont {Suhai}},\ }\bibfield  {title} {\bibinfo {title} {Dielectric effects due to the environment on the structure and proton affinity of retinal {S}chiff base models},\ }\href@noop {} {\bibfield  {journal} {\bibinfo  {journal} {Chemical Physics Letters}\ }\textbf {\bibinfo {volume} {299}},\ \bibinfo {pages} {457} (\bibinfo {year} {1999}{\natexlab{b}})}\BibitemShut {NoStop}%
\bibitem [{\citenamefont {Tajkhorshid}\ \emph {et~al.}(1999)\citenamefont {Tajkhorshid}, \citenamefont {Paizs},\ and\ \citenamefont {Suhai}}]{tajkhorshid1999role}%
  \BibitemOpen
  \bibfield  {author} {\bibinfo {author} {\bibfnamefont {E.}~\bibnamefont {Tajkhorshid}}, \bibinfo {author} {\bibfnamefont {B.}~\bibnamefont {Paizs}},\ and\ \bibinfo {author} {\bibfnamefont {S.}~\bibnamefont {Suhai}},\ }\bibfield  {title} {\bibinfo {title} {Role of isomerization barriers in the p{K}a control of the retinal {S}chiff base: a density functional study},\ }\href@noop {} {\bibfield  {journal} {\bibinfo  {journal} {The Journal of Physical Chemistry B}\ }\textbf {\bibinfo {volume} {103}},\ \bibinfo {pages} {4518} (\bibinfo {year} {1999})}\BibitemShut {NoStop}%
\bibitem [{\citenamefont {Zhou}\ \emph {et~al.}(2002)\citenamefont {Zhou}, \citenamefont {Tajkhorshid}, \citenamefont {Frauenheim}, \citenamefont {Suhai},\ and\ \citenamefont {Elstner}}]{zhou2002performance}%
  \BibitemOpen
  \bibfield  {author} {\bibinfo {author} {\bibfnamefont {H.}~\bibnamefont {Zhou}}, \bibinfo {author} {\bibfnamefont {E.}~\bibnamefont {Tajkhorshid}}, \bibinfo {author} {\bibfnamefont {T.}~\bibnamefont {Frauenheim}}, \bibinfo {author} {\bibfnamefont {S.}~\bibnamefont {Suhai}},\ and\ \bibinfo {author} {\bibfnamefont {M.}~\bibnamefont {Elstner}},\ }\bibfield  {title} {\bibinfo {title} {Performance of the {AM1}, {PM3}, and {SCC-DFTB} methods in the study of conjugated {S}chiff base molecules},\ }\href@noop {} {\bibfield  {journal} {\bibinfo  {journal} {Chemical Physics}\ }\textbf {\bibinfo {volume} {277}},\ \bibinfo {pages} {91} (\bibinfo {year} {2002})}\BibitemShut {NoStop}%
\bibitem [{\citenamefont {De~Vico}\ \emph {et~al.}(2002)\citenamefont {De~Vico}, \citenamefont {Page}, \citenamefont {Garavelli}, \citenamefont {Bernardi}, \citenamefont {Basosi},\ and\ \citenamefont {Olivucci}}]{de2002reaction}%
  \BibitemOpen
  \bibfield  {author} {\bibinfo {author} {\bibfnamefont {L.}~\bibnamefont {De~Vico}}, \bibinfo {author} {\bibfnamefont {C.~S.}\ \bibnamefont {Page}}, \bibinfo {author} {\bibfnamefont {M.}~\bibnamefont {Garavelli}}, \bibinfo {author} {\bibfnamefont {F.}~\bibnamefont {Bernardi}}, \bibinfo {author} {\bibfnamefont {R.}~\bibnamefont {Basosi}},\ and\ \bibinfo {author} {\bibfnamefont {M.}~\bibnamefont {Olivucci}},\ }\bibfield  {title} {\bibinfo {title} {Reaction path analysis of the “tunable” photoisomerization selectivity of free and locked retinal chromophores},\ }\href@noop {} {\bibfield  {journal} {\bibinfo  {journal} {Journal of the American Chemical Society}\ }\textbf {\bibinfo {volume} {124}},\ \bibinfo {pages} {4124} (\bibinfo {year} {2002})}\BibitemShut {NoStop}%
\bibitem [{\citenamefont {Gozem}\ \emph {et~al.}(2017)\citenamefont {Gozem}, \citenamefont {Luk}, \citenamefont {Schapiro},\ and\ \citenamefont {Olivucci}}]{gozem2017theory}%
  \BibitemOpen
  \bibfield  {author} {\bibinfo {author} {\bibfnamefont {S.}~\bibnamefont {Gozem}}, \bibinfo {author} {\bibfnamefont {H.~L.}\ \bibnamefont {Luk}}, \bibinfo {author} {\bibfnamefont {I.}~\bibnamefont {Schapiro}},\ and\ \bibinfo {author} {\bibfnamefont {M.}~\bibnamefont {Olivucci}},\ }\bibfield  {title} {\bibinfo {title} {Theory and simulation of the ultrafast double-bond isomerization of biological chromophores},\ }\href@noop {} {\bibfield  {journal} {\bibinfo  {journal} {Chemical Reviews}\ }\textbf {\bibinfo {volume} {117}},\ \bibinfo {pages} {13502} (\bibinfo {year} {2017})}\BibitemShut {NoStop}%
\bibitem [{\citenamefont {Zen}\ \emph {et~al.}(2015)\citenamefont {Zen}, \citenamefont {Coccia}, \citenamefont {Gozem}, \citenamefont {Olivucci},\ and\ \citenamefont {Guidoni}}]{zen2015quantum}%
  \BibitemOpen
  \bibfield  {author} {\bibinfo {author} {\bibfnamefont {A.}~\bibnamefont {Zen}}, \bibinfo {author} {\bibfnamefont {E.}~\bibnamefont {Coccia}}, \bibinfo {author} {\bibfnamefont {S.}~\bibnamefont {Gozem}}, \bibinfo {author} {\bibfnamefont {M.}~\bibnamefont {Olivucci}},\ and\ \bibinfo {author} {\bibfnamefont {L.}~\bibnamefont {Guidoni}},\ }\bibfield  {title} {\bibinfo {title} {Quantum {M}onte {C}arlo treatment of the charge transfer and diradical electronic character in a retinal chromophore minimal model},\ }\href@noop {} {\bibfield  {journal} {\bibinfo  {journal} {Journal of Chemical Theory and Computation}\ }\textbf {\bibinfo {volume} {11}},\ \bibinfo {pages} {992} (\bibinfo {year} {2015})}\BibitemShut {NoStop}%
\bibitem [{\citenamefont {Aradi}\ \emph {et~al.}(2007)\citenamefont {Aradi}, \citenamefont {Hourahine},\ and\ \citenamefont {Frauenheim}}]{aradi2007dftb+}%
  \BibitemOpen
  \bibfield  {author} {\bibinfo {author} {\bibfnamefont {B.}~\bibnamefont {Aradi}}, \bibinfo {author} {\bibfnamefont {B.}~\bibnamefont {Hourahine}},\ and\ \bibinfo {author} {\bibfnamefont {T.}~\bibnamefont {Frauenheim}},\ }\bibfield  {title} {\bibinfo {title} {{DFTB+}, a sparse matrix-based implementation of the {DFTB} method},\ }\href@noop {} {\bibfield  {journal} {\bibinfo  {journal} {The Journal of Physical Chemistry A}\ }\textbf {\bibinfo {volume} {111}},\ \bibinfo {pages} {5678} (\bibinfo {year} {2007})}\BibitemShut {NoStop}%
\bibitem [{\citenamefont {Hourahine}\ \emph {et~al.}(2007)\citenamefont {Hourahine}, \citenamefont {Sanna}, \citenamefont {Aradi}, \citenamefont {K{\"o}hler}, \citenamefont {Niehaus},\ and\ \citenamefont {Frauenheim}}]{hourahine2007self}%
  \BibitemOpen
  \bibfield  {author} {\bibinfo {author} {\bibfnamefont {B.}~\bibnamefont {Hourahine}}, \bibinfo {author} {\bibfnamefont {S.}~\bibnamefont {Sanna}}, \bibinfo {author} {\bibfnamefont {B.}~\bibnamefont {Aradi}}, \bibinfo {author} {\bibfnamefont {C.}~\bibnamefont {K{\"o}hler}}, \bibinfo {author} {\bibfnamefont {T.}~\bibnamefont {Niehaus}},\ and\ \bibinfo {author} {\bibfnamefont {T.}~\bibnamefont {Frauenheim}},\ }\bibfield  {title} {\bibinfo {title} {Self-interaction and strong correlation in {DFTB}},\ }\href@noop {} {\bibfield  {journal} {\bibinfo  {journal} {The Journal of Physical Chemistry A}\ }\textbf {\bibinfo {volume} {111}},\ \bibinfo {pages} {5671} (\bibinfo {year} {2007})}\BibitemShut {NoStop}%
\bibitem [{\citenamefont {Melix}\ \emph {et~al.}(2016)\citenamefont {Melix}, \citenamefont {Oliveira}, \citenamefont {R{\"u}ger},\ and\ \citenamefont {Heine}}]{melix2016spin}%
  \BibitemOpen
  \bibfield  {author} {\bibinfo {author} {\bibfnamefont {P.}~\bibnamefont {Melix}}, \bibinfo {author} {\bibfnamefont {A.~F.}\ \bibnamefont {Oliveira}}, \bibinfo {author} {\bibfnamefont {R.}~\bibnamefont {R{\"u}ger}},\ and\ \bibinfo {author} {\bibfnamefont {T.}~\bibnamefont {Heine}},\ }\bibfield  {title} {\bibinfo {title} {Spin polarization in {SCC-DFTB}},\ }\href@noop {} {\bibfield  {journal} {\bibinfo  {journal} {Theoretical Chemistry Accounts}\ }\textbf {\bibinfo {volume} {135}},\ \bibinfo {pages} {1} (\bibinfo {year} {2016})}\BibitemShut {NoStop}%
\bibitem [{\citenamefont {Sugihara}\ \emph {et~al.}(2002)\citenamefont {Sugihara}, \citenamefont {Buss}, \citenamefont {Entel}, \citenamefont {Elstner},\ and\ \citenamefont {Frauenheim}}]{sugihara2002}%
  \BibitemOpen
  \bibfield  {author} {\bibinfo {author} {\bibfnamefont {M.}~\bibnamefont {Sugihara}}, \bibinfo {author} {\bibfnamefont {V.}~\bibnamefont {Buss}}, \bibinfo {author} {\bibfnamefont {P.}~\bibnamefont {Entel}}, \bibinfo {author} {\bibfnamefont {M.}~\bibnamefont {Elstner}},\ and\ \bibinfo {author} {\bibfnamefont {T.}~\bibnamefont {Frauenheim}},\ }\bibfield  {title} {\bibinfo {title} {11-cis-retinal protonated {S}chiff base: influence of the protein environment on the geometry of the rhodopsin chromophore},\ }\href@noop {} {\bibfield  {journal} {\bibinfo  {journal} {Biochemistry}\ }\textbf {\bibinfo {volume} {41}},\ \bibinfo {pages} {15259} (\bibinfo {year} {2002})}\BibitemShut {NoStop}%
\bibitem [{\citenamefont {Bondar}\ \emph {et~al.}(2004)\citenamefont {Bondar}, \citenamefont {Elstner}, \citenamefont {Suhai}, \citenamefont {Smith},\ and\ \citenamefont {Fischer}}]{bondar2004mechanism}%
  \BibitemOpen
  \bibfield  {author} {\bibinfo {author} {\bibfnamefont {A.-N.}\ \bibnamefont {Bondar}}, \bibinfo {author} {\bibfnamefont {M.}~\bibnamefont {Elstner}}, \bibinfo {author} {\bibfnamefont {S.}~\bibnamefont {Suhai}}, \bibinfo {author} {\bibfnamefont {J.~C.}\ \bibnamefont {Smith}},\ and\ \bibinfo {author} {\bibfnamefont {S.}~\bibnamefont {Fischer}},\ }\bibfield  {title} {\bibinfo {title} {Mechanism of primary proton transfer in bacteriorhodopsin},\ }\href@noop {} {\bibfield  {journal} {\bibinfo  {journal} {Structure}\ }\textbf {\bibinfo {volume} {12}},\ \bibinfo {pages} {1281} (\bibinfo {year} {2004})}\BibitemShut {NoStop}%
\bibitem [{\citenamefont {Elghobashi-Meinhardt}\ \emph {et~al.}(2018)\citenamefont {Elghobashi-Meinhardt}, \citenamefont {Phatak}, \citenamefont {Bondar}, \citenamefont {Elstner},\ and\ \citenamefont {Smith}}]{elghobashi2018catalysis}%
  \BibitemOpen
  \bibfield  {author} {\bibinfo {author} {\bibfnamefont {N.}~\bibnamefont {Elghobashi-Meinhardt}}, \bibinfo {author} {\bibfnamefont {P.}~\bibnamefont {Phatak}}, \bibinfo {author} {\bibfnamefont {A.-N.}\ \bibnamefont {Bondar}}, \bibinfo {author} {\bibfnamefont {M.}~\bibnamefont {Elstner}},\ and\ \bibinfo {author} {\bibfnamefont {J.~C.}\ \bibnamefont {Smith}},\ }\bibfield  {title} {\bibinfo {title} {Catalysis of ground state cis-trans isomerization of bacteriorhodopsin’s retinal chromophore by a hydrogen-bond network},\ }\href@noop {} {\bibfield  {journal} {\bibinfo  {journal} {The Journal of Membrane Biology}\ }\textbf {\bibinfo {volume} {251}},\ \bibinfo {pages} {315} (\bibinfo {year} {2018})}\BibitemShut {NoStop}%
\bibitem [{\citenamefont {Torrie}\ and\ \citenamefont {Valleau}(1977)}]{torrie1977nonphysical}%
  \BibitemOpen
  \bibfield  {author} {\bibinfo {author} {\bibfnamefont {G.~M.}\ \bibnamefont {Torrie}}\ and\ \bibinfo {author} {\bibfnamefont {J.~P.}\ \bibnamefont {Valleau}},\ }\bibfield  {title} {\bibinfo {title} {Nonphysical sampling distributions in {M}onte {C}arlo free-energy estimation: Umbrella sampling},\ }\href@noop {} {\bibfield  {journal} {\bibinfo  {journal} {Journal of Computational Physics}\ }\textbf {\bibinfo {volume} {23}},\ \bibinfo {pages} {187} (\bibinfo {year} {1977})}\BibitemShut {NoStop}%
\bibitem [{\citenamefont {Laio}\ and\ \citenamefont {Parrinello}(2002)}]{laio2002escaping}%
  \BibitemOpen
  \bibfield  {author} {\bibinfo {author} {\bibfnamefont {A.}~\bibnamefont {Laio}}\ and\ \bibinfo {author} {\bibfnamefont {M.}~\bibnamefont {Parrinello}},\ }\bibfield  {title} {\bibinfo {title} {Escaping free-energy minima},\ }\href@noop {} {\bibfield  {journal} {\bibinfo  {journal} {Proceedings of the National Academy of Sciences}\ }\textbf {\bibinfo {volume} {99}},\ \bibinfo {pages} {12562} (\bibinfo {year} {2002})}\BibitemShut {NoStop}%
\bibitem [{\citenamefont {Bussi}\ and\ \citenamefont {Tribello}(2019)}]{bussi2019analyzing}%
  \BibitemOpen
  \bibfield  {author} {\bibinfo {author} {\bibfnamefont {G.}~\bibnamefont {Bussi}}\ and\ \bibinfo {author} {\bibfnamefont {G.~A.}\ \bibnamefont {Tribello}},\ }\bibfield  {title} {\bibinfo {title} {Analyzing and biasing simulations with {PLUMED}},\ }in\ \href@noop {} {\emph {\bibinfo {booktitle} {Biomolecular Simulations}}}\ (\bibinfo  {publisher} {Springer},\ \bibinfo {year} {2019})\ pp.\ \bibinfo {pages} {529--578}\BibitemShut {NoStop}%
\bibitem [{\citenamefont {Grimme}\ \emph {et~al.}(2010)\citenamefont {Grimme}, \citenamefont {Antony}, \citenamefont {Ehrlich},\ and\ \citenamefont {Krieg}}]{grimme2010consistent}%
  \BibitemOpen
  \bibfield  {author} {\bibinfo {author} {\bibfnamefont {S.}~\bibnamefont {Grimme}}, \bibinfo {author} {\bibfnamefont {J.}~\bibnamefont {Antony}}, \bibinfo {author} {\bibfnamefont {S.}~\bibnamefont {Ehrlich}},\ and\ \bibinfo {author} {\bibfnamefont {H.}~\bibnamefont {Krieg}},\ }\bibfield  {title} {\bibinfo {title} {A consistent and accurate ab initio parametrization of density functional dispersion correction ({DFT-D}) for the 94 elements {H}-{Pu}},\ }\href@noop {} {\bibfield  {journal} {\bibinfo  {journal} {The Journal of Chemical Physics}\ }\textbf {\bibinfo {volume} {132}} (\bibinfo {year} {2010})}\BibitemShut {NoStop}%
\bibitem [{\citenamefont {Hourahine}\ \emph {et~al.}(2020)\citenamefont {Hourahine}, \citenamefont {Aradi}, \citenamefont {Blum}, \citenamefont {Bonaf{\'e}}, \citenamefont {Buccheri}, \citenamefont {Camacho}, \citenamefont {Cevallos}, \citenamefont {Deshaye}, \citenamefont {Dumitric{\u{a}}}, \citenamefont {Dominguez} \emph {et~al.}}]{hourahine2020dftb+}%
  \BibitemOpen
  \bibfield  {author} {\bibinfo {author} {\bibfnamefont {B.}~\bibnamefont {Hourahine}}, \bibinfo {author} {\bibfnamefont {B.}~\bibnamefont {Aradi}}, \bibinfo {author} {\bibfnamefont {V.}~\bibnamefont {Blum}}, \bibinfo {author} {\bibfnamefont {F.}~\bibnamefont {Bonaf{\'e}}}, \bibinfo {author} {\bibfnamefont {A.}~\bibnamefont {Buccheri}}, \bibinfo {author} {\bibfnamefont {C.}~\bibnamefont {Camacho}}, \bibinfo {author} {\bibfnamefont {C.}~\bibnamefont {Cevallos}}, \bibinfo {author} {\bibfnamefont {M.}~\bibnamefont {Deshaye}}, \bibinfo {author} {\bibfnamefont {T.}~\bibnamefont {Dumitric{\u{a}}}}, \bibinfo {author} {\bibfnamefont {A.}~\bibnamefont {Dominguez}}, \emph {et~al.},\ }\bibfield  {title} {\bibinfo {title} {{DFTB+}, a software package for efficient approximate density functional theory based atomistic simulations},\ }\href@noop {} {\bibfield  {journal} {\bibinfo  {journal} {The Journal of Chemical Physics}\ }\textbf {\bibinfo {volume} {152}} (\bibinfo {year} {2020})}\BibitemShut {NoStop}%
\bibitem [{\citenamefont {Gaus}\ \emph {et~al.}(2013)\citenamefont {Gaus}, \citenamefont {Goez},\ and\ \citenamefont {Elstner}}]{gaus2013parametrization}%
  \BibitemOpen
  \bibfield  {author} {\bibinfo {author} {\bibfnamefont {M.}~\bibnamefont {Gaus}}, \bibinfo {author} {\bibfnamefont {A.}~\bibnamefont {Goez}},\ and\ \bibinfo {author} {\bibfnamefont {M.}~\bibnamefont {Elstner}},\ }\bibfield  {title} {\bibinfo {title} {Parametrization and benchmark of {DFTB3} for organic molecules},\ }\href@noop {} {\bibfield  {journal} {\bibinfo  {journal} {Journal of Chemical Theory and Computation}\ }\textbf {\bibinfo {volume} {9}},\ \bibinfo {pages} {338} (\bibinfo {year} {2013})}\BibitemShut {NoStop}%
\bibitem [{\citenamefont {Larsen}\ \emph {et~al.}(2017)\citenamefont {Larsen}, \citenamefont {Mortensen}, \citenamefont {Blomqvist}, \citenamefont {Castelli}, \citenamefont {Christensen}, \citenamefont {Du{\l}ak}, \citenamefont {Friis}, \citenamefont {Groves}, \citenamefont {Hammer}, \citenamefont {Hargus} \emph {et~al.}}]{larsen2017atomic}%
  \BibitemOpen
  \bibfield  {author} {\bibinfo {author} {\bibfnamefont {A.~H.}\ \bibnamefont {Larsen}}, \bibinfo {author} {\bibfnamefont {J.~J.}\ \bibnamefont {Mortensen}}, \bibinfo {author} {\bibfnamefont {J.}~\bibnamefont {Blomqvist}}, \bibinfo {author} {\bibfnamefont {I.~E.}\ \bibnamefont {Castelli}}, \bibinfo {author} {\bibfnamefont {R.}~\bibnamefont {Christensen}}, \bibinfo {author} {\bibfnamefont {M.}~\bibnamefont {Du{\l}ak}}, \bibinfo {author} {\bibfnamefont {J.}~\bibnamefont {Friis}}, \bibinfo {author} {\bibfnamefont {M.~N.}\ \bibnamefont {Groves}}, \bibinfo {author} {\bibfnamefont {B.}~\bibnamefont {Hammer}}, \bibinfo {author} {\bibfnamefont {C.}~\bibnamefont {Hargus}}, \emph {et~al.},\ }\bibfield  {title} {\bibinfo {title} {The atomic simulation environment—a {P}ython library for working with atoms},\ }\href@noop {} {\bibfield  {journal} {\bibinfo  {journal} {Journal of Physics: Condensed Matter}\ }\textbf {\bibinfo {volume} {29}},\ \bibinfo {pages} {273002} (\bibinfo {year} {2017})}\BibitemShut {NoStop}%
\bibitem [{\citenamefont {Fletcher}(2000)}]{fletcher2000practical}%
  \BibitemOpen
  \bibfield  {author} {\bibinfo {author} {\bibfnamefont {R.}~\bibnamefont {Fletcher}},\ }\href@noop {} {\emph {\bibinfo {title} {Practical methods of optimization}}}\ (\bibinfo  {publisher} {John Wiley \& Sons},\ \bibinfo {year} {2000})\BibitemShut {NoStop}%
\bibitem [{\citenamefont {Frisch}\ \emph {et~al.}(2016)\citenamefont {Frisch}, \citenamefont {Trucks}, \citenamefont {Schlegel}, \citenamefont {Scuseria}, \citenamefont {Robb}, \citenamefont {Cheeseman}, \citenamefont {Scalmani}, \citenamefont {Barone}, \citenamefont {Petersson}, \citenamefont {Nakatsuji}, \citenamefont {Li}, \citenamefont {Caricato}, \citenamefont {Marenich}, \citenamefont {Bloino}, \citenamefont {Janesko}, \citenamefont {Gomperts}, \citenamefont {Mennucci}, \citenamefont {Hratchian}, \citenamefont {Ortiz}, \citenamefont {Izmaylov}, \citenamefont {Sonnenberg}, \citenamefont {Williams-Young}, \citenamefont {Ding}, \citenamefont {Lipparini}, \citenamefont {Egidi}, \citenamefont {Goings}, \citenamefont {Peng}, \citenamefont {Petrone}, \citenamefont {Henderson}, \citenamefont {Ranasinghe}, \citenamefont {Zakrzewski}, \citenamefont {Gao}, \citenamefont {Rega}, \citenamefont {Zheng}, \citenamefont {Liang}, \citenamefont {Hada}, \citenamefont {Ehara}, \citenamefont {Toyota}, \citenamefont {Fukuda},
  \citenamefont {Hasegawa}, \citenamefont {Ishida}, \citenamefont {Nakajima}, \citenamefont {Honda}, \citenamefont {Kitao}, \citenamefont {Nakai}, \citenamefont {Vreven}, \citenamefont {Throssell}, \citenamefont {Montgomery}, \citenamefont {Peralta}, \citenamefont {Ogliaro}, \citenamefont {Bearpark}, \citenamefont {Heyd}, \citenamefont {Brothers}, \citenamefont {Kudin}, \citenamefont {Staroverov}, \citenamefont {Keith}, \citenamefont {Kobayashi}, \citenamefont {Normand}, \citenamefont {Raghavachari}, \citenamefont {Rendell}, \citenamefont {Burant}, \citenamefont {Iyengar}, \citenamefont {Tomasi}, \citenamefont {Cossi}, \citenamefont {Millam}, \citenamefont {Klene}, \citenamefont {Adamo}, \citenamefont {Cammi}, \citenamefont {Ochterski}, \citenamefont {Martin}, \citenamefont {Morokuma}, \citenamefont {Farkas}, \citenamefont {Foresman},\ and\ \citenamefont {Fox}}]{g16}%
  \BibitemOpen
  \bibfield  {author} {\bibinfo {author} {\bibfnamefont {M.~J.}\ \bibnamefont {Frisch}}, \bibinfo {author} {\bibfnamefont {G.~W.}\ \bibnamefont {Trucks}}, \bibinfo {author} {\bibfnamefont {H.~B.}\ \bibnamefont {Schlegel}}, \bibinfo {author} {\bibfnamefont {G.~E.}\ \bibnamefont {Scuseria}}, \bibinfo {author} {\bibfnamefont {M.~A.}\ \bibnamefont {Robb}}, \bibinfo {author} {\bibfnamefont {J.~R.}\ \bibnamefont {Cheeseman}}, \bibinfo {author} {\bibfnamefont {G.}~\bibnamefont {Scalmani}}, \bibinfo {author} {\bibfnamefont {V.}~\bibnamefont {Barone}}, \bibinfo {author} {\bibfnamefont {G.~A.}\ \bibnamefont {Petersson}}, \bibinfo {author} {\bibfnamefont {H.}~\bibnamefont {Nakatsuji}}, \bibinfo {author} {\bibfnamefont {X.}~\bibnamefont {Li}}, \bibinfo {author} {\bibfnamefont {M.}~\bibnamefont {Caricato}}, \bibinfo {author} {\bibfnamefont {A.~V.}\ \bibnamefont {Marenich}}, \bibinfo {author} {\bibfnamefont {J.}~\bibnamefont {Bloino}}, \bibinfo {author} {\bibfnamefont {B.~G.}\ \bibnamefont {Janesko}}, \bibinfo {author}
  {\bibfnamefont {R.}~\bibnamefont {Gomperts}}, \bibinfo {author} {\bibfnamefont {B.}~\bibnamefont {Mennucci}}, \bibinfo {author} {\bibfnamefont {H.~P.}\ \bibnamefont {Hratchian}}, \bibinfo {author} {\bibfnamefont {J.~V.}\ \bibnamefont {Ortiz}}, \bibinfo {author} {\bibfnamefont {A.~F.}\ \bibnamefont {Izmaylov}}, \bibinfo {author} {\bibfnamefont {J.~L.}\ \bibnamefont {Sonnenberg}}, \bibinfo {author} {\bibfnamefont {D.}~\bibnamefont {Williams-Young}}, \bibinfo {author} {\bibfnamefont {F.}~\bibnamefont {Ding}}, \bibinfo {author} {\bibfnamefont {F.}~\bibnamefont {Lipparini}}, \bibinfo {author} {\bibfnamefont {F.}~\bibnamefont {Egidi}}, \bibinfo {author} {\bibfnamefont {J.}~\bibnamefont {Goings}}, \bibinfo {author} {\bibfnamefont {B.}~\bibnamefont {Peng}}, \bibinfo {author} {\bibfnamefont {A.}~\bibnamefont {Petrone}}, \bibinfo {author} {\bibfnamefont {T.}~\bibnamefont {Henderson}}, \bibinfo {author} {\bibfnamefont {D.}~\bibnamefont {Ranasinghe}}, \bibinfo {author} {\bibfnamefont {V.~G.}\ \bibnamefont
  {Zakrzewski}}, \bibinfo {author} {\bibfnamefont {J.}~\bibnamefont {Gao}}, \bibinfo {author} {\bibfnamefont {N.}~\bibnamefont {Rega}}, \bibinfo {author} {\bibfnamefont {G.}~\bibnamefont {Zheng}}, \bibinfo {author} {\bibfnamefont {W.}~\bibnamefont {Liang}}, \bibinfo {author} {\bibfnamefont {M.}~\bibnamefont {Hada}}, \bibinfo {author} {\bibfnamefont {M.}~\bibnamefont {Ehara}}, \bibinfo {author} {\bibfnamefont {K.}~\bibnamefont {Toyota}}, \bibinfo {author} {\bibfnamefont {R.}~\bibnamefont {Fukuda}}, \bibinfo {author} {\bibfnamefont {J.}~\bibnamefont {Hasegawa}}, \bibinfo {author} {\bibfnamefont {M.}~\bibnamefont {Ishida}}, \bibinfo {author} {\bibfnamefont {T.}~\bibnamefont {Nakajima}}, \bibinfo {author} {\bibfnamefont {Y.}~\bibnamefont {Honda}}, \bibinfo {author} {\bibfnamefont {O.}~\bibnamefont {Kitao}}, \bibinfo {author} {\bibfnamefont {H.}~\bibnamefont {Nakai}}, \bibinfo {author} {\bibfnamefont {T.}~\bibnamefont {Vreven}}, \bibinfo {author} {\bibfnamefont {K.}~\bibnamefont {Throssell}}, \bibinfo {author}
  {\bibfnamefont {J.~A.}\ \bibnamefont {Montgomery}, \bibfnamefont {{Jr.}}}, \bibinfo {author} {\bibfnamefont {J.~E.}\ \bibnamefont {Peralta}}, \bibinfo {author} {\bibfnamefont {F.}~\bibnamefont {Ogliaro}}, \bibinfo {author} {\bibfnamefont {M.~J.}\ \bibnamefont {Bearpark}}, \bibinfo {author} {\bibfnamefont {J.~J.}\ \bibnamefont {Heyd}}, \bibinfo {author} {\bibfnamefont {E.~N.}\ \bibnamefont {Brothers}}, \bibinfo {author} {\bibfnamefont {K.~N.}\ \bibnamefont {Kudin}}, \bibinfo {author} {\bibfnamefont {V.~N.}\ \bibnamefont {Staroverov}}, \bibinfo {author} {\bibfnamefont {T.~A.}\ \bibnamefont {Keith}}, \bibinfo {author} {\bibfnamefont {R.}~\bibnamefont {Kobayashi}}, \bibinfo {author} {\bibfnamefont {J.}~\bibnamefont {Normand}}, \bibinfo {author} {\bibfnamefont {K.}~\bibnamefont {Raghavachari}}, \bibinfo {author} {\bibfnamefont {A.~P.}\ \bibnamefont {Rendell}}, \bibinfo {author} {\bibfnamefont {J.~C.}\ \bibnamefont {Burant}}, \bibinfo {author} {\bibfnamefont {S.~S.}\ \bibnamefont {Iyengar}}, \bibinfo {author}
  {\bibfnamefont {J.}~\bibnamefont {Tomasi}}, \bibinfo {author} {\bibfnamefont {M.}~\bibnamefont {Cossi}}, \bibinfo {author} {\bibfnamefont {J.~M.}\ \bibnamefont {Millam}}, \bibinfo {author} {\bibfnamefont {M.}~\bibnamefont {Klene}}, \bibinfo {author} {\bibfnamefont {C.}~\bibnamefont {Adamo}}, \bibinfo {author} {\bibfnamefont {R.}~\bibnamefont {Cammi}}, \bibinfo {author} {\bibfnamefont {J.~W.}\ \bibnamefont {Ochterski}}, \bibinfo {author} {\bibfnamefont {R.~L.}\ \bibnamefont {Martin}}, \bibinfo {author} {\bibfnamefont {K.}~\bibnamefont {Morokuma}}, \bibinfo {author} {\bibfnamefont {O.}~\bibnamefont {Farkas}}, \bibinfo {author} {\bibfnamefont {J.~B.}\ \bibnamefont {Foresman}},\ and\ \bibinfo {author} {\bibfnamefont {D.~J.}\ \bibnamefont {Fox}},\ }\href@noop {} {\bibinfo {title} {Gaussian~16 {R}evision {C}.01}} (\bibinfo {year} {2016}),\ \bibinfo {note} {gaussian Inc. Wallingford CT}\BibitemShut {NoStop}%
\bibitem [{\citenamefont {Berendsen}\ \emph {et~al.}(1984)\citenamefont {Berendsen}, \citenamefont {Postma}, \citenamefont {Van~Gunsteren}, \citenamefont {DiNola},\ and\ \citenamefont {Haak}}]{berendsen1984molecular}%
  \BibitemOpen
  \bibfield  {author} {\bibinfo {author} {\bibfnamefont {H.~J.}\ \bibnamefont {Berendsen}}, \bibinfo {author} {\bibfnamefont {J.~v.}\ \bibnamefont {Postma}}, \bibinfo {author} {\bibfnamefont {W.~F.}\ \bibnamefont {Van~Gunsteren}}, \bibinfo {author} {\bibfnamefont {A.}~\bibnamefont {DiNola}},\ and\ \bibinfo {author} {\bibfnamefont {J.~R.}\ \bibnamefont {Haak}},\ }\bibfield  {title} {\bibinfo {title} {Molecular dynamics with coupling to an external bath},\ }\href@noop {} {\bibfield  {journal} {\bibinfo  {journal} {The Journal of Chemical Physics}\ }\textbf {\bibinfo {volume} {81}},\ \bibinfo {pages} {3684} (\bibinfo {year} {1984})}\BibitemShut {NoStop}%
\bibitem [{\citenamefont {Nos{\'e}}(1984)}]{nose1984unified}%
  \BibitemOpen
  \bibfield  {author} {\bibinfo {author} {\bibfnamefont {S.}~\bibnamefont {Nos{\'e}}},\ }\bibfield  {title} {\bibinfo {title} {A unified formulation of the constant temperature molecular dynamics methods},\ }\href@noop {} {\bibfield  {journal} {\bibinfo  {journal} {The Journal of Chemical Physics}\ }\textbf {\bibinfo {volume} {81}},\ \bibinfo {pages} {511} (\bibinfo {year} {1984})}\BibitemShut {NoStop}%
\bibitem [{\citenamefont {Hoover}(1985)}]{hoover1985canonical}%
  \BibitemOpen
  \bibfield  {author} {\bibinfo {author} {\bibfnamefont {W.~G.}\ \bibnamefont {Hoover}},\ }\bibfield  {title} {\bibinfo {title} {Canonical dynamics: {E}quilibrium phase-space distributions},\ }\href@noop {} {\bibfield  {journal} {\bibinfo  {journal} {Physical Review A}\ }\textbf {\bibinfo {volume} {31}},\ \bibinfo {pages} {1695} (\bibinfo {year} {1985})}\BibitemShut {NoStop}%
\bibitem [{\citenamefont {Martyna}\ \emph {et~al.}(1992)\citenamefont {Martyna}, \citenamefont {Klein},\ and\ \citenamefont {Tuckerman}}]{martyna1992nose}%
  \BibitemOpen
  \bibfield  {author} {\bibinfo {author} {\bibfnamefont {G.~J.}\ \bibnamefont {Martyna}}, \bibinfo {author} {\bibfnamefont {M.~L.}\ \bibnamefont {Klein}},\ and\ \bibinfo {author} {\bibfnamefont {M.}~\bibnamefont {Tuckerman}},\ }\bibfield  {title} {\bibinfo {title} {Nos{\'e}--{H}oover chains: The canonical ensemble via continuous dynamics},\ }\href@noop {} {\bibfield  {journal} {\bibinfo  {journal} {The Journal of Chemical Physics}\ }\textbf {\bibinfo {volume} {97}},\ \bibinfo {pages} {2635} (\bibinfo {year} {1992})}\BibitemShut {NoStop}%
\bibitem [{\citenamefont {Tribello}\ \emph {et~al.}(2014)\citenamefont {Tribello}, \citenamefont {Bonomi}, \citenamefont {Branduardi}, \citenamefont {Camilloni},\ and\ \citenamefont {Bussi}}]{tribello2014plumed}%
  \BibitemOpen
  \bibfield  {author} {\bibinfo {author} {\bibfnamefont {G.~A.}\ \bibnamefont {Tribello}}, \bibinfo {author} {\bibfnamefont {M.}~\bibnamefont {Bonomi}}, \bibinfo {author} {\bibfnamefont {D.}~\bibnamefont {Branduardi}}, \bibinfo {author} {\bibfnamefont {C.}~\bibnamefont {Camilloni}},\ and\ \bibinfo {author} {\bibfnamefont {G.}~\bibnamefont {Bussi}},\ }\bibfield  {title} {\bibinfo {title} {{PLUMED} 2: New feathers for an old bird},\ }\href@noop {} {\bibfield  {journal} {\bibinfo  {journal} {Computer physics communications}\ }\textbf {\bibinfo {volume} {185}},\ \bibinfo {pages} {604} (\bibinfo {year} {2014})}\BibitemShut {NoStop}%
\bibitem [{\citenamefont {Schlegel}(1982)}]{schlegel1982optimization}%
  \BibitemOpen
  \bibfield  {author} {\bibinfo {author} {\bibfnamefont {H.~B.}\ \bibnamefont {Schlegel}},\ }\bibfield  {title} {\bibinfo {title} {Optimization of equilibrium geometries and transition structures},\ }\href@noop {} {\bibfield  {journal} {\bibinfo  {journal} {Journal of Computational Chemistry}\ }\textbf {\bibinfo {volume} {3}},\ \bibinfo {pages} {214} (\bibinfo {year} {1982})}\BibitemShut {NoStop}%
\bibitem [{\citenamefont {Peng}\ and\ \citenamefont {Bernhard~Schlegel}(1993)}]{peng1993combining}%
  \BibitemOpen
  \bibfield  {author} {\bibinfo {author} {\bibfnamefont {C.}~\bibnamefont {Peng}}\ and\ \bibinfo {author} {\bibfnamefont {H.}~\bibnamefont {Bernhard~Schlegel}},\ }\bibfield  {title} {\bibinfo {title} {Combining synchronous transit and quasi-{N}ewton methods to find transition states},\ }\href@noop {} {\bibfield  {journal} {\bibinfo  {journal} {Israel Journal of Chemistry}\ }\textbf {\bibinfo {volume} {33}},\ \bibinfo {pages} {449} (\bibinfo {year} {1993})}\BibitemShut {NoStop}%
\bibitem [{\citenamefont {Peng}\ \emph {et~al.}(1996)\citenamefont {Peng}, \citenamefont {Ayala}, \citenamefont {Schlegel},\ and\ \citenamefont {Frisch}}]{peng1996using}%
  \BibitemOpen
  \bibfield  {author} {\bibinfo {author} {\bibfnamefont {C.}~\bibnamefont {Peng}}, \bibinfo {author} {\bibfnamefont {P.~Y.}\ \bibnamefont {Ayala}}, \bibinfo {author} {\bibfnamefont {H.~B.}\ \bibnamefont {Schlegel}},\ and\ \bibinfo {author} {\bibfnamefont {M.~J.}\ \bibnamefont {Frisch}},\ }\bibfield  {title} {\bibinfo {title} {Using redundant internal coordinates to optimize equilibrium geometries and transition states},\ }\href@noop {} {\bibfield  {journal} {\bibinfo  {journal} {Journal of Computational Chemistry}\ }\textbf {\bibinfo {volume} {17}},\ \bibinfo {pages} {49} (\bibinfo {year} {1996})}\BibitemShut {NoStop}%
\bibitem [{\citenamefont {Ochterski}(2000)}]{ochterski2000thermochemistry}%
  \BibitemOpen
  \bibfield  {author} {\bibinfo {author} {\bibfnamefont {J.~W.}\ \bibnamefont {Ochterski}},\ }\bibfield  {title} {\bibinfo {title} {Thermochemistry in {G}aussian},\ }\href@noop {} {\bibfield  {journal} {\bibinfo  {journal} {Gaussian Inc.}\ } (\bibinfo {year} {2000})}\BibitemShut {NoStop}%
\bibitem [{\citenamefont {Ochterski}(1999)}]{ochterski1999vibrational}%
  \BibitemOpen
  \bibfield  {author} {\bibinfo {author} {\bibfnamefont {J.~W.}\ \bibnamefont {Ochterski}},\ }\bibfield  {title} {\bibinfo {title} {Vibrational analysis in {G}aussian},\ }\href@noop {} {\bibfield  {journal} {\bibinfo  {journal} {Gaussian Inc.}\ } (\bibinfo {year} {1999})}\BibitemShut {NoStop}%
\bibitem [{\citenamefont {Paizs}\ \emph {et~al.}(1999)\citenamefont {Paizs}, \citenamefont {Tajkhorshid},\ and\ \citenamefont {Suhai}}]{paizs1999electronic}%
  \BibitemOpen
  \bibfield  {author} {\bibinfo {author} {\bibfnamefont {B.}~\bibnamefont {Paizs}}, \bibinfo {author} {\bibfnamefont {E.}~\bibnamefont {Tajkhorshid}},\ and\ \bibinfo {author} {\bibfnamefont {S.}~\bibnamefont {Suhai}},\ }\bibfield  {title} {\bibinfo {title} {Electronic effects on the ground-state rotational barrier of polyene {S}chiff bases: a molecular orbital study},\ }\href@noop {} {\bibfield  {journal} {\bibinfo  {journal} {The Journal of Physical Chemistry B}\ }\textbf {\bibinfo {volume} {103}},\ \bibinfo {pages} {5388} (\bibinfo {year} {1999})}\BibitemShut {NoStop}%
\bibitem [{\citenamefont {Okada}\ \emph {et~al.}(2004)\citenamefont {Okada}, \citenamefont {Sugihara}, \citenamefont {Bondar}, \citenamefont {Elstner}, \citenamefont {Entel},\ and\ \citenamefont {Buss}}]{okada2004retinal}%
  \BibitemOpen
  \bibfield  {author} {\bibinfo {author} {\bibfnamefont {T.}~\bibnamefont {Okada}}, \bibinfo {author} {\bibfnamefont {M.}~\bibnamefont {Sugihara}}, \bibinfo {author} {\bibfnamefont {A.-N.}\ \bibnamefont {Bondar}}, \bibinfo {author} {\bibfnamefont {M.}~\bibnamefont {Elstner}}, \bibinfo {author} {\bibfnamefont {P.}~\bibnamefont {Entel}},\ and\ \bibinfo {author} {\bibfnamefont {V.}~\bibnamefont {Buss}},\ }\bibfield  {title} {\bibinfo {title} {The retinal conformation and its environment in rhodopsin in light of a new 2.2 {{\AA}} crystal structure},\ }\href@noop {} {\bibfield  {journal} {\bibinfo  {journal} {Journal of Molecular Biology}\ }\textbf {\bibinfo {volume} {342}},\ \bibinfo {pages} {571} (\bibinfo {year} {2004})}\BibitemShut {NoStop}%
\bibitem [{\citenamefont {Buda}\ \emph {et~al.}(2000)\citenamefont {Buda}, \citenamefont {Giannozzi},\ and\ \citenamefont {Mauri}}]{buda2000density}%
  \BibitemOpen
  \bibfield  {author} {\bibinfo {author} {\bibfnamefont {F.}~\bibnamefont {Buda}}, \bibinfo {author} {\bibfnamefont {P.}~\bibnamefont {Giannozzi}},\ and\ \bibinfo {author} {\bibfnamefont {F.}~\bibnamefont {Mauri}},\ }\bibfield  {title} {\bibinfo {title} {Density functional theory study of the structure and 13{C} chemical shifts of retinylidene iminium salts},\ }\href@noop {} {\bibfield  {journal} {\bibinfo  {journal} {The Journal of Physical Chemistry B}\ }\textbf {\bibinfo {volume} {104}},\ \bibinfo {pages} {9048} (\bibinfo {year} {2000})}\BibitemShut {NoStop}%
\bibitem [{\citenamefont {Bondar}\ \emph {et~al.}(2010)\citenamefont {Bondar}, \citenamefont {Smith},\ and\ \citenamefont {Elstner}}]{bondar2010mechanism}%
  \BibitemOpen
  \bibfield  {author} {\bibinfo {author} {\bibfnamefont {A.-N.}\ \bibnamefont {Bondar}}, \bibinfo {author} {\bibfnamefont {J.~C.}\ \bibnamefont {Smith}},\ and\ \bibinfo {author} {\bibfnamefont {M.}~\bibnamefont {Elstner}},\ }\bibfield  {title} {\bibinfo {title} {Mechanism of a proton pump analyzed with computer simulations},\ }\href@noop {} {\bibfield  {journal} {\bibinfo  {journal} {Theoretical Chemistry Accounts}\ }\textbf {\bibinfo {volume} {125}},\ \bibinfo {pages} {353} (\bibinfo {year} {2010})}\BibitemShut {NoStop}%
\bibitem [{\citenamefont {Vanden-Eijnden}\ \emph {et~al.}(2010)\citenamefont {Vanden-Eijnden} \emph {et~al.}}]{vanden2010transition}%
  \BibitemOpen
  \bibfield  {author} {\bibinfo {author} {\bibfnamefont {E.}~\bibnamefont {Vanden-Eijnden}} \emph {et~al.},\ }\bibfield  {title} {\bibinfo {title} {Transition-path theory and path-finding algorithms for the study of rare events.},\ }\href@noop {} {\bibfield  {journal} {\bibinfo  {journal} {Annual Review of Physical Chemistry}\ }\textbf {\bibinfo {volume} {61}},\ \bibinfo {pages} {391} (\bibinfo {year} {2010})}\BibitemShut {NoStop}%
\bibitem [{\citenamefont {Roux}(2021)}]{roux2021string}%
  \BibitemOpen
  \bibfield  {author} {\bibinfo {author} {\bibfnamefont {B.}~\bibnamefont {Roux}},\ }\bibfield  {title} {\bibinfo {title} {String method with swarms-of-trajectories, mean drifts, lag time, and committor},\ }\href@noop {} {\bibfield  {journal} {\bibinfo  {journal} {The Journal of Physical Chemistry A}\ }\textbf {\bibinfo {volume} {125}},\ \bibinfo {pages} {7558} (\bibinfo {year} {2021})}\BibitemShut {NoStop}%
\bibitem [{\citenamefont {Jung}\ \emph {et~al.}(2023)\citenamefont {Jung}, \citenamefont {Covino}, \citenamefont {Arjun}, \citenamefont {Leitold}, \citenamefont {Dellago}, \citenamefont {Bolhuis},\ and\ \citenamefont {Hummer}}]{jung2023machine}%
  \BibitemOpen
  \bibfield  {author} {\bibinfo {author} {\bibfnamefont {H.}~\bibnamefont {Jung}}, \bibinfo {author} {\bibfnamefont {R.}~\bibnamefont {Covino}}, \bibinfo {author} {\bibfnamefont {A.}~\bibnamefont {Arjun}}, \bibinfo {author} {\bibfnamefont {C.}~\bibnamefont {Leitold}}, \bibinfo {author} {\bibfnamefont {C.}~\bibnamefont {Dellago}}, \bibinfo {author} {\bibfnamefont {P.~G.}\ \bibnamefont {Bolhuis}},\ and\ \bibinfo {author} {\bibfnamefont {G.}~\bibnamefont {Hummer}},\ }\bibfield  {title} {\bibinfo {title} {Machine-guided path sampling to discover mechanisms of molecular self-organization},\ }\href@noop {} {\bibfield  {journal} {\bibinfo  {journal} {Nature Computational Science}\ ,\ \bibinfo {pages} {1}} (\bibinfo {year} {2023})}\BibitemShut {NoStop}%
\bibitem [{\citenamefont {Lie}\ \emph {et~al.}(2013)\citenamefont {Lie}, \citenamefont {Fackeldey},\ and\ \citenamefont {Weber}}]{Lie2013}%
  \BibitemOpen
  \bibfield  {author} {\bibinfo {author} {\bibfnamefont {H.~C.}\ \bibnamefont {Lie}}, \bibinfo {author} {\bibfnamefont {K.}~\bibnamefont {Fackeldey}},\ and\ \bibinfo {author} {\bibfnamefont {M.}~\bibnamefont {Weber}},\ }\bibfield  {title} {\bibinfo {title} {A square root approximation of transition rates for a {M}arkov {S}tate {M}odel},\ }\href@noop {} {\bibfield  {journal} {\bibinfo  {journal} {SIAM Journal on Matrix Analysis and Applications}\ }\textbf {\bibinfo {volume} {34}},\ \bibinfo {pages} {738–756} (\bibinfo {year} {2013})}\BibitemShut {NoStop}%
\bibitem [{\citenamefont {Donati}\ \emph {et~al.}(2021)\citenamefont {Donati}, \citenamefont {Weber},\ and\ \citenamefont {Keller}}]{donati2021markov}%
  \BibitemOpen
  \bibfield  {author} {\bibinfo {author} {\bibfnamefont {L.}~\bibnamefont {Donati}}, \bibinfo {author} {\bibfnamefont {M.}~\bibnamefont {Weber}},\ and\ \bibinfo {author} {\bibfnamefont {B.~G.}\ \bibnamefont {Keller}},\ }\bibfield  {title} {\bibinfo {title} {Markov models from the square root approximation of the {F}okker-{P}lanck equation: calculating the grid-dependent flux},\ }\href@noop {} {\bibfield  {journal} {\bibinfo  {journal} {Journal of Physics: Condensed Matter}\ }\textbf {\bibinfo {volume} {33}},\ \bibinfo {pages} {115902} (\bibinfo {year} {2021})}\BibitemShut {NoStop}%
\bibitem [{\citenamefont {Bolhuis}\ \emph {et~al.}(2002)\citenamefont {Bolhuis}, \citenamefont {Chandler}, \citenamefont {Dellago},\ and\ \citenamefont {Geissler}}]{bolhuis2002transition}%
  \BibitemOpen
  \bibfield  {author} {\bibinfo {author} {\bibfnamefont {P.~G.}\ \bibnamefont {Bolhuis}}, \bibinfo {author} {\bibfnamefont {D.}~\bibnamefont {Chandler}}, \bibinfo {author} {\bibfnamefont {C.}~\bibnamefont {Dellago}},\ and\ \bibinfo {author} {\bibfnamefont {P.~L.}\ \bibnamefont {Geissler}},\ }\bibfield  {title} {\bibinfo {title} {Transition path sampling: Throwing ropes over rough mountain passes, in the dark},\ }\href@noop {} {\bibfield  {journal} {\bibinfo  {journal} {Annual Review of Physical Chemistry}\ }\textbf {\bibinfo {volume} {53}},\ \bibinfo {pages} {291} (\bibinfo {year} {2002})}\BibitemShut {NoStop}%
\bibitem [{\citenamefont {Zuckerman}\ and\ \citenamefont {Chong}(2017)}]{zuckerman2017weighted}%
  \BibitemOpen
  \bibfield  {author} {\bibinfo {author} {\bibfnamefont {D.~M.}\ \bibnamefont {Zuckerman}}\ and\ \bibinfo {author} {\bibfnamefont {L.~T.}\ \bibnamefont {Chong}},\ }\bibfield  {title} {\bibinfo {title} {Weighted ensemble simulation: review of methodology, applications, and software},\ }\href@noop {} {\bibfield  {journal} {\bibinfo  {journal} {Annual Review of Biophysics}\ }\textbf {\bibinfo {volume} {46}},\ \bibinfo {pages} {43} (\bibinfo {year} {2017})}\BibitemShut {NoStop}%
\bibitem [{\citenamefont {Palacio-Rodriguez}\ and\ \citenamefont {Pietrucci}(2022)}]{palacio2022free}%
  \BibitemOpen
  \bibfield  {author} {\bibinfo {author} {\bibfnamefont {K.}~\bibnamefont {Palacio-Rodriguez}}\ and\ \bibinfo {author} {\bibfnamefont {F.}~\bibnamefont {Pietrucci}},\ }\bibfield  {title} {\bibinfo {title} {Free energy landscapes, diffusion coefficients, and kinetic rates from transition paths},\ }\href@noop {} {\bibfield  {journal} {\bibinfo  {journal} {Journal of Chemical Theory and Computation}\ }\textbf {\bibinfo {volume} {18}},\ \bibinfo {pages} {4639} (\bibinfo {year} {2022})}\BibitemShut {NoStop}%
\bibitem [{\citenamefont {Donati}\ and\ \citenamefont {Keller}(2018)}]{donati2018girsanov}%
  \BibitemOpen
  \bibfield  {author} {\bibinfo {author} {\bibfnamefont {L.}~\bibnamefont {Donati}}\ and\ \bibinfo {author} {\bibfnamefont {B.~G.}\ \bibnamefont {Keller}},\ }\bibfield  {title} {\bibinfo {title} {Girsanov reweighting for metadynamics simulations},\ }\href@noop {} {\bibfield  {journal} {\bibinfo  {journal} {The Journal of Chemical Physics}\ }\textbf {\bibinfo {volume} {149}},\ \bibinfo {pages} {072335} (\bibinfo {year} {2018})}\BibitemShut {NoStop}%
\bibitem [{\citenamefont {Shmilovich}\ and\ \citenamefont {Ferguson}(2023)}]{shmilovich2023girsanov}%
  \BibitemOpen
  \bibfield  {author} {\bibinfo {author} {\bibfnamefont {K.}~\bibnamefont {Shmilovich}}\ and\ \bibinfo {author} {\bibfnamefont {A.~L.}\ \bibnamefont {Ferguson}},\ }\bibfield  {title} {\bibinfo {title} {Girsanov reweighting enhanced sampling technique ({GREST}): On-the-fly data-driven discovery of and enhanced sampling in slow collective variables},\ }\href@noop {} {\bibfield  {journal} {\bibinfo  {journal} {The Journal of Physical Chemistry A}\ } (\bibinfo {year} {2023})}\BibitemShut {NoStop}%
\end{thebibliography}%


\begin{thebibliography}{10}

\bibitem{mortimer2000physical}
Robert~G Mortimer.
\newblock {\em Physical chemistry}.
\newblock Academic Press, 2000.

\bibitem{peters2017reaction}
Baron Peters.
\newblock {\em Reaction rate theory and rare events}.
\newblock Elsevier, 2017.

\bibitem{mcquarrie1997physical}
Donald~Allan McQuarrie and John~Douglas Simon.
\newblock {\em Physical chemistry: a molecular approach}, volume~1.
\newblock University Science Books Sausalito, CA, 1997.

\bibitem{ochterski1999vibrational}
Joseph~W Ochterski.
\newblock Vibrational analysis in {G}aussian.
\newblock {\em Gaussian Inc.}, 1999.

\bibitem{ochterski2000thermochemistry}
Joseph~W Ochterski.
\newblock Thermochemistry in {G}aussian.
\newblock {\em Gaussian Inc.}, 2000.

\bibitem{laidler1987chemical}
Keith~J. Laidler.
\newblock {\em Chemical Kinetics}.
\newblock Pearson Education Inc., 1987.

\bibitem{grubmuller1995predicting}
Helmut Grubm{\"u}ller.
\newblock Predicting slow structural transitions in macromolecular systems: conformational flooding.
\newblock {\em Physical Review E}, 52(3):2893, 1995.

\bibitem{voter1997hyperdynamics}
Arthur~F Voter.
\newblock Hyperdynamics: Accelerated molecular dynamics of infrequent events.
\newblock {\em Physical Review Letters}, 78(20):3908, 1997.

\bibitem{huber1994local}
Thomas Huber, Andrew~E Torda, and Wilfred~F Van~Gunsteren.
\newblock Local elevation: a method for improving the searching properties of molecular dynamics simulation.
\newblock {\em Journal of Computer-Aided Molecular Design}, 8(6):695--708, 1994.

\bibitem{laio2002escaping}
Alessandro Laio and Michele Parrinello.
\newblock Escaping free-energy minima.
\newblock {\em Proceedings of the National Academy of Sciences}, 99(20):12562--12566, 2002.

\bibitem{tiwary2013metadynamics}
Pratyush Tiwary and Michele Parrinello.
\newblock From metadynamics to dynamics.
\newblock {\em Physical Review Letters}, 111(23):230602, 2013.

\bibitem{khan2020fluxional}
Salman~A Khan, Bradley~M Dickson, and Baron Peters.
\newblock How fluxional reactants limit the accuracy/efficiency of infrequent metadynamics.
\newblock {\em The Journal of Chemical Physics}, 153(5):054125, 2020.

\bibitem{palacio2022free}
Karen Palacio-Rodriguez and Fabio Pietrucci.
\newblock Free energy landscapes, diffusion coefficients, and kinetic rates from transition paths.
\newblock {\em Journal of Chemical Theory and Computation}, 18(8):4639--4648, 2022.

\bibitem{gv6}
Roy Dennington, Todd~A. Keith, and John~M. Millam.
\newblock Gauss{V}iew {V}ersion {6}, 2019.
\newblock {S}emichem Inc. Shawnee Mission KS.

\bibitem{gaus2013parametrization}
Michael Gaus, Albrecht Goez, and Marcus Elstner.
\newblock Parametrization and benchmark of {DFTB3} for organic molecules.
\newblock {\em Journal of Chemical Theory and Computation}, 9(1):338--354, 2013.

\bibitem{zhou2002performance}
Hongyi Zhou, Emad Tajkhorshid, Thomas Frauenheim, S{\'a}ndor Suhai, and Marcus Elstner.
\newblock Performance of the {AM1}, {PM3}, and {SCC-DFTB} methods in the study of conjugated {S}chiff base molecules.
\newblock {\em Chemical Physics}, 277(2):91--103, 2002.

\bibitem{bondar2011ground}
Ana-Nicoleta Bondar, Michaela Knapp-Mohammady, S{\'a}ndor Suhai, Stefan Fischer, and Jeremy~C Smith.
\newblock Ground-state properties of the retinal molecule: from quantum mechanical to classical mechanical computations of retinal proteins.
\newblock {\em Theoretical Chemistry Accounts}, 130:1169--1183, 2011.

\bibitem{hourahine2020dftb+}
Ben Hourahine, B{\'a}lint Aradi, Volker Blum, F~Bonaf{\'e}, A~Buccheri, Cristopher Camacho, Caterina Cevallos, MY~Deshaye, T~Dumitric{\u{a}}, A~Dominguez, et~al.
\newblock {DFTB+}, a software package for efficient approximate density functional theory based atomistic simulations.
\newblock {\em The Journal of Chemical Physics}, 152(12), 2020.

\bibitem{larsen2017atomic}
Ask~Hjorth Larsen, Jens~J{\o}rgen Mortensen, Jakob Blomqvist, Ivano~E Castelli, Rune Christensen, Marcin Du{\l}ak, Jesper Friis, Michael~N Groves, Bj{\o}rk Hammer, Cory Hargus, et~al.
\newblock The atomic simulation environment—a {P}ython library for working with atoms.
\newblock {\em Journal of Physics: Condensed Matter}, 29(27):273002, 2017.

\bibitem{fletcher2000practical}
Roger Fletcher.
\newblock {\em Practical methods of optimization}.
\newblock John Wiley \& Sons, 2000.

\bibitem{henkelman2000improved}
Graeme Henkelman and Hannes J{\'o}nsson.
\newblock Improved tangent estimate in the nudged elastic band method for finding minimum energy paths and saddle points.
\newblock {\em The Journal of Chemical Physics}, 113(22):9978--9985, 2000.

\bibitem{henkelman2000climbing}
Graeme Henkelman, Blas~P Uberuaga, and Hannes J{\'o}nsson.
\newblock A climbing image nudged elastic band method for finding saddle points and minimum energy paths.
\newblock {\em The Journal of Chemical Physics}, 113(22):9901--9904, 2000.

\bibitem{g16}
M.~J. Frisch, G.~W. Trucks, H.~B. Schlegel, G.~E. Scuseria, M.~A. Robb, J.~R. Cheeseman, G.~Scalmani, V.~Barone, G.~A. Petersson, H.~Nakatsuji, X.~Li, M.~Caricato, A.~V. Marenich, J.~Bloino, B.~G. Janesko, R.~Gomperts, B.~Mennucci, H.~P. Hratchian, J.~V. Ortiz, A.~F. Izmaylov, J.~L. Sonnenberg, D.~Williams-Young, F.~Ding, F.~Lipparini, F.~Egidi, J.~Goings, B.~Peng, A.~Petrone, T.~Henderson, D.~Ranasinghe, V.~G. Zakrzewski, J.~Gao, N.~Rega, G.~Zheng, W.~Liang, M.~Hada, M.~Ehara, K.~Toyota, R.~Fukuda, J.~Hasegawa, M.~Ishida, T.~Nakajima, Y.~Honda, O.~Kitao, H.~Nakai, T.~Vreven, K.~Throssell, J.~A. Montgomery, {Jr.}, J.~E. Peralta, F.~Ogliaro, M.~J. Bearpark, J.~J. Heyd, E.~N. Brothers, K.~N. Kudin, V.~N. Staroverov, T.~A. Keith, R.~Kobayashi, J.~Normand, K.~Raghavachari, A.~P. Rendell, J.~C. Burant, S.~S. Iyengar, J.~Tomasi, M.~Cossi, J.~M. Millam, M.~Klene, C.~Adamo, R.~Cammi, J.~W. Ochterski, R.~L. Martin, K.~Morokuma, O.~Farkas, J.~B. Foresman, and D.~J. Fox.
\newblock Gaussian~16 {R}evision {C}.01, 2016.
\newblock Gaussian Inc. Wallingford CT.

\bibitem{berendsen1984molecular}
Herman~JC Berendsen, JPM~van Postma, Wilfred~F Van~Gunsteren, ARHJ DiNola, and Jan~R Haak.
\newblock Molecular dynamics with coupling to an external bath.
\newblock {\em The Journal of Chemical Physics}, 81(8):3684--3690, 1984.

\bibitem{nose1984unified}
Shuichi Nos{\'e}.
\newblock A unified formulation of the constant temperature molecular dynamics methods.
\newblock {\em The Journal of Chemical Physics}, 81(1):511--519, 1984.

\bibitem{hoover1985canonical}
William~G Hoover.
\newblock Canonical dynamics: {E}quilibrium phase-space distributions.
\newblock {\em Physical Review A}, 31(3):1695, 1985.

\bibitem{martyna1992nose}
Glenn~J Martyna, Michael~L Klein, and Mark Tuckerman.
\newblock Nos{\'e}--{H}oover chains: The canonical ensemble via continuous dynamics.
\newblock {\em The Journal of Chemical Physics}, 97(4):2635--2643, 1992.

\bibitem{barducci2008well}
Alessandro Barducci, Giovanni Bussi, and Michele Parrinello.
\newblock Well-tempered metadynamics: a smoothly converging and tunable free-energy method.
\newblock {\em Physical Review Letters}, 100(2):020603, 2008.

\bibitem{bonomi2009plumed}
Massimiliano Bonomi, Davide Branduardi, Giovanni Bussi, Carlo Camilloni, Davide Provasi, Paolo Raiteri, Davide Donadio, Fabrizio Marinelli, Fabio Pietrucci, Ricardo~A Broglia, et~al.
\newblock {PLUMED}: A portable plugin for free-energy calculations with molecular dynamics.
\newblock {\em Computer Physics Communications}, 180(10):1961--1972, 2009.

\bibitem{tribello2014plumed}
Gareth~A Tribello, Massimiliano Bonomi, Davide Branduardi, Carlo Camilloni, and Giovanni Bussi.
\newblock {PLUMED} 2: New feathers for an old bird.
\newblock {\em Computer physics communications}, 185(2):604--613, 2014.

\bibitem{plumed2019promoting}
{The PLUMED consortium}.
\newblock Promoting transparency and reproducibility in enhanced molecular simulations.
\newblock {\em Nature methods}, 16(8):670--673, 2019.

\bibitem{branduardi2012metadynamics}
Davide Branduardi, Giovanni Bussi, and Michele Parrinello.
\newblock Metadynamics with adaptive {G}aussians.
\newblock {\em Journal of Chemical Theory and Computation}, 8(7):2247--2254, 2012.

\bibitem{bussi2019analyzing}
Giovanni Bussi and Gareth~A Tribello.
\newblock Analyzing and biasing simulations with {PLUMED}.
\newblock In {\em Biomolecular Simulations}, pages 529--578. Springer, 2019.

\bibitem{bussi2020using}
Giovanni Bussi and Alessandro Laio.
\newblock Using metadynamics to explore complex free-energy landscapes.
\newblock {\em Nature Reviews Physics}, 2(4):200--212, 2020.

\bibitem{torrie1977nonphysical}
Glenn~M Torrie and John~P Valleau.
\newblock Nonphysical sampling distributions in {M}onte {C}arlo free-energy estimation: Umbrella sampling.
\newblock {\em Journal of Computational Physics}, 23(2):187--199, 1977.

\bibitem{hourahine2007self}
B~Hourahine, S~Sanna, B~Aradi, C~K{\"o}hler, Th~Niehaus, and Th~Frauenheim.
\newblock Self-interaction and strong correlation in {DFTB}.
\newblock {\em The Journal of Physical Chemistry A}, 111(26):5671--5677, 2007.

\bibitem{tan2012theory}
Zhiqiang Tan, Emilio Gallicchio, Mauro Lapelosa, and Ronald~M Levy.
\newblock Theory of binless multi-state free energy estimation with applications to protein-ligand binding.
\newblock {\em The Journal of Chemical Physics}, 136(14):04B608, 2012.

\bibitem{efron1982jackknife}
Bradley Efron.
\newblock {\em The jackknife, the bootstrap and other resampling plans}.
\newblock SIAM, 1982.

\bibitem{gatz1995standard}
Donald~F Gatz and Luther Smith.
\newblock The standard error of a weighted mean concentration—-{I}. {B}ootstrapping vs other methods.
\newblock {\em Atmospheric Environment}, 29(11):1185--1193, 1995.

\bibitem{hub2010g_wham}
Jochen~S Hub, Bert~L De~Groot, and David Van Der~Spoel.
\newblock g\_wham--{A} free weighted histogram analysis implementation including robust error and autocorrelation estimates.
\newblock {\em Journal of Chemical Theory and Computation}, 6(12):3713--3720, 2010.

\bibitem{salvalaglio2014assessing}
Matteo Salvalaglio, Pratyush Tiwary, and Michele Parrinello.
\newblock Assessing the reliability of the dynamics reconstructed from metadynamics.
\newblock {\em Journal of Chemical Theory and Computation}, 10(4):1420--1425, 2014.

\bibitem{becke1988density}
Axel~D Becke.
\newblock Density-functional exchange-energy approximation with correct asymptotic behavior.
\newblock {\em Physical Review A}, 38(6):3098, 1988.

\bibitem{lee1988development}
Chengteh Lee, Weitao Yang, and Robert~G Parr.
\newblock Development of the {C}olle-{S}alvetti correlation-energy formula into a functional of the electron density.
\newblock {\em Physical Review B}, 37(2):785, 1988.

\bibitem{tajkhorshid1999influence}
Emadeddin Tajkhorshid and Sandor Suhai.
\newblock Influence of the methyl groups on the structure, charge distribution, and proton affinity of the retinal {S}chiff base.
\newblock {\em The Journal of Physical Chemistry B}, 103(26):5581--5590, 1999.

\bibitem{tajkhorshid1999role}
Emadeddin Tajkhorshid, B{\'e}la Paizs, and Sandor Suhai.
\newblock Role of isomerization barriers in the p{K}a control of the retinal {S}chiff base: a density functional study.
\newblock {\em The Journal of Physical Chemistry B}, 103(21):4518--4527, 1999.

\bibitem{grimme2010consistent}
Stefan Grimme, Jens Antony, Stephan Ehrlich, and Helge Krieg.
\newblock A consistent and accurate ab initio parametrization of density functional dispersion correction ({DFT-D}) for the 94 elements {H}-{Pu}.
\newblock {\em The Journal of Chemical Physics}, 132(15), 2010.

\bibitem{schlegel1982optimization}
H~Bernhard Schlegel.
\newblock Optimization of equilibrium geometries and transition structures.
\newblock {\em Journal of Computational Chemistry}, 3(2):214--218, 1982.

\bibitem{peng1993combining}
Chunyang Peng and H~Bernhard~Schlegel.
\newblock Combining synchronous transit and quasi-{N}ewton methods to find transition states.
\newblock {\em Israel Journal of Chemistry}, 33(4):449--454, 1993.

\bibitem{peng1996using}
Chunyang Peng, Philippe~Y Ayala, H~Bernhard Schlegel, and Michael~J Frisch.
\newblock Using redundant internal coordinates to optimize equilibrium geometries and transition states.
\newblock {\em Journal of Computational Chemistry}, 17(1):49--56, 1996.

\end{thebibliography}

\makeatletter\@input{xx.tex}\makeatother
\end{document}


\preprint{AIP/123-QED}

\title[]{Supplementary Information: Thermal Isomerization Rates in Retinal Analogues using Ab-Initio Molecular Dynamics}%
\author{Simon Ghysbrecht}
\author{Bettina G.~Keller}%
\email{bettina.keller@fu-berlin.de}

\date{\today}

\maketitle

\section{Transition state theories}
\label{chapter:Theoretical Background}

The introduction of activated complex theory introduced here largely follows the textbook by Mortimer \cite{mortimer2000physical}. Alternative derivations can be found in textbooks by Peters \cite{peters2017reaction} or McQuarrie~\&~Simon \cite{mcquarrie1997physical}.
\subsection*{Eyring transition state theory}
The cis-trans isomerization is a unimolecular reaction $A \rightarrow B$ which, according to the theory of the activated complex, is modeled as
\begin{equation}
    	A \,\ce{<=>}\, AB^{\ddagger} \rightarrow B
\end{equation}
where the reactant $A$ is the cis state, $AB^{\ddagger}$ is the activated complex, and $B$ is the trans state. One assumes (1) that $A$ and $AB^{\ddagger}$ are in equilibrium with equilibrium constant
\begin{eqnarray}
	K^{\ddagger} &=& \frac{[AB^{\ddagger}]}{[A]}
\end{eqnarray}
where $[...]$ denotes concentrations, and (2) that the rate-determining step in the reaction is the decay of $AB^{\ddagger}$ to the product state $B$. 
Then the reaction rate is
\begin{eqnarray}
\label{eq:general_rate}
    r_{AB} &=& -\frac{d [A]}{dt} = k_{AB}[A] = \nu_r [AB^{\ddagger}] = \nu_r K^{\ddagger} [A]
\end{eqnarray}
where $k_{AB}$ is the rate constant with units $\mathrm{s}^{-1}$, $\nu_r$ is the rate at which the TS conformation decays, and the last equality arises from the definition of the equilibrium constant.
Note that for a unimolecular reaction, equilibrium constant $K^{\ddagger}$ and dimensionless equilibrium constant $\widetilde{K}^{\ddagger}$ are identical, because
\begin{eqnarray}
	K^{\ddagger} = \frac{[AB^{\ddagger}]}{[A]}	=  \frac{[AB^{\ddagger}]/c^{\circ}}{[A]/c^{\circ}} = \widetilde{K}^{\ddagger}\, ,
\end{eqnarray}
where $c^{\circ}$ is the standard concentration.

The rate constant of the reaction then is
\begin{eqnarray}
\label{eq:k_AB_TST}
	k_{AB}^\mathrm{Eyr} = \nu_r K^{\ddagger} 
        &=& \nu_r \cdot \frac{q_{AB^{\ddagger}}}{q_A} \exp\left(-\frac{E_b}{RT}\right)
\end{eqnarray}
where we related $K^{\ddagger}$ to its definition in terms of molecular partition functions $q_{AB^{\ddagger}}$ and $q_A$. 
$q_A$ is calculated with respect to the energy $E_A$ at the minimum energy conformation of $A$, whereas $\widetilde{q}_{AB^{\ddagger}}$ is calculated with respect to the energy at the saddle point of the Born-Oppenheimer potential energy surface $E_{AB^{\ddagger}}$, i.e.~both energies are measured at a single point in conformational space. 
$E_b = E_{AB^{\ddagger}} - E_A$ accounts for the the difference in reference energies and can be interpreted as the energetic barrier of the reaction.
$R$ is the ideal gas constant, and $T$ is the temperature. 
We calculate and report potential and free energies in units of J/mol, correspondingly the thermal energy is also reported as a molar quantity: $RT$. 
Likewise masses and reduced masses are treated as molar masses: kg/mol.
If potential and free energies are treated with units of J, and masses with units of kg, $R$ should be replaced by the Boltzmann constant $k_B = R/N_A$ in eq.~\ref{eq:k_AB_TST} and all of the following equations. 
$N_A$ is the Avogadro constant.
Assuming that the electronic, translational, rotational and vibrational degrees of freedom are sufficiently decoupled, one can decompose the molecular partition function 
\begin{equation}
	q_i = q_{i, \mathrm{el}} \cdot q_{i, \mathrm{tr}} \cdot q_{i, \mathrm{rot}} \cdot q_{i, \mathrm{vib}} \qquad i = A, AB^{\dagger}
\end{equation}
into electronic partition function $q_{i, \mathrm{el}}$, translational partition function $q_{i, \mathrm{tr}}$, rotational partition function $q_{i, \mathrm{rot}}$, and vibrational partition function $q_{i, \mathrm{vib}}$. The ratio of partition functions in eq.~\ref{eq:k_AB_TST} then decomposes into factors
\begin{equation}
\label{eq:factorized_partition_function_ratio}
	\frac{q_{AB^{\ddagger}}}{q_A} = 
	\frac{q_{AB^{\ddagger}, \mathrm{el}}}{q_{A, \mathrm{el}}}\cdot
	\frac{q_{AB^{\ddagger}, \mathrm{tr}}}{q_{A, \mathrm{tr}}}\cdot
	\frac{q_{AB^{\ddagger}, \mathrm{rot}}}{q_{A, \mathrm{rot}}}\cdot
	\frac{q_{AB^{\ddagger}, \mathrm{vib}}}{q_{A, \mathrm{vib}}} \, .
\end{equation}
In retinal, the electronically excited states are only populated when the molecule is excited by light. 
At room temperature and without any external excitation, only the electronic ground state is populated. 
Therefore $q_{AB^{\ddagger}, \mathrm{el}} = q_{A, \mathrm{el}} = 1$, and the electronic factor in eq.~\ref{eq:factorized_partition_function_ratio} equals 1.  
The translational partition function is modelled using the quantum mechanical treatment of a particle in a box. 
The translational factor then reduces to the ratio of masses $q_{AB^{\ddagger}, \mathrm{tr}}/q_{A, \mathrm{tr}} = \left(M_{AB^{\ddagger}}/M_{A}\right)^{3/2}$ = 1, since in a unimolecular reaction the molar mass of the reactant $M_{A}$ and the molar mass of the activated complex $M_{AB^{\ddagger}}$ are equal.
The rotational partition function is modelled using the quantum mechanical treatment of a rigid rotor 
\begin{equation}
    q_{i, rot} = \dfrac{\pi^2}{\sigma_i} \sqrt{\dfrac{8 \pi \,I_{i,a}\, RT}{h^2}} \sqrt{\dfrac{8 \pi \,I_{i,b}\, RT}{h^2}} \sqrt{\dfrac{8 \pi \,I_{i,c}\, RT}{h^2}} 
    \qquad
    i = A, AB^{\dagger}
\end{equation}
where $I_{i, a}$, $I_{i, b}$, and $I_{i, c}$ are the three moments of inertia in the two retinal conformations (measured at the minimum of the reactant state and at the saddle point of the transition state and in molar units), $h$ is the molar Planck constant, and $\sigma_i$ is the symmetry number which is 1 for both conformations.
The rotational factor thus reduces to
\begin{equation}
\label{eq:rot_part_ratio}
\frac{q_{AB^{\ddagger}, \mathrm{rot}}}{q_{A, \mathrm{rot}}}
= \sqrt{\frac{I_{AB^{\ddagger}, a}}{I_{A, a}} \cdot \frac{I_{AB^{\ddagger}, b}}{I_{A, b}} \cdot \frac{I_{AB^{\ddagger}, c}}{I_{A, c}}} \, .
\end{equation}
We assume that the dynamics of the remaining $3N-6$ vibrational degrees of freedom can be modelled by a multidimensional harmonic oscillators (harmonic approximation), where the frequencies $\nu_{i,k}$, with $i = A, AB^{\ddagger}$, along each direction are given by the square rooted eigenvalues of the mass-weighted Hessian matrices at the minimum of the reactant state for $q_{A,\mathrm{vib}}$, and at the saddle point for $q_{AB^{\ddagger},\mathrm{vib}}$ \cite{ochterski1999vibrational,ochterski2000thermochemistry}. 
We define $q_{A; \mathrm{vib}}$ and $q_{AB^\ddagger; \mathrm{vib}}$ relative to the PES minimum at $A$ and relative to the saddle point of the PES minimum at $AB^\ddagger$, i.e.~the vibrational partition functions include the zero-point energies.
The vibrational partition function at the reactant state is
\begin{eqnarray}
	q_{A; \mathrm{vib}} 
	&=& \prod_{k=1}^{3N-6} \frac{\exp\left(-\frac{h\nu_{A,k}}{2RT}\right)}{1-\exp\left(-\frac{h\nu_{A,k}}{RT}\right)} \, .
\label{eq:q_A_vib} 
\end{eqnarray}
Molar masses are used for the mass-weighted Hessian matrices, and the force-constant of the potential is given in molar units.
The mass-weighted Hessian matrix at the saddle point has one negative eigenvalue and accordingly one imaginary frequency $\nu_{AB^{\ddagger},r}^*$. 
This imaginary frequency cannot be properly treated as harmonic vibration, and one therefore defines a reduced vibrational partition function from which the imaginary frequency is excluded 
\begin{eqnarray}
	\widetilde{q}_{AB^{\ddagger}; \mathrm{vib}} 
	&=& \prod_{k=1, k\ne r}^{3N-6} \frac{\exp\left(-\frac{h\nu_{AB^{\ddagger},k}}{2RT}\right)}{1-\exp\left(-\frac{h\nu_{AB^{\ddagger},k}}{RT}\right)}  \, .
\label{eq:q_AB_vib}  
\end{eqnarray}
The reduced vibrational partition function $\widetilde{q}_{AB^{\ddagger}; \mathrm{vib}}$ is related to the vibrational partition function $q_{AB^{\ddagger}; \mathrm{vib}}$ by a multiplicative factor $q_{AB^{\ddagger}; \mathrm{vib}} = f_r \cdot \widetilde{q}_{AB^{\ddagger}; \mathrm{vib}}$, which is approximated as
\begin{eqnarray}
    f_r &\approx& \frac{RT}{h\nu_{AB^{\ddagger},r}} =  \frac{RT}{h\nu_r}\, .
\label{eq:nu_r_approx}    
\end{eqnarray}
where the absolute value of the imaginary frequency  is set equal to the decay rate of the activated complex: $|\nu_{AB^{\ddagger},r}^*| = \nu_r$.
The vibrational factor then is
\begin{eqnarray}
	\frac{q_{AB^{\ddagger}, \mathrm{vib}}}{q_{A, \mathrm{vib}}}
	=	\frac{RT}{h\nu_r} \cdot \frac{\widetilde{q}_{AB^{\ddagger}, \mathrm{vib}}}{q_{A, \mathrm{vib}}}
\end{eqnarray}

With these models and approximations, the reaction rate in eq.~\ref{eq:k_AB_TST} can be calculated as
\begin{eqnarray}
	k_{AB}^{\mathrm{Eyr}} 
	&=& \nu_r \cdot \frac{RT}{h\nu_r} \cdot 
	\frac{\widetilde{q}_{AB^{\ddagger}, \mathrm{vib}}}{q_{A, \mathrm{vib}}}
	\cdot \frac{q_{AB^{\ddagger}, \mathrm{rot}}}{q_{A, \mathrm{rot}}} \cdot \exp\left(-\frac{E_b}{RT}\right) \cr
	&=& \frac{RT}{h} \cdot 
	\frac{\widetilde{q}_{AB^{\ddagger}, \mathrm{vib}}}{q_{A, \mathrm{vib}}}
	\cdot \frac{q_{AB^{\ddagger}, \mathrm{rot}}}{q_{A, \mathrm{rot}}} \cdot \exp\left(-\frac{E_b}{RT}\right) 
\label{eq:EyringTST_SI} 
 \end{eqnarray}
where the frequency of the reactive mode $\nu_r$ cancels. This equation for the reaction rate is often called \textbf{Eyring TST} or \textbf{harmonic TST}. 
By setting
\begin{eqnarray}
\label{eq:deltaG_ddagger2}
    \Delta F^\ddagger &=& E_b - RT \ln \left[ \frac{\widetilde{q}_{AB^{\ddagger}, \mathrm{vib}}}{q_{A, \mathrm{vib}}}
	\cdot \frac{q_{AB^{\ddagger}, \mathrm{rot}}}{q_{A, \mathrm{rot}}}\right]
\end{eqnarray}
eq.~\ref{eq:EyringTST_SI} becomes consistent with eq.~\ref{main-eq:EyringTST}.
There are two ways to justify the treatment of the imaginary frequency \cite{laidler1987chemical}.
In both derivations, one realizes that the eigenvector associated to $\nu_r^*$ corresponds to the direction of the reaction coordinate $s$ at the saddle point.
In the first derivation \cite{laidler1987chemical, mortimer2000physical}, one treats the motion along this coordinate as a vibrational motion with frequency $\nu_r^*$. 
Applying the high-temperature approximation yields eq.~\ref{eq:nu_r_approx}.
One further assumes that the TS decays with $\nu_r$, i.e. there is no restoring force and with the first vibration along this coordinate the $AB^\ddagger$ falls apart.
In the second derivation \cite{laidler1987chemical, peters2017reaction}, the frequency at which the $AB^\ddagger$ decays, $\nu_r$, is modelled as the reactive flux across the transition state region.
The reactive degree of freedom is then treated as a translational degree of freedom, rather than a vibrational degree of freedom. 
To obtain the reactive flux, one additionally calculates the expected value of the absolute velocity at TS in the classical approximation. 
This approach also leads to eq.~\ref{eq:nu_r_approx}.
\subsection*{High-temperature approximation}
In the high-temperature approximation, one assumes that the rate is dominated by the energy barrier and the vibrational contribution, and thus
\begin{eqnarray}
    \frac{q_{AB^{\ddagger}, \mathrm{rot}}}{q_{A, \mathrm{rot}}}  \approx 1\, .
\end{eqnarray}
Then 
\begin{eqnarray}
	k_{AB}^{\mathrm{ht}}
	&=& \frac{RT}{h} \cdot 
	\frac{\widetilde{q}_{AB^{\ddagger}, \mathrm{vib}}}{q_{A, \mathrm{vib}}} \cdot \exp\left(-\frac{E_b}{RT}\right)  \, .
\label{eq:k_ht_intermediate_step} 
\end{eqnarray}
Using eqs.~\ref{eq:q_A_vib} and \ref{eq:q_AB_vib}, the ration of vibriational partition functions can be reformulated as
\begin{align}
    \frac{\widetilde{q}_{AB^{\ddagger}, \mathrm{vib}}}{q_{A, \mathrm{vib}}} 
    = e^{-\frac{h\nu_{A,r}}{2RT}} 
    \cdot 
    \prod_{k=1, k\ne r}^{3N-6} \frac{\exp\left(-\frac{h\nu_{A,k}}{2RT}\right)}{\exp\left(-\frac{h\nu_{AB^\ddagger,k}}{2RT}\right)} 
    \cdot 
    \frac{\prod_{k=1, k\ne r}^{3N-6} 1-\exp\left(-\frac{h\nu_{AB^{\dagger},k}}{RT}\right)}{\prod_{k=1}^{3N-6} 1-\exp\left(-\frac{h\nu_{A,k}}{RT}\right)}   \, .
\end{align}
The first factor is the contribution of the specific mode in $A$ that corresponds the reactive mode at $AB^\ddagger$. 
The second factor represents the contribution due to the zero-point energies of the vibrational modes (excluding the reactive mode). 
The third factor is the ratio the vibrational partition functions relative to the zero-point energies. 
At high temperatures, the ratio in the first term can be approximated as 
\begin{eqnarray}
    \prod_{k=1, k\ne r}^{3N-6}    \frac{\exp\left(-\frac{h\nu_{A,k}}{2RT}\right)}{\exp\left(-\frac{h\nu_{AB^\ddagger,k}}{2RT}\right)}
    &=&
    \prod_{k=1, k\ne r}^{3N-6} \exp\left(-\frac{h\, \Delta\nu_k}{2RT}\right)\cr
    &\approx& 1\, 
\label{eq:ht_term1}    
\end{eqnarray}
where $\Delta\nu_k = \nu_{A,k}- \nu_{AB^\ddagger,k}$ is the frequency difference between $A$ and $AB^\ddagger$ of the $k$th mode, 
We assumed that $h\Delta\nu_k \ll RT$, and hence $\exp\left(-\frac{h\, \Delta\nu_k}{2RT}\right) \approx \exp(0) = 1$.
We furthermore assume that for the mode $\nu_{A,r}$ has a low frequency, such that $h\nu_{A,r} \ll RT$ and hence $e^{-\frac{h\nu_{A,r}}{2RT}} \approx \exp(0) = 1$.
Then the entire first factor can be approximated by $1$.
In making this approximation we neglect the contribution of the zero-point energies to the free energy difference $\Delta F^\ddagger$.
For the high-temperature approximation of the second term, one uses a Maclaurin series expansion of the exponential function and truncates it after the linear term
\begin{eqnarray}
1-\exp\left(-\frac{h\nu_{A,k}}{RT}\right)
\approx 1- 1 + \frac{h\nu_{A,k}}{RT} = \frac{h\nu_{A,k}}{RT}
\end{eqnarray}
which is justified if $h\nu_{A,k} \ll RT$.
Then 
\begin{align}
\frac{\prod_{k=1, k\ne r}^{3N-6} 1-\exp\left(-\frac{h\nu_{AB^{\dagger},k}}{RT}\right)}{\prod_{k=1}^{3N-6} 1-\exp\left(-\frac{h\nu_{A,k}}{RT}\right)} 
&\approx  \frac{\prod_{k=1, k\ne r}^{3N-6} \frac{RT}{h\nu_{AB^{\ddagger},k}}}{\prod_{k=1}^{3N-6}  \frac{RT}{h\nu_{A,k}}} \cr
&=\frac{h}{RT}  \cdot \frac{\prod_{k=1}^{3N-6}  \nu_{A,k}} 	{\prod_{k=1, k\ne r}^{3N-6} \nu_{AB^{\ddagger},k}}	\, .
\label{eq:ht_term2}    
\end{align}
Inserting eq.~\ref{eq:ht_term1} and \ref{eq:ht_term2} into eq.~\ref{eq:k_ht_intermediate_step} yields the high-temperature approximation to the rate constant form Eyring TST
\begin{eqnarray}
	k_{AB}^{\mathrm{ht}} 
	= \frac{\prod_{k=1}^{3N-6}  \nu_{A,k}} 	{\prod_{k=1, k\ne r}^{3N-6} \nu_{AB^{\ddagger},k}} \cdot \exp\left(-\frac{E_b}{RT}\right)\, . 
\label{eq:k_AB_ht_app} 
\end{eqnarray}
\subsection*{Classical harmonic approximation}
Eq.~\ref{eq:k_AB_ht_app} can also be obtained from the classical treatment of the vibrational partition function. 
The classical Hamilton function of a one-dimensional harmonic vibration is $\mathcal{H}(x,p) = \frac{\kappa}{2} x^2 + \frac{p^2}{2M}$, where $M$ is the reduced molar mass, $x$ is the position, $p$ is the momentum, and $\kappa$ is the molar force constant for the harmonic approximation of the potential.
The corresponding classical partition function is
\begin{align}
	q_{\mathrm{classical}, \mathrm{vib}, 1D} 
	&=\frac{1}{h}
	\int_{-\infty}^{\infty} \mathrm{d}x
	\int_{-\infty}^{\infty} \mathrm{d}p  \exp \left(-\frac{1}{RT} \mathcal{H}(x,p) \right)\cr
	&=\frac{1}{h}
	\int_{-\infty}^{\infty} \mathrm{d}x \exp \left(-\frac{1}{RT}  \frac{\kappa}{2} x^2 \right)
	\int_{-\infty}^{\infty} \mathrm{d}p \exp \left(-\frac{1}{RT} \frac{p^2}{2M}\right)\cr
	&=\frac{1}{h} \cdot
	\sqrt{\pi RT  \frac{2}{\kappa}} \cdot
	\sqrt{\pi RT 2M} \cr
	&=\frac{RT}{h}  \cdot 2\pi \cdot
	\sqrt{  \frac{M}{\kappa}} \cr
	&=\frac{RT}{h\nu}
\end{align}
where we used that the two integrals are Gaussian integrals, and that the frequency of a harmonic oscillator is $\nu = \frac{1}{2\pi}\cdot \sqrt{\kappa/M}$. The vibrational partition function of a system with $N$ atoms then is
\begin{eqnarray}
	q_{\mathrm{classical},\mathrm{vib}} = \prod_{k=1}^{3N-6}\frac{RT}{h\nu_k}
\end{eqnarray}
The ratio of classical vibrational partition functions then is
\begin{eqnarray}
 \frac{\widetilde{q}_{AB^{\ddagger}, \mathrm{classical}, \mathrm{vib}}}{q_{A, \mathrm{classical}, \mathrm{vib}}} 
 &=& \frac{\prod_{k=1, k\ne r}^{3N-6}\frac{RT}{h\nu_{AB^\ddagger,k}}}{\prod_{k=1}^{3N-6}\frac{RT}{h\nu_{A,k}}} \cr
 &=& \frac{h}{RT}  \cdot \frac{\prod_{k=1}^{3N-6}  \nu_{A,k}} 	{\prod_{k=1, k\ne r}^{3N-6} \nu_{AB^{\ddagger},k}}	\, .
\end{eqnarray} 
Using this ratio in eq.~\ref{eq:k_ht_intermediate_step} yields the high-temperature approximation for rate constant from Eyring TST (eq.~\ref{eq:k_AB_ht_app}).
This result shows that, in the limit of high temperature, classical and quantum mechanical treatment of transition state theory coincide.
At $T=300 \, \mathrm{K}$, this is approximation is valid for $\nu \ll k_BT/h = 6.25\cdot10^{12}\, \mathrm{s}^{-1}$ or $208.5\,\mathrm{cm}^{-1}$. 
Since many vibrational modes in our two compounds have higher frequencies, the high-temperature approximation does not yet strictly apply. 
However, the two molecular scaffolds are very rigid and most vibrational modes likely have similar frequencies in $A$ and in $AB^\ddagger$.
The corresponding factor for these modes in $\frac{\widetilde{q}_{AB^{\ddagger}, \mathrm{vib}}}{q_{A, \mathrm{vib}}}$ then is about 1 and its contribution to the free-energy difference $\Delta F^\ddagger$ is approximately zero.
The one covalent bond that is altered during the reaction, the C$_{13}$=C$_{14}$ double bond, is represented by $\nu_r$ and is treated separately in the theory of the activated complex. 
Thus, the high-temperature approximation of the transition state rate can be a valid approximation, even if the high-temperature limit for the individual vibrational partition functions is not yet reached.

\subsection*{Sampling anharmonic vibrations}
The dynamics of the a system with $N$ atoms evolve in the $6N$-dimensional phase space $\gamma = (x, p)\in  \mathbb{R}^{6N}$, where $x\in \mathbb{R}^{3N}$ are the atomic positions and $p\in \mathbb{R}^{3N}$ are the atomic momenta. 
$V(x)$ is the molecular potential energy function and represents the Born-Oppenheimer surface. 
In the present study, these dynamics are simulated in the NVT ensemble at $T=300\, \mathrm{K}$.
The classical Hamilton function of such a system is $\mathcal{H}(x,p) = V(x) + \sum_{i=1}^{3N} \frac{p_j^2}{2M_j}$, where $M_j$ is the molar mass associated to the $j$th degree of freedom. 
The classical partition functions of the reactant state $A$ and the activated complex $AB^{\ddagger}$ are
\begin{eqnarray}
q_{\mathrm{classical, i}} &=& q_{\mathrm{classical},\,x,i} \cdot q_{\mathrm{classical},\,p,i} \qquad  i = A, AB^{\ddagger}
\end{eqnarray}
with 
\begin{eqnarray}
q_{\mathrm{classical},\,x, i} &=&  \frac{1}{h^{3N}}\int_{x \in i} \mathrm{d}x \exp\left(-\frac{1}{RT}V(x)\right) \cr
q_{\mathrm{classical},\,p, i} &=&  \int \mathrm{d}p \exp\left(-\frac{1}{RT} \sum_{i=j}^{3N} \frac{p_j^2}{2M_i}\right) \cr
                           && i = A, AB^{\ddagger}\, .
\end{eqnarray}
We assumed that all $N$ atoms in the system are distinguishable, which is a reasonable assumption for a single molecule like retinal. 
It is a matter of convention whether the factor $\frac{1}{h^{3N}}$ is included in the configurational partition function $q_{\mathrm{classical},\,x}$ or not.
Using classical partition functions in eq.~\ref{eq:k_AB_TST} yields
\begin{eqnarray}
	k_{AB}^{\mathrm{Eyr}} &=& \nu_r \cdot K^\ddagger 
            = \nu_r \cdot \frac{q_{\mathrm{classical},AB^{\ddagger}}}{q_{\mathrm{classical},A}} \,.
\end{eqnarray}
Note that in the classical treatment of the partition function, $A$ and $AB^{\ddagger}$ have the same reference energy and the factor $E_b$ is absorbed into the partition function. 
The factor  $\exp\left(-\frac{E_b}{RT}\right)$ arises when the integration for $q_{\mathrm{classical},\,x, i}$ is carried out.
Classical partition functions and reaction rates can be estimated by sampling the dynamics of the molecule using MD simulations. 
In this sense, MD simulation is a numerical integration technique to solve the high-dimensional integrals that appear in the classical partition function, without resorting to a harmonic approximation.
However, owing to the large barriers involved in chemical reactions, the numerical estimates converge poorly when the simulation is carried out at $V(x)$.
Instead one uses importance sampling and reweighting. 
Here, we discuss metadynamics \cite{grubmuller1995predicting, voter1997hyperdynamics, huber1994local, laio2002escaping} and infrequent metadynamics \cite{tiwary2013metadynamics} as importance sampling methods to obtain the classical approximation of Eyring TST. 
In this method, one samples the dynamics at a biased potential 
\begin{eqnarray}
	V^{\mathrm{InMetaD}}(x, t) = V(x) + U(x, t)
\end{eqnarray}
where $U(x,t)$ is a time-dependent bias.
The bias is chosen such that the rate constant $k_{AB}^{\mathrm{InMetaD}}$ at $V^{\mathrm{InMetaD}}(x)$ is increased to the rate constant $k_{AB}$ at the molecular potential: $k_{AB}^{\mathrm{InMetaD}} = \alpha k_{AB}$, where the acceleration factor is
\begin{eqnarray}
\label{eq:alpha_inmetad}
	\alpha 
	&=&\frac{k_{AB}^{\mathrm{InMetaD}}}{k_{AB}} \cr
	&=& \nu_r^{\mathrm{InMetaD}}  \frac{q_{\mathrm{classical},AB^{\ddagger}}^{\mathrm{InMetaD}}}{q_{\mathrm{classical},A}^{\mathrm{InMetaD}}} \cdot 
	\frac{1}{\nu_r}  \frac{q_{\mathrm{classical},A}}{q_{\mathrm{classical}, AB^{\ddagger}}}\cr
	&\approx& \frac{q_{\mathrm{classical},A}}{q_{\mathrm{classical},A}^{\mathrm{InMetaD}}} \, ,
\end{eqnarray}
where $q_{\mathrm{classical}, A}^{\mathrm{InMetaD}}$ and $q_{\mathrm{classical},AB^{\ddagger}}^{\mathrm{InMetaD}}$ are the classical partition functions for $A$ and $AB^\ddagger$ at the biased potential $V^{\mathrm{InMetaD}}(x)$.
In the last line in eq.~\ref{eq:alpha_inmetad}, we assumed that 
$\nu_r \approx \nu_r^{\mathrm{InMetaD}}$ and
$q_{\mathrm{classical},AB^{\ddagger}} \approx q_{\mathrm{classical},AB^{\ddagger}}^{\mathrm{InMetaD}}$.
This is the case if the bias is only deposited in the reactant state, and the potential in the region around $AB^\ddagger$ remains unmodified. 
At constant bias $U(x)$, this acceleration factor is essentially the free energy difference in the reactant state $A$ and can be estimated as
\begin{eqnarray}
\alpha
&=& \frac{q_{\mathrm{classical},A}}{q_{\mathrm{classical},A}^{\mathrm{InMetaD}}} \cr
&=& \frac{\frac{1}{h^{3N}}\int_{x \in A} \mathrm{d}x \exp\left(-\frac{1}{RT}V(x)\right)}{\frac{1}{h^{3N}}\int_{x \in A} \mathrm{d}x \exp\left(-\frac{1}{RT}\left[V(x)+U(x)\right]\right)} \cr
&=& \frac{\int_{x \in A} \mathrm{d}x \exp\left(-\frac{1}{RT}\left[V(x)+ U(x)\right]\right)\cdot \exp\left(+\frac{1}{RT}U(x)\right)}{\int_{x \in A} \mathrm{d}x \exp\left(-\frac{1}{RT}\left[V(x)+U(x)\right]\right)} \cr
&=& \left\langle \exp\left(+\frac{1}{RT}U(x)\right) \right\rangle_A
\label{eq:alpha_01}
\end{eqnarray}
where $\langle ...\rangle_A$ denotes an ensemble average restricted to the reactant state $A$ and is measured at $V^{\mathrm{InMetaD}}(x)$. 
To estimate the rate from the simulation data, one uses that the rate is related to the dynamics via
\begin{eqnarray}
    k_{AB} = \frac{1}{\tau_{AB}} = \frac{1}{\alpha} \, k_{AB}^{\mathrm{InMetaD}} = \frac{1}{\alpha\tau_{AB}^{\mathrm{InMetaD}}} 
\end{eqnarray}
where $\tau_{AB}^{\mathrm{InMetaD}}$ is the mean first passage time at $V^{\mathrm{InMetaD}}(x)$. 
The estimator for the mean first passage time is the arithmetic mean of the sampled first passage times $\tau_{AB,i}^{\mathrm{InMetaD}} $
\begin{eqnarray}
	\tau_{AB}^{\mathrm{InMetaD}} = \lim_{N_{AB} \rightarrow \infty}\,\frac{1}{N_{AB}} \sum_{i=1}^{N_{AB}} \tau_{AB, i}^{\mathrm{InMetaD}}
\end{eqnarray}
where $N_{AB}$ is the number of transition events observed during the simulation.
Alternatively, one can estimate $\tau_{AB}^{\mathrm{InMetaD}}$ by fitting the cumulative distribution function of the simulated first passage times to the cumulative distribution of a Poisson process.
In infrequent metadynamics \cite{tiwary2013metadynamics}, a time-dependent bias $U(x,t)$ is used. 
One assumes $U(x,t)$ changes slowly enough that at any time $t$ the assumptions of Eyring TST are met, and that no bias is deposited in the transition state region.
One starts a simulation $x^{(i)}$ in reactant state $A$ and stops the simulation as soon as it crosses $AB^{\ddagger}$. 
Because of the time-dependent bias, the first passage times at $V^{\mathrm{InMetaD}}(x)$ and at $V(x)$ are not simply related by constant acceleration function, but by the following time integral 
\begin{eqnarray}
    \tau_{AB,i} &=&  \int_{t=0}^{\tau_{AB,i}^{\mathrm{InMetaD}}} \mathrm{d}t\, \exp\left(+\frac{1}{k_BT}U(x^{(i)}_t)\right)\, ,
\label{eq:InMetaD_reweighted_tau01}    
\end{eqnarray}
where we used the result from eq.~\ref{eq:alpha_01} to approximate the instantaneous acceleration.
Discretizing the integral yields \cite{khan2020fluxional, palacio2022free}
\begin{eqnarray}
	\tau_{AB,i} &\approx& \Delta t \sum_{t=1}^{N_{i,t}} \exp\left(+\frac{1}{k_BT}U(x^{(i)}_t)\right) \, ,
\end{eqnarray}
where 
$x^{(i)}_t$ is the $t$th time step in trajectory $x^{(i)}$, 
$\Delta t$ is the time difference between subsequent frames in the trajectory
$U(x^{(i)}_t)$ is the bias at time $t\Delta t$ at position $x^{(i)}_t$,
and $N_t$ is the number of frames in the trajectory. 
From these rescaled first passage times, one can estimate the mean first passage time $\tau_{AB}$ by fitting to a Poisson distribution.

Eq.~\ref{eq:InMetaD_reweighted_tau01} can be cast in terms of an acceleration factor by multiplying with $1 =\tau_{AB, i}^{\mathrm{InMetaD}}/ \tau_{AB,i}^{\mathrm{InMetaD}}$
\begin{eqnarray}
    \tau_{AB,i} &=&  \tau_{AB,i}^{\mathrm{InMetaD}} \cdot \alpha(\tau_{AB,i}^{\mathrm{InMetaD}})
\label{eq:InMetaD_reweighted_tau02}    
\end{eqnarray}
where
\begin{equation}
\alpha(\tau_{AB,i}^{\mathrm{InMetaD}})
=\frac{1}{\tau_{AB,i}^{\mathrm{InMetaD}}} \int_{t=0}^{\tau_{AB,i}^{\mathrm{InMetaD}}}\mathrm{d}t \, \exp\left(+\frac{1}{k_BT}U(x^{(i)}_{t})\right) \, .
\end{equation}

\clearpage

\section{Computational Details}
\label{chapter:DFTB_computational_details}

\subsection{Geometry optimization on the DFTB3 PES}
\label{sec:method_DFTB3}
Starting structures of the cis and the trans conformation of pSb5 and pSb1 were generated using Gaussian's graphical interface, GaussView 6 \cite{gv6}, and converted to \textit{.xyz} file format.
These structures were energy minimized on the level of the self-consistent-charge density-functional tight-binding method including the third order correction (DFTB3) and using the 3ob-3-1 Slater–Koster parameter set \cite{gaus2013parametrization}.
Applicability of self-consistent charge DFTB to the retinal cofactor has been extensively documented \cite{zhou2002performance,bondar2011ground}.
Energy minimizations and constrained optimizations were carried out by interfacing DFTB+ software package  \cite{hourahine2020dftb+} with the Atomic Simulation Environment (ASE) \cite{larsen2017atomic}, and using the Broyden–Fletcher–Goldfarb–Shanno (BFGS) algorithm \cite{fletcher2000practical} with a  maximum force of $10^{-5}\,\mathrm{eV/\AA}$ as convergence criterion.
Transition states were optimized using the Nudged Elastic Band (NEB) method \cite{henkelman2000improved} with climbing image \cite{henkelman2000climbing} using 22 nodes starting from the optimized reactant and product states (trans and cis respectively) in combination with BFGS for numerical optimization.
For pSb5, the spring constants of the bands were chosen to be $0.8\,\mathrm{eV/\AA}^2$, while a maximal force of $0.02\,\mathrm{eV/\AA}$ was used as convergence criterion.
For pSb1, the transition state search was considerably more sensitive to the parameters used.
Here, a two-step optimization was performed, first using NEB without climbing image with spring constants of $0.3\,\mathrm{eV/\AA}^2$ and a maximum force of $0.01\,\mathrm{eV/\AA}$, followed by a second NEB optimization with climbing image using the same spring constants and maximum force.
Moments of inertia for the reactant states (cis or trans) as well as for transition states were obtained by entering the configurations into the Gaussian 16 software \cite{g16}, which calculates rotational temperatures $\Theta_{\mathrm{rot},i,k}=h^2/8\pi^2k_BI_{i,k}$, where $i$ designates the configuration ($A$ or $AB^\ddagger$) and $k$ designates the axis of inertia ($a$, $b$ or $c$, see Appendix A).
These were used to calculate the ratio of rotational partition functions $q_{AB^\ddagger,\mathrm{rot}}/q_{A,\mathrm{rot}}$ as in eq.~\ref{eq:rot_part_ratio}, from which the rotational contributions $\Delta F_{\mathrm{rot}}$ (eq.~\ref{main-eq:DeltaF_rot}) to the free energy difference $\Delta F^\ddagger$ (eq.~\ref{main-eq:deltaG_Eyr_2}) was computed, see Tables \ref{main-tab:pSb5_rates} and \ref{main-tab:pSb1_rates}.
Vibrational mode analyses were carried out for the reactant states (cis or trans) as well as for transition states using DFTB+.
The smallest six frequencies 
correspond to the translational and rotational degrees of freedom, and were excluded from the subsequent calculation.
After removing the smallest frequencies, reactant state configurations did not have imaginary frequencies, and transition state configurations only had one large imaginary frequency, as is expected for optimized structures.

\begin{figure*}[h]
\centering
\includegraphics[width=\textwidth]{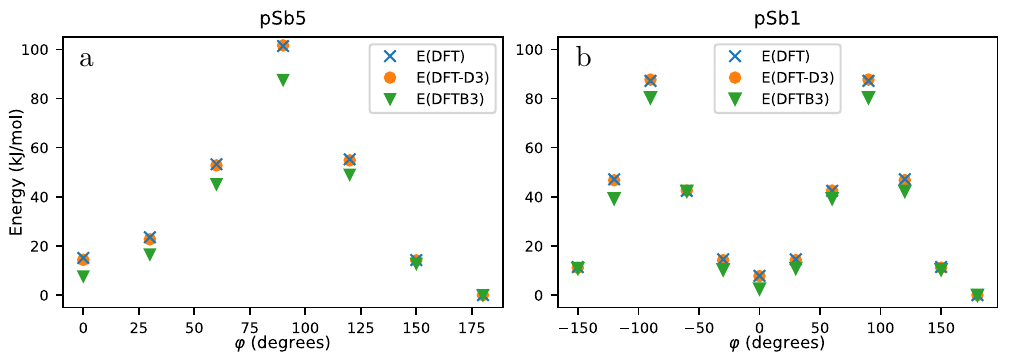} 
\caption{
Electronic energies from constrained optimization calculations using DFTB3 ($E(\mathrm{DFTB3})$) as well as unrestricted DFT/B3LYP with ($E(\mathrm{DFT-D3})$) and without ($E(\mathrm{DFT})$) Grimme's D3 correction for \textbf{a:} pSb5 and \textbf{b:} pSb1. 
Energies have been taken relative to the trans structure (180 degrees) within the respective level of theory. 
}
\label{fig:D3_comparison}
\centering
\end{figure*}

From the resulting frequencies, the vibrational contributions $\Delta F_{\mathrm{vib}}$ (eq.~\ref{main-eq:DeltaF_vib}) to the free energy differences for Eyring TST ($\Delta F^\ddagger$, eq.~\ref{main-eq:deltaG_Eyr_2}) and $\Delta F^\mathrm{ht}_{\mathrm{vib}}$ for the high-temperature limit ($\Delta F^{\ddagger,\mathrm{ht}}$, eq.~\ref{main-eq:deltaG_high_T}) was calculated.
From the free energy differences $\Delta F^\ddagger$ and $\Delta F^{\ddagger,\mathrm{ht}}$ Eyring TST rate constants and and high-temperature TST constants were calculated using eqs.~\ref{main-eq:EyringTST} and \ref{main-eq:high_T_TST}, respectively.
To obtain the energy scan in Fig.~\ref{fig:D3_comparison}, twelve constrained geometry optimizations were carried out for each compound, where the dihedral angle $\varphi$ was constrained at values in 30 degree intervals over the whole 360 degree range, that is at -150, -120, -90, -60, -30, 0, 30, 60, 90, 120, 150 and 180 degrees.
%

\subsection{Well-tempered metadynamics with DFTB3}
Starting structures of the cis and the trans conformation of pSb5 and pSb1 were generated as described in section \ref{sec:method_DFTB3} of the supplement.
MD simulations (equilibration and production) were carried out using the DFTB+ software package \cite{hourahine2020dftb+}. 
The potential energy and the resulting forces were calculated with the self-consistent-charge density-functional tight-binding method including the third order correction (DFTB3) in combination with the 3ob-3-1 Slater–Koster parameter set \cite{gaus2013parametrization}. 
The equations of motions were integrated using the velocity-Verlet integrator with a time step of $\Delta t = 1\,\mathrm{fs}$.
The $1\,\mathrm{fs}$ time step was validated by comparison to the periods of the fastest vibrational components obtained from vibrational mode analysis.
These were $9.7\,\mathrm{fs}$ and $10.0\,\mathrm{fs}$ in the reactant states, and $9.6$ and $10.0\,\mathrm{fs}$  in the transition states, for pSb5 and pSb1 respectively.
Thus, our time step is about an order of magnitude smaller then the fastest vibrational mode. 
The system was equilibrated for $50\,\mathrm{ps}$ using Berendsen thermostat \cite{berendsen1984molecular} at $T = 300\,\mathrm{K}$ with a coupling strength of $\Delta t/ \tau = 5 \times 10^{-4}$, corresponding to a the coupling time of $\tau = 2\,\mathrm{ps}$.
This was followed by a second equilibration of $50\,\mathrm{ps}$ at $T=300\,\mathrm{K}$ using a Nos\'e-Hoover thermostat \cite{nose1984unified,hoover1985canonical,martyna1992nose} of chain length three and coupling frequency of $0.5\,\mathrm{THz}$, corresponding to a the coupling time of $\tau = 2\,\mathrm{ps}$.
Well-tempered metadynamics \cite{barducci2008well} were carried out by plugging the PLUMED software package \cite{bonomi2009plumed,tribello2014plumed,plumed2019promoting} with the DFTB+ package \cite{hourahine2020dftb+}.
Potential energy, integrator settings and settings of the Nos\'e-Hoover thermostat were the same as in the second equilibration run. 
For each of the two systems, we carried out three different well-tempered metadynamics simulations (MetaD1, MetaD2, MetaD3).
As biased collective variable, we used the C$_{12}$-C$_{13}$=C$_{14}$-C$_{15}$ dihedral angle $\varphi$.
Height and width of the Gaussian bias potentials, deposition rate, and bias factor, as well as simulation time are reported in Table \ref{main-tab:metad_parameters}.
Sporadically, the self-consistent charge calculation of the DFTB3 force evaluation would fail to converge for a specific configuration along a longer metadynamics run.
In that case, a small perturbation was enforced to the velocities of the corresponding configuration, after which the metadynamics simulation was resumed.
Unbiasing weights for the trajectory were calculated using the bias potential obtained at the end as described in Ref.~\citenum{branduardi2012metadynamics}.
Free energy surfaces were calculated after building a weighted histogram from the trajectory starting at a simulation time where the bias can be considered converged. 
\subsection{Error estimates for metadynamics FES}
Error estimates for free energy profiles obtained from metadynamics reweighting can be determined using the block analysis technique \cite{bussi2019analyzing} on the reweighted trajectory. 
Block analysis was carried out using the example code on the PLUMED website. 

The free energy difference at a certain simulation time is calculated by determining the FES corresponding to the bias at that time (i.e.~from the scaled upside-down bias, see Refs.~\citenum{branduardi2012metadynamics,bussi2019analyzing,bussi2020using}). 
This FES is used to calculate the relative probabilities of being in cis versus being in trans.
Using eq.~\ref{main-eq:FES}:
\begin{equation}
    \pi_\mathrm{cis} = \int_{-\pi/2}^{\pi/2}\mathrm{d}\varphi \, \pi(\varphi) = \int_{-\pi/2}^{\pi/2}\mathrm{d}\varphi \, \exp\left(-\frac{F(\varphi)}{k_BT}\right)
\end{equation}
and equivalent for trans in $\varphi<-\pi/2$ and $\varphi>\pi/2$.
The free energy of a state can then be calculated using $F_\mathrm{cis}=-k_BT\ln \pi_\mathrm{cis}$ and equivalent for trans, and the free energy difference
\begin{align}
    \Delta F = F_\mathrm{cis} - F_\mathrm{trans} = -k_BT\ln\frac{\pi_\mathrm{cis}}{\pi_\mathrm{trans}}  \, .
\end{align}
\subsection{Umbrella Sampling}

Umbrella Sampling \cite{torrie1977nonphysical} has been run for thermal isomerization over the C$_{13}$=C$_{14}$ double bond for both pSb5 and pSb1, constraining the C$_{12}$-C$_{13}$=C$_{14}$-C$_{15}$ dihedral angle $\varphi$.
Biasing of the CV was carried out by plugging the PLUMED software package \cite{bonomi2009plumed,tribello2014plumed,plumed2019promoting} with the DFTB+ package \cite{hourahine2007self}.
For both pSb5 and pSb1, three sets of umbrella sampling simulations have been performed, i.e.~sets US1, US2 and US3.
Each set has the same parameter setup and consisted of 83 trajectories.
The parameters (number of windows, the region of $\varphi$ in which they are distributed at a regular interval, and the force constant of the umbrella potential) are reported in Table \ref{main-tab:us_parameters}.
Each window was initialized starting from the structure obtained from constrained optimization at the $\varphi$-value closest to the position of the window.
For every window, an initial equilibration of $25\,\mathrm{ps}$ was performed from the respective starting structure using a Berendsen thermostat of coupling strength $5 \times 10^{-4}$.
The initial velocities were generated randomly from an initial Maxwell-Boltzmann distribution of atomic velocities.
This is followed by a second equilibration run of $25\,\mathrm{ps}$ using the same Nos\'e-Hoover chain setup that is used during production runs.
Every production trajectory was $2\,\mathrm{ns}$ making for a total simulation time of $166\,\mathrm{ns}$ per set.
Binless WHAM \cite{tan2012theory,bussi2019analyzing} was used to establish the free energy profile from the trajectory data. 
Error estimates for the free energy profiles obtained from umbrella sampling can be computed using the bootstrapping method \cite{efron1982jackknife}. 
For each umbrella, the trajectory was split in 20 blocks of equal length. 
A `new' trajectory of the same length as the original is then constructed by taking combinations of these 20 blocks with the possibility of repetition. 
After doing this for all umbrellas, the free energy surface is recalculated using WHAM. 
This procedure is repeated 200 times, producing 200 free energy surfaces which allows calculation of standard deviations which can be shown to be good estimates of standard errors on the free energy surface \cite{gatz1995standard}. 
Notice the standard errors might be underestimated because of correlations between blocks within each trajectory \cite{hub2010g_wham}.
Because of important correlated motion in orthogonal degrees of freedom, we expect this to be the case for US of pSb5.
%


\subsection{Unbiased MD simulations with DFTB3}
Separate unbiased ab-initio MD simulations were performed in the cis and the trans state for both pSb5 and pSb1, i.e.~generating a total of four unbiased trajectories.
Dynamics were simulated using the DFTB+ software package \cite{hourahine2007self} at the DFTB3 level, with starting structures of the cis and the trans conformation of pSb5 and pSb1 generated similarly as before (section \ref{sec:method_DFTB3} of the supplement).
For each simulation, first a $25\,\mathrm{ps}$ equilibration using the Berendsen thermostat \cite{berendsen1984molecular} at $300\,\mathrm{K}$ with a coupling time of $2\,\mathrm{ps}$ was performed,
followed by a second equilibration of $25\,\mathrm{ps}$ at $300\,\mathrm{K}$ using a Nos\'e-Hoover thermostat \cite{nose1984unified,hoover1985canonical,martyna1992nose} with chain length 3 and a coupling time of $2\,\mathrm{ps}$.
Once equilibrated, production runs of $2\,\mathrm{ns}$ were performed.

\subsection{Infrequent metadynamics with DFTB3}
Infrequent metadynamics \cite{tiwary2013metadynamics,salvalaglio2014assessing} was used to study a total of four transitions: cis $\rightarrow$ trans and trans $\rightarrow$ cis for both pSb5 and pSb1.
For each of these four transitions, two separate sets of infrequent metadynamics (InMetaD1 and InMetaD2) were carried out, giving rise to a total of eight sets and eight corresponding rates (four for pSb5 in Table \ref{main-tab:pSb5_rates} and four for pSb1 in Table \ref{main-tab:pSb1_rates}).
Each set consists of a number of trajectories (also referred to as `runs' $i$) starting in the reactant state and ending in the product state. 
Trajectories were generated by plugging the PLUMED software package \cite{tribello2014plumed} with the DFTB+ package \cite{hourahine2007self}.
Potential energy, integrator settings and settings of the Nos\'e-Hoover thermostat were the same as in the metadynamics simulations described above. 
As biased collective variable, we used the C$_{12}$-C$_{13}$=C$_{14}$-C$_{15}$ dihedral angle $\varphi$.
Height and width of the Gaussian bias potentials, deposition rate, and bias factor, number of simulations are reported in Table \ref{main-tab:metad_parameters}.

Notice that, because DFTB+ implements deterministic dynamics (velocity-Verlet + Nos\'e-Hoover), additional care needs to be taken in generating the correct initial conditions, i.e.~to start each run from uncorrelated starting configurations and velocities according to local equilibrium in the reactant state.
Generating good initial starting states was taken care of in the equilibration phase of each separate run.
To obtain uncorrelated starting configurations for the infrequent metadynamics runs, we equilibrated the starting reactant state configuration with randomized velocities before each run.
Additionally, the total equilibration time of each run was randomized to be anywhere between 10 and 100$\,\mathrm{ps}$.
Trajectories for runs from trans to cis were terminated once a value of $\varphi\in\left[-\pi/5,\pi/5\right]$ was reached, where the molecule is definitely in the cis state.
The biased transition time $t^{(i)}_{\mathrm{t}\rightarrow \mathrm{c}}$ was then taken to be the time of the last trajectory point where the configuration can still be considered at the trans side, i.e.~the last trajectory point where $\varphi<-\pi/2$ or $\varphi>\pi/2$.
The unbiased transition times $\tau_{\mathrm{t}\rightarrow \mathrm{c},i}$ can then be calculated from eq.~\ref{main-eq:InMetaD}.
Trajectories for runs from cis to trans were stopped once a value of $\varphi<-4\pi/5$ or $\varphi>4\pi/5$ was reached, where the molecule is definitely in the trans state.
The biased transition time $t^{(i)}_{\mathrm{c}\rightarrow \mathrm{t}}$ was then taken to be the time of the last trajectory point where the configuration can still be considered at the cis side, i.e.~the last trajectory point where $\varphi\in\left[-\pi/2,\pi/2\right]$, and unbiased transition times $\tau_{\mathrm{c}\rightarrow \mathrm{t},i}$ can be calculated from eq.~\ref{main-eq:InMetaD}.
To get an idea of the actual simulation times, the minimum and maximum biased transition times for each set are given in Table \ref{tab:InMetaD_parameters}.
Notice that acceleration factors $\alpha$ were very large for our particular application of infrequent metadynamics, with numbers up to the order of $\alpha\left(t^{(i)}_{t\rightarrow c}\right) \sim 10^{14}$ in set InMetaD1 for trans to cis runs for pSb5.
The estimated mean first-passage times can then be obtained by fitting the reweighted transition times of all runs within a set to a Poisson distribution (eq.~\ref{main-eq:TCDF}), and rates can be calculated directly using eq.~\ref{main-eq:MFPT}.
Parameters for the infrequent metadynamics simulations, $p$-values of the Kolmogorov-Smirnoff test, as well as ranges of the simulated first-passage times are reported in Table \ref{tab:InMetaD_parameters}.

\begin{table*}
    \centering
    \begin{tabular}{|c|c|c|c|c|c|c|c|c|c|}
    \hline\hline
        \textbf{Molecule} & \textbf{Set} & \textbf{pace}  & \textbf{runs} & $k_{t\rightarrow c}$ [s$^{-1}$] & $k_{c\rightarrow t}$ [s$^{-1}$] & $\displaystyle\min_it^{(i)}_{t\rightarrow c}$ & $\displaystyle\max_i t^{(i)}_{t\rightarrow c}$ & $\displaystyle\min_i t^{(i)}_{c\rightarrow t}$& $\displaystyle\max_i t^{(i)}_{c\rightarrow t}$\\
         &  & [ps] & & ($p$-value) & ($p$-value) & [ns] & [ns] & [ns] & [ns]  \\
        \hline\hline
        \textbf{pSb5} & InMetaD1 & 5 & 25 & $1.94\times10^{-6}$  & $1.04\times10^{-4}$& 10.8 & 18.2 & 8.4 & 15.4 \\ 
         & &  & & (0.58) & (0.77) & &&&\\
        \textbf{pSb5} & InMetaD2 & 10 & 30 & $2.63\times10^{-6}$ & $9.22\times10^{-5}$ & 22.0 & 32.6 & 18.2 & 28.0\\ 
         & &  & & (0.93) & (0.78) & &&& \\
        \textbf{pSb1} & InMetaD1 & 5 & 25 & $2.97\times10^{-3}$  & $9.30\times10^{-3}$& 5.6 & 11.8 & 6.7 & 10.6\\ 
         & &  & & (0.77) & (0.56) & & & & \\
        \textbf{pSb1} & InMetaD2  & 10 & 30 & $3.09\times10^{-3}$ & $1.22\times10^{-2}$ & 13.6 & 20.1 & 11.8 & 19.9\\ 
         & &  & & (0.75) & (0.44) & &&& \\
         \hline\hline
    \end{tabular}
    \caption{Parameters and rates for infrequent metadynamics simulations of thermal cis-trans isomerization around the C$_{13}$=C$_{14}$ double bond in pSb5 and pSb1.
    Parameters of the metadynamics biasing can be found in Table \ref{main-tab:metad_parameters}.
    Rates are repeated from Tables \ref{main-tab:pSb5_rates} and \ref{main-tab:pSb1_rates}.
    The pace and the amount of runs used to fit the TCDF are given.
    Additionally, the minimum and maximum biased transition times are given for each set.
    For each rate calculation, the $p$-value is shown as well.
    }
    \label{tab:InMetaD_parameters}
\end{table*}
%

%
\subsection{Pontryagin and Kramers Rate Calculations}
\label{chapter:Pontryagin_Kramers}
Rates from Pontryagin and Kramers' rate equations (eqs.~\ref{main-eq:Pontryagin} and \ref{main-eq:Kramers} respectively) were calculated using free energy surfaces MetaD1 for both pSb5 and pSb1, see Fig.~\ref{main-fig:FES_diff}.
Diffusion profiles for both pSb5 and pSb1 were taken from the umbrella windows of the corresponding US3 set.
Calculation of rates from the Pontryagin equation in eq.~\ref{main-eq:Pontryagin} entails a nested integration over the free energy profile $F(\varphi)$ and the position-dependent diffusion $D(\varphi)$. 
The integrals were calculated numerically, where the inner integral was carried out starting from the barrier peak on the other side of the reactant state. 
Rates were initially calculated separately in the two possible direction, i.e.~going clockwise or counterclockwise, for both trans-to-cis as cis-to-trans isomerizations.
The actual rates are then obtained by summing:
\begin{subequations}
\label{eq:sum_rates}
\begin{align}
     k_{\mathrm{trans}\rightarrow\mathrm{cis}} &= k_{\mathrm{t}\rightarrow\mathrm{c,left}} +  k_{\mathrm{t}\rightarrow\mathrm{c,right}} \\
     k_{\mathrm{cis}\rightarrow\mathrm{trans}} &= k_{\mathrm{c}\rightarrow\mathrm{t,left}} +  k_{\mathrm{c}\rightarrow\mathrm{t,right}}
\end{align}
\end{subequations}
where $k_{\mathrm{t}\rightarrow\mathrm{c,left}}$ indicates trans-to-cis isomerization over the left free energy barrier, and analogous for $k_{\mathrm{t}\rightarrow\mathrm{c,right}}$, $k_{\mathrm{c}\rightarrow\mathrm{t,left}}$ and $k_{\mathrm{c}\rightarrow\mathrm{t,right}}$.
For calculation of Kramers' rate equation (eq.~\ref{main-eq:Kramers}), free energy barriers $F^\ddagger$ were determined directly from the free energy profile by subtracting the maximum free energy value at the corresponding peak (left or right) by the minimum value at the reactant side under consideration. 
In this way, four energy barriers per free energy surface $F^\ddagger_{\mathrm{t}\rightarrow\mathrm{c,left}}$, $F^\ddagger_{\mathrm{t}\rightarrow\mathrm{c,right}}$, $F^\ddagger_{\mathrm{c}\rightarrow\mathrm{t,left}}$ and $F^\ddagger_{\mathrm{c}\rightarrow\mathrm{t,right}}$ are obtained. 
Masses in reduced dimensions for reactant states $\mu_\mathrm{trans}$ and $\mu_\mathrm{cis}$ were calculated by running unbiased $2\,\mathrm{ns}$ runs in the corresponding states, calculating the average kinetic energy in the reduced dimension (i.e.~the dihedral angle) and comparing to the temperature using the equipartition theorem:
\begin{equation}
\label{eq:reduced_mass}
    \mu_A = \frac{k_B T}{\left< v_\varphi^2 \right>_A}.
\end{equation}
The reactant state dihedral velocities $\omega_A$ (where $A$ denotes cis or trans) can then be calculating using 
\begin{equation}
\label{eq:omega_A}
    \omega_A = \sqrt{\frac{\kappa_A}{\mu_A}}
\end{equation}
where spring constant $\kappa_A$ is obtained by fitting the free energy surface to a harmonic potential $\frac{1}{2}\kappa_A(\varphi-\varphi_A)^2$ where $\varphi_A$ corresponds to the free energy minimum at the corresponding reactant state $A$. 
Fits for the trans and cis free energy wells show close agreement with harmonic potentials at the bottom, which validates the harmonic assumptions of the reactant and product states in the formulations for Kramers' equation.
The friction coefficient $\xi$ in eq.~\ref{main-eq:Kramers} was taken to be the friction coefficient $\xi^\ddagger$ at the barrier top, which can be calculated directly from the diffusion profile using the Einstein-Stokes relation:
\begin{equation}
\label{eq:gamma_ddagger}
    \xi^\ddagger = \frac{k_BT}{\mu^\ddagger D^\ddagger}
\end{equation} 
where $D^\ddagger$ is the value of the diffusion coefficient at the barrier top and $\mu^\ddagger$ has been approximated by averaging $\mu_\mathrm{cis}$ and $\mu_\mathrm{trans}$. 
The angular frequency at the barrier top $\omega^\ddagger$ has been calculated in a similar way as at the reactant states using:
\begin{equation}
\label{eq:omega_ddagger}
    \omega^\ddagger = \sqrt{\frac{\kappa^\ddagger}{\mu^\ddagger}} 
\end{equation} 
where $\kappa^\ddagger$ was obtained using a parabolic fit to the free energy surface at the barrier top. 
Again, total rates are obtained by summing rates for both barriers as in eqs.~\ref{eq:sum_rates}.

The resulting parameters for Kramers' rate constant are reported in Tables \ref{tab:RC_based_rates_pSb5} and \ref{tab:RC_based_rates_pSb1}.

\begin{table*}
    \centering
    \begin{tabular}{ c  c  c  c  c  c}
    &  &  \multicolumn{2}{c}{trans$\rightarrow$cis} & \multicolumn{2}{c}{cis$\rightarrow$trans} \\ 
    & \textbf{units}  & $TS$ & $TS'$  & $TS$ & $TS'$ \\ 
    \hline
    \hline
    $\mu_\mathrm{A}$            & [$\mathrm{kg}.\mathrm{m}^2.\mathrm{rad}^{-2}$]
                                & $ 7.03 \times 10^{-47} $
                                & $ 7.03 \times 10^{-47} $
                                & $ 2.35 \times 10^{-47} $
                                & $ 2.35 \times 10^{-47} $
                                \\

    $D^\ddagger$              &[$\mathrm{rad}^2.\mathrm{ps}^{-1}$]
                                & $ 0.68 $
                                & $ 0.67 $
                                & $ 0.68 $
                                & $ 0.67 $\\
                                
    $\xi^\ddagger$         &[$\mathrm{ps}^{-1}$] 
                                & $ 1.31 \times 10^{2} $
                                & $ 1.31 \times 10^{2} $
                                & $ 1.31 \times 10^{2} $
                                & $ 1.31 \times 10^{2} $\\

    $\omega_\mathrm{A}$         &[$\mathrm{ps}^{-1}$]
                                & $ 4.67 \times 10^{1} $
                                & $ 4.67 \times 10^{1} $
                                & $ 6.68 \times 10^{1} $
                                & $ 6.68 \times 10^{1} $ \\
                                
    $\omega^\ddagger$         &[$\mathrm{ps}^{-1}$]
                                & $ 3.08 \times 10^{2} $
                                & $ 2.80 \times 10^{2} $
                                & $ 3.08 \times 10^{2} $
                                & $ 2.80 \times 10^{2} $ \\
                                
    $F^{\ddagger}$              &[$\mathrm{kJ}.\mathrm{mol}^{-1}$]
                                & $ 89.0 $ 
                                & $ 88.7 $
                                & $ 80.5 $
                                & $ 80.2 $\\ 
    \hline
    \hline
    \end{tabular}
    \caption{Parameters for one-dimensional rate theories calculated for $F(\varphi)$ for pSb5 using metadynamics (MetaD1) for trans$\rightarrow$cis and cis$\rightarrow$trans transitions.
    Values are given separately for transitions over the left ($TS$) and right ($TS'$) torsional barrier.
    }
    \label{tab:RC_based_rates_pSb5}
\end{table*}

\begin{table*}
    \centering
    \begin{tabular}{ c  c  c  c  c  c}
    &  &  \multicolumn{2}{c}{trans$\rightarrow$cis} & \multicolumn{2}{c}{cis$\rightarrow$trans} \\ 
    & \textbf{units}  & $TS$ & $TS'$  & $TS$ & $TS'$ \\ 
    \hline
    \hline
    $\mu_\mathrm{A}$            & [$\mathrm{kg}.\mathrm{m}^2.\mathrm{rad}^{-2}$]
                                & $ 6.80 \times 10^{-47} $
                                & $ 6.80 \times 10^{-47} $
                                & $ 2.49 \times 10^{-47} $
                                & $ 2.49 \times 10^{-47} $
                                \\

    $D^\ddagger$              &[$\mathrm{rad}^2.\mathrm{ps}^{-1}$]
                                & $ 0.62 $
                                & $ 0.64 $
                                & $ 0.62 $
                                & $ 0.64 $\\
                                
    $\xi^\ddagger$         &[$\mathrm{ps}^{-1}$] 
                                & $ 1.45 \times 10^{2} $
                                & $ 1.39 \times 10^{2} $
                                & $ 1.45 \times 10^{2} $
                                & $ 1.39 \times 10^{2} $\\

    $\omega_\mathrm{A}$         &[$\mathrm{ps}^{-1}$]
                                & $ 4.34 \times 10^{1} $
                                & $ 4.34 \times 10^{1} $
                                & $ 6.24 \times 10^{1} $
                                & $ 6.24 \times 10^{1} $ \\
                                
    $\omega^\ddagger$         &[$\mathrm{ps}^{-1}$]
                                & $ 2.08 \times 10^{2} $
                                & $ 2.26 \times 10^{2} $
                                & $ 2.08 \times 10^{2} $
                                & $ 2.26 \times 10^{2} $ \\
                                
    $F^{\ddagger}$              &[$\mathrm{kJ}.\mathrm{mol}^{-1}$]
                                & $ 78.7 $ 
                                & $ 78.9 $
                                & $ 74.8 $
                                & $ 75.0 $\\ 
    \hline
    \hline
    \end{tabular}
    \caption{Parameters for one-dimensional rate theories calculated for $F(\varphi)$ for pSb1 using metadynamics (MetaD1) for trans$\rightarrow$cis and cis$\rightarrow$trans transitions.
    Values are given separately for transitions over the left ($TS$) and right ($TS'$) torsional barrier.
    }
    \label{tab:RC_based_rates_pSb1}
\end{table*}

\subsection{Geometry Optimization using unrestricted DFT and DFT-D3}
All calculations for pSb5 and pSb1 at the DFT level were performed with the Gaussian 16 software \cite{g16} using unrestricted DFT with the B3LYP functional \cite{becke1988density,lee1988development} and 6-31G* basis set, same as in Refs.~\citenum{tajkhorshid1999influence,tajkhorshid1999role,bondar2011ground}.
All calculations were done in parallel with and without Grimme's empirical dispersion corrections with the D3 damping function \cite{grimme2010consistent}, while using the same functional and basis functions.
Adding D3 corrections in Gaussian was done by including the \textit{EmpiricalDispersion=GD3} keyword.
Calculations including D3 corrections are designated as DFT-D3 calculations.
First, full geometry optimizations were performed in the cis and trans states, where the \textit{opt=tight} keyword was used to control the convergence criterion.
Furthermore, geometry optimizations with constrained dihedral angle $\varphi$ were performed using the \textit{ModRedundant} keyword with default cutoffs for the convergence criterions of the optimization algorithm.
Gaussian uses the Berny optimization algorithm \cite{schlegel1982optimization} as default for both minimizations and optimizations.

Transition state search was performed using the Synchronous Transit-guided Quasi-Newton (STQN) method \cite{peng1993combining,peng1996using} with the \textit{opt=qst3} keyword, where the reactant and product state input configurations were chosen to be the geometry optimized structures in the trans and cis states.
Separate transition states were optimized for different directions of trans-cis rotation, i.e.~clockwise or counterclockwise.
This was done by providing an approximate transition state configuration with dihedral angle of $\varphi=90^\circ$ or $\varphi=-90^\circ$ respectively.
Gaussian performs a full thermochemical analysis including calculation of the translational, rotational and vibrational partition functions and corresponding energies and entropies \cite{ochterski2000thermochemistry}.
This allows for straightforward calculation of rates using Eyring's equation (eq.~\ref{main-eq:EyringTST}).
Rotational characteristic temperatures are calculated by Gaussian by default, and can be used to calculate the rotational contributions to the free energy difference ($\Delta F^\ddagger$, eq.~\ref{main-eq:deltaG_Eyr}), similar as was done for DFTB3.
Vibrational mode analysis was performed on the reactant and transition state structures using the \textit{freq} keyword in Gaussian.
Notice Gaussian automatically disregards the frequencies associated with the translational and rotational degrees of freedom.
The remaining frequencies can be used to calculate the vibrational contributions to the free energy differences, both fully quantum mechanical ($\Delta F^\ddagger$, eq.~\ref{main-eq:deltaG_Eyr}) as in the high-temperature limit ($\Delta F^{\ddagger,\mathrm{ht}}$, eq.~\ref{main-eq:deltaG_high_T}), same as was done for DFTB3.
Again, the imaginary frequency needs to be left out for transition states.
Subsequently, rates for high-temperature TST can be calculated according to eq.~\ref{main-eq:high_T_TST}.
Rotational and vibrational contributions to the free energy differences as well as rates can be found in Tables \ref{main-tab:pSb5_rates} and \ref{main-tab:pSb1_rates}.
\clearpage

\section{Supplementary figures}
\label{chapter:supplementary_figures}

\begin{figure*}[h]
\centering
\includegraphics[width=\textwidth]{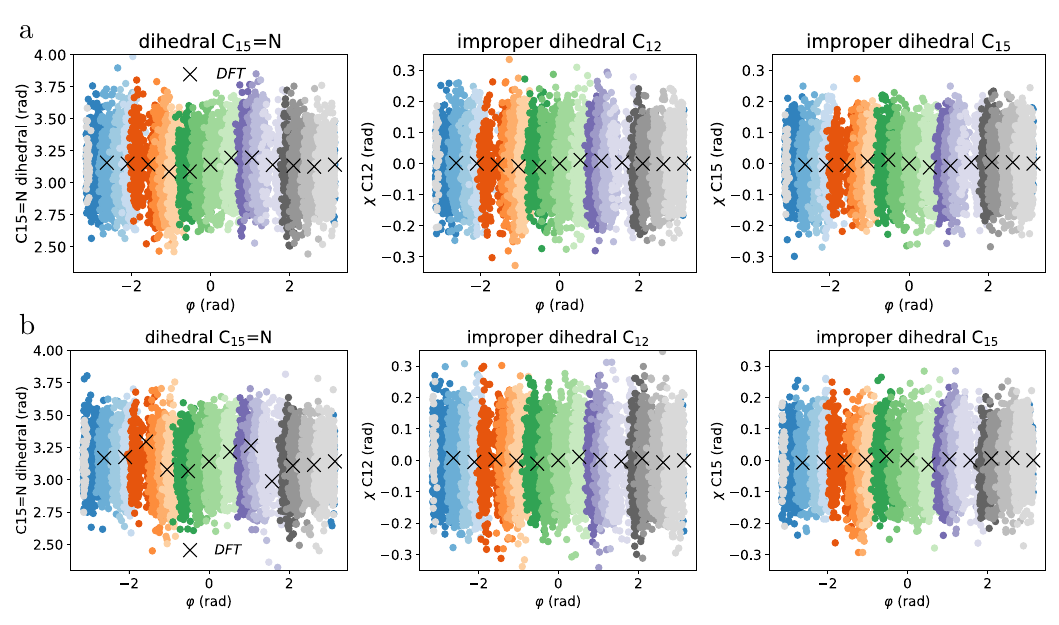}
\caption{
\textbf{a:} Sampling using DFTB3 set US3 for \textbf{pSb5} showing correlation between C$_{13}$=C$_{14}$ dihedral $\varphi$ and improper dihedrals of substituents on C$_{13}$, C$_{14}$, C$_{12}$ and C$_{15}$ atoms.
Configurations taken from 83 trajectories from harmonic restraints at different values of $\varphi$, with configurations of each trajectory colored with a different color.  
Black crosses correspond to configurations obtained from geometry optimization using unrestricted DFT/B3LYP with constrained dihedral $\varphi$. \newline
\textbf{b:} Sampling using DFTB3 set US1 for \textbf{pSb1} showing correlation between C$_{13}$=C$_{14}$ dihedral $\varphi$ and improper dihedrals of substituents on C$_{13}$, C$_{14}$, C$_{12}$ and C$_{15}$ atoms.
Configurations taken from 83 trajectories from harmonic restraints at different values of $\varphi$, with configurations of each trajectory colored with a different color.
Black crosses correspond to configurations obtained from geometry optimization using unrestricted DFT/B3LYP with constrained dihedral $\varphi$.
}
\label{fig:rest_correlation}
\centering
\end{figure*}

\clearpage 
\section{References}
\bibliographystyle{unsrt}
\bibliography{literature}

\makeatletter\@input{yy.tex}\makeatother